\documentstyle[seceq,mbf,amsfonts,amssymb]{ptptex}

\def\boldsymbol#1{{\mbf #1}}
\def\eqref#1{(\ref{#1})}
\renewenvironment{pmatrix}[1]{\left( 
\begin{array}{#1}}{\end{array}\right)}
\def\text#1{\hbox{#1}}
\def\paragraph#1{\noindent{\bf #1}}
\newenvironment{proof}{\par\noindent {\it Proof.
 }}{\hfill$\Box$\par\medskip}

\def\C{{\cal C}}

\def\LL{{\cal L}}
\def\M{{\cal M}}
\def\N{{\cal N}}

\def\linebreak{\hfill\break}

\def\mod{{\rm mod}}

\def\tend{\rightarrow}

\def\equivalent{\quad\Leftrightarrow\quad}
\def\therefore{\mbox{\setbox0=\hbox{X}\hbox{$\ldotp$}\raise0.7\ht0\hbox{$\ldotp$}\hbox{$\ldotp$}} \quad }
\def\because{\mbox{\setbox0=\hbox{X}\raise0.7\ht0\hbox{$\ldotp$}\hbox{$\ldotp$}\raise0.7\ht0\hbox{$\ldotp$}}\kern0pt }

\def\r#1{{\rm #1}}

\def\bm#1{\boldsymbol{#1}}

\def\maps{\rightarrow}
\def\mapsnamed#1{\stackrel{#1}{\longrightarrow}}
\def\Set#1{\left\{#1\right\}}
\def\SetDef#1#2{\left\{#1 \mid #2\right\}}

\def\Frac(#1/#2){\left(\frac{#1}{#2}\right)}

\def\ad{{\rm ad}}
\def\Tr{{\rm Tr}}
\def\Tp#1{\,{}^t\! #1}

\def\Aut{{\rm Aut}}

\def\Eq#1{\begin{equation} #1 \end{equation}}
\def\Eqr#1{\begin{eqnarray} #1 \end{eqnarray}}

\def\Bitm{\begin{itemize}}
\def\Eitm{\end{itemize}}
\newtheorem{proposition}{Proposition}[section]

\def\NN{{{\mathbb N}}}
\def\ZR{{{\mathbb Z}}}

\def\RF{{{\mathbb R}}}
\def\CF{{{\mathbb C}}}

\def\sdp{\rtimes}
\def\Mono{{\rm Mono}}
\def\DiffG{{\rm Diff}}
\def\Aut{{\rm Aut}}

\def\LieA{{\LL}}
\def\Ad{{\rm Ad}}

\def\Isom{{\rm Isom}}
\def\Nil{\hbox{\it Nil}}
\def\Sol{\hbox{\it Sol}}
\def\PSL{\hbox{\it PSL}}
\def\OG{\hbox{\it O}}
\def\SO{\hbox{\it SO}}

\def\ISO{\hbox{\it ISO}}
\def\IO{\hbox{\it IO}}
\def\IGL{\hbox{\it IGL}}
\def\GL{\hbox{\it GL}}
\def\SL{\hbox{\it SL}}
\def\TSL{\widetilde{\SL_2\RF}}
\def\HSL{\hbox{\it TSL}}
\def\SU{\hbox{\it SU}}
\def\DG{\hbox{\it D}}
\def\BTI{{\rm I}}
\def\BTII{{\rm II}}
\def\BTIII{{\rm III}}
\def\BTIV{{\rm IV}}
\def\BTV{{\rm V}}
\def\BTVI{{\rm VI}}
\def\BTVII{{\rm VII}}
\def\BTVIII{{\rm VIII}}
\def\BTIX{{\rm IX}}
\def\Gmax{ G_{\rm max}}

\title{Phase Space of Compact Bianchi Models with Fluid}

\author{Hideo Kodama}

\inst{Yukawa Institute for Theoretical Physics\\
Kyoto University, Kyoto 606-8502, Japan}

\abst{
The structure of phase space is determined for spatially compact and 
locally homogeneous universe models with fluid. Analysis covers 
models with all possible space topologies except for those covered 
by $S^3$, $H^3$ or $S^2\times\RF$ which have no moduli freedom. We 
show that space topology significantly affects the number of 
dynamical degrees of freedom of the system. In particular, we give a 
detailed proof of the result that for the systems modeled on the 
Thurston types $H^2\times\RF$ and $\TSL$, which have locally the 
Bianchi type III or VIII symmetry, the number of dynamical degrees of 
freedom increases without bound when the space topology becomes more 
and more complicated, which was first pointed out by Koike, Tanimoto 
and Hosoya in an incomplete form. }

\begin{document}

\maketitle

\section{Introduction}

Bianchi models provide the basis of modern cosmological models as the  
zeroth approximation. They have been also frequently used as the 
simplest models called minisuperspace models in the study of quantum 
gravity and quantum cosmology. In many of the investigations in these 
fields, however, topological features of models were neglected, and 
models with the simplest space topology were treated. Such treatments 
are allowed if one is interested only in local structures and dynamics 
of models. However, if one wants to discuss global structures and 
dynamics of models, one must take into account the effects of space 
topology. In particular, when one constructs a minisuperspace model in 
quantum gravity, one has to use a Bianchi model with compact space in 
order to make the canonical variables well-defined.

When one considers Bianchi models with non-trivial topologies, various 
new features arise. First, space topology is strongly restricted by the 
Bianchi type. In particular, there exists no spatially compact model 
which has the local symmetry of the Bianchi type $\BTIV$ or 
$\BTVI_h$($h\not=0,-1$)%
\cite{Ashtekar.A&Samuel1991,Fujiwara.Y&Kodama&Ishihara1993}. 
Second, there appear new dynamical degrees of freedom called the 
moduli, which describe the degrees of freedom in deforming the global 
structure of space preserving the local structure and the 
topology%
\cite{Ashtekar.A&Samuel1991,Koike.T&Tanimoto&Hosoya1994}. 
This is a higher-dimensional analogue of the well-known moduli 
freedom for the flat 2-dimensional torus. This appearance of the 
moduli freedom is very important in the arguments of minisuperspace 
models based on Bianchi models, because it significantly alters the 
canonical structure and dynamics of the systems. From this 
standpoint, we developed a formalism to construct the 
diffeomorphism-invariant phase space and determine its canonical 
structure for locally homogeneous systems on compact closed 
3-manifolds on the basis of the Thurston classification of the 
maximal geometries with compact quotients, and applied it to vacuum 
systems of the Thurston types $E^3,\Nil$ and $\Sol$, corresponding 
to the Bianchi types $\BTI,\BTVII_0,\BTII$ and $\BTVI_0$, in our 
previous paper (Paper I)\cite{Kodama.H1998}.

One of the main purpose of the present paper is to extend the analysis 
in Paper I to systems on spaces of the Thurston types $H^2\times\RF$ 
and $\TSL$ corresponding to the Bianchi types $\BTIII$ and $\BTVIII$. 
By this the analysis of the moduli freedom for all compact Bianchi 
models is completed, because the compact Bianchi models of the types 
$\BTV,\BTVII_h$ and $\BTIX$ have no moduli 
freedom%
\cite{Fujiwara.Y&Kodama&Ishihara1993,%
Koike.T&Tanimoto&Hosoya1994,%
Barrow.J&Kodama2001,Barrow.J&Kodama2001A}. 
Another purpose of the present paper is to extend the analysis of the 
vacuum systems in Paper I to systems with perfect fluid. Its main 
motivation is the generality problem of Bianchi models. As is well 
known, among the Bianchi models with simply connected space, the types 
$\BTVI_h(h\not=0,-1),\BTVII_h(h\not=0),\BTVIII$ and $\BTIX$ are the 
most generic models both for the vacuum and the fluid cases in the 
sense that the parameter count, i.e,. the dimension of the solution 
space, for them are the largest: $4$ for the vacuum case and $8$ for 
the fluid case%
\cite{Ellis.G&MacCallum1969,Siklos.S1978}. Further, the Bianchi type 
$\BTVII_0$ is more general than the type $\BTI$ in the same sense. 
This generality argument based on the parameter count is the 
starting point in the isotropization problem in the spatially 
homogeneous 
cosmology%
\cite{Collins.C&Hawking1973,Doroshkevich.A&Lukash&Novikov1973,%
Lukash.V1975,Barrow.J1982,Barrow.J&Sonoda1986,%
Wainwright.J&Hancock&Uggla1999,Nilsson.U&Hancock&Wainwright2000}, 
and also plays the basic role in the mathematical cosmology%
\cite{Wainwright.J&Ellis1997B}. 
However, the situation drastically changes when the space is 
compactified. In particular, the type $\BTI$ becomes the most generic 
among the types $\BTI,\BTII,\BTV,\BTVI_0,\BTVII_0,\BTVII_h$ and $\BTIX$ 
for the vacuum case as was shown in Paper I. Hence it is interesting to 
see how the introduction of fluid alters the parameter count and to 
extend the analysis to the Bianchi type $\BTIII$ and $\BTVIII$. Partial 
answers to these problems were published in the joint work by the 
author and John D. 
Barrow\cite{Barrow.J&Kodama2001,Barrow.J&Kodama2001A}. In the 
present paper we give complete answers with detailed accounts on 
their derivations.

The paper is organized as follows. In the next section, we give the 
outline of the formalism developed in Paper I to determine the 
diffeomorphism-invariant phase space of a locally homogeneous system 
on a compact and closed 3-manifold. Then we apply it to the fluid 
model on orientable compact 3-manifolds of all possible  topologies 
with non-trivial moduli freedom in the order of the Thurston types 
$E^3$(\S3), $\Nil$(\S4), $\Sol$(\S5), $H^2\times\RF$(\S6) and 
$\TSL$(\S7). Since the fluid system cannot be put in the canonical 
form, we do not discuss the canonical structure of the phase space. 
Section 8 is devoted to summary and discussion.

\paragraph{Notations:} Throughout the paper, $R(\theta)$ and  
$R_i(\theta)$($i=1,2,3$) denote the rotation matrix of angle $\theta$ 
in the 2-plane and that around the $i$-th coordinate axis in the 
3-dimensional Euclidean space, respectively. $R_i(\pi)$ is often simply 
written as $I_i$. The block diagonal matrix with block matrices $A, B, 
\cdots$ is denoted as 
$D(A,B,\cdots)$. For a three dimensional object $X$ such as a vector 
and a square matrix, its two dimensional part is often written as $\hat 
X$. In particular, for a 3-vector $\bm{v}=(v^1,v^2,v^3)$, 
$\hat{\bm{v}}=(v^1,v^2)$. This symbol is also used to denote the 
3-vector $(v^1,v^2,0)$ if there is no possibility of confusion. 
Further, the row vector expression and the column vector expression 
of a vector with the same components are often denoted by the same 
symbol. Finally, the linear transformation $\bm{x}\mapsto 
A\bm{x}+\bm{a}$ is often denoted as $(\bm{a},A)$.

\section{General Framework}

As in Paper I, we define the spatially locally homogeneous system as a 
globally hyperbolic spacetime whose universal covering is spatially 
homogeneous, and its invariance group $G$ as the invariance group of 
its universal covering. We often call such a system with compact space 
simply {\it a compact Bianchi model}. Under this definition, a compact 
Bianchi model is specified by topology of the compact space, an 
invariance group of transformations on the universal covering 
spacetime, and a set of fields on it. In the present paper, we mainly 
consider systems with perfect fluid, for which the fields consist of 
the spacetime metric and the 4-velocity and density of each component 
of fluid. We further restrict considerations to spacetimes with 
orientable spaces as in Paper I. 

\subsection{Compact Bianchi models}

\begin{table}
\caption{\label{table:ThurstonBianchi}Thurston types and Bianchi types}
\begin{tabular}{lllll}
\hline
\hline
{\bf Space} & $\Gmax $ & $G^+_\r{max}$ & $G_\r{min}$ 
& {\bf Bianchi type} \\\hline
\\
$E^3$ 
& $\IO(3)$ & $\ISO(3)$ & $\RF^3(=\BTI)$ 
&$\BTI$ \\
&         &          & $\BTVII_0^{(\pm)}$ 
&$\BTVII_0$ \\
\\
$S^3$ 
& \OG(4) & $\SO(4)$ & $\SU(2)(=\BTIX)$ 
&$\BTIX$ \\
\\
$H^3$ 
& $\OG_+(3,1)$& $\PSL_2\CF$ & $\Gmax^+$ 
&$\BTV$, $\BTVII_h (h\not=0)$ \\  
\\
$\Nil$   
& $\Nil\sdp\OG(2)$ &$\Gmax$ & $\Nil(=\BTII)$ 
& $\BTII$ \\
\\
$\Sol$  
& $\Sol\sdp\DG_4$  & $\Sol\sdp \DG_2$ & $\Sol(=\BTVI_0)$ 
& $\BTVI_0$ \\
\\
$\widetilde{\SL_2\RF}$ 
& $\widetilde{\SL_2\RF}\sdp \OG(2)$ & $\Gmax$ &  
$\widetilde{\SL_2\RF}(=\BTVIII)$& $\BTVIII$ \\
&&&$\Gmax$ & $\BTIII$\\
\\
$H^2\times E^1$ 
& $\OG_+(2,1)\times \IO(1)$ & $(\PSL_2\RF\times\RF)\sdp\ZR_2$ & 
$\PSL_2\RF\times\RF$ & $\BTIII$ \\
\\
$S^2\times E^1$ 
& \OG(3)$\times$ $\IO(1)$ & $(\SO(3)\times\RF)\sdp\ZR_2$ & 
$\SO(3)\times \RF$ &Kantowski-Sachs \\
&&&& models\\
\hline\hline
\end{tabular}
\end{table}

At present, topologies of generic compact and closed 3-manifolds are 
not classified yet, but those which admit locally homogeneous metrics 
are well classified. According to Thurston%
\cite{Thurston.W1979B,Thurston.W1982,Scott.P1983}, 
such a 3-manifold $\Sigma$ is specified by a maximal geometry 
$(\tilde\Sigma,\Gmax)$ on its universal covering $\tilde\Sigma$ and a 
discrete subgroup $K$ of $\Gmax$ isomorphic to the fundamental group 
$\pi_1(\Sigma)$. Here, $\Gmax$ is a maximal 
transformation group on $\tilde \Sigma$ among all possible isometry 
groups of metrics on $\tilde\Sigma$. There exist only 8 maximal 
geometries for which the quotient $\tilde \Sigma/K$ is a smooth compact 
closed 3-manifold for some $K\subset \Gmax$; $E^3, S^3, H^3, \Nil, 
\Sol, 
H^2\times\RF, \TSL$ and $S^2\times\RF$(see Table 
\ref{table:ThurstonBianchi}). Possible choices of the discrete subgroup 
$K$ are completely determined for all Thurston types except for 
$H^3$\cite{Scott.P1983}. Their explicit expressions for the Thurston 
types $E^3, \Nil$ and $\Sol$ were given in Paper I and those for 
$H^2\times\RF$ and $\TSL$ will be given in \S\ref{sec:FundamentalGroup} 
and 
\S\ref{sec:FundamentalGroup2}.

The invariance group $G$ of a locally homogeneous system contains some 
Bianchi group by definition, and at the same time it is a subgroup of 
one of the 8 Thurston type maximal symmetry groups. The latter maximal 
geometry is uniquely determined by the topology of the space. However, 
the invariance group does not uniquely determine the Thurston type. 
Hence the correspondence between the Thurston types and the Bianchi 
types is not one-to-one as shown in Table \ref{table:ThurstonBianchi}. 
That is, the Thurston type $E^3$ contains the Bianchi types $\BTI$ and 
$\BTVII_0$ as subgeometries and the 
Thurston type $H^3$ contains the Bianchi types $\BTV$ and 
$\BTVII_h$($h\not=0$). Further, the Bianchi type III belongs to the two 
Thurston types $H^2\times\RF$ and $\TSL$. The former degeneracy in the 
correspondence implies that locally homogeneous systems with different 
Bianchi symmetries can be implemented on spacetimes with the same space 
topology. The same feature appears in non-compact Bianchi models, in 
which the eight Bianchi types, $\BTI\sim\BTVIII$, are realized on the 
same space $\RF^3$. Compared with this non-compact case, the degeneracy 
in the 
correspondence between the topology and the Bianchi type for the 
compact case is rather small. In other words, the space topology 
strongly restricts the local symmetry of a system in compact Bianchi 
models. On the other hand, the degeneracy in the correspondence 
concerning the Bianchi type $\BTIII$ is a new feature for compact 
Bianchi models, and indicates the possibility that two spacetimes 
with different space topologies may have the same local symmetry. 
However, since such systems always have symmetries larger than the 
Bianchi $\BTIII$ symmetry as will be shown in 
\S\ref{sec:H^2xR:InvG}, such possibility is not realized. 

Another important feature in the correspondence between the Thurston 
types and the Bianchi types is the fact that the Bianchi groups $\BTIV$ 
and $\BTVI_h$ do not belong to any Thurston type as shown in Table 
\ref{table:ThurstonBianchi}. This implies that the Bianchi types 
$\BTIV$ and $\BTVI_h$ cannot be compactified. From a geometrical 
viewpoint, this follows from  the fact that each of $\BTIV$ and 
$\BTVI_h$ is really the maximal connected subgroup of a maximal 
symmetry, but any transformation preserving this symmetry group leaves 
invariant a vector field with non-vanishing divergence (see the 
discussion in \S\ref{sec:H^2xR:InvG} on this special vector). For a 
similar reason, compact Bianchi models with the type $\BTV$ or 
$\BTVII_h$($h\not=0$) symmetry are always isotropic, hence their spaces 
are quotients of the constant curvature space $H^3$. Further, such 
space has no moduli freedom due to Mostow's rigidity theorem%
\cite{Fujiwara.Y&Kodama&Ishihara1993,Barrow.J&Kodama2001,%
Barrow.J&Kodama2001A}. 
Hence the phase space of a system on such a space is trivial.  Spaces 
obtained as quotients of $S^3$ do not have the moduli freedom either, 
although they allow local anisotropy. The phase space of a system on 
such a space coincides with that for its universal covering.%
\footnote{A compact quotient of $S^2\times\RF$ does not have the moduli 
freedom in our definition given in \S\ref{sec:InvPhaseSpace}. This 
definition is different from that adopted in Ref. 
\citen{Koike.T&Tanimoto&Hosoya1994}. In their definition, compact 
quotients of $S^2\times\RF$ have non-trivial moduli degree of 
freedom. Although it is stated in Ref. 
\citen{Koike.T&Tanimoto&Hosoya1994} that $S^2\times S^1$ has two 
moduli degrees of freedom in their definition, it is not not 
correct. The correct number is one.}

From these observations we see that the phase space of a compact 
Bianchi model has a non-trivial structure different from that of its 
universal covering only for the Bianchi types 
$\BTI,\BTII,\BTIII,\BTVI_0,\BTVII_0$ and $\BTVIII$, or equivalently for 
the Thurston types $E^3$, $\Nil$, $\Sol$, $H^2\times\RF$ and $\TSL$.

\subsection{Invariant phase space}\label{sec:InvPhaseSpace}

Let $(M,g,\cdots)$ be a compact Bianchi model with matter satisfying 
the Einstein equations. Then its universal covering $(\tilde 
M,\tilde g,\cdots)$ is invariant under some transformation group $G$ 
whose orbits are spatial hypersurfaces $\tilde \Sigma(t)$ in $\tilde 
M$ where $t$ is a label representing the time coordinate. In this 
paper we only consider the case in which $G$ contains some subgroup 
$G_s$ which acts simply transitively on each $\tilde \Sigma(t)$. As 
is well-known, we can then always find the space coordinates 
$\bm{x}$ of $\tilde M$  such that the action of $G$ is represented 
by transformations of the $\bm{x}$ coordinates independent of $t$. In 
other words, $\tilde M$ is written as $\tilde\Sigma\times\RF$, where 
$\tilde\Sigma\approx \tilde\Sigma(t)$, and the action of $G$ on 
$\tilde M$ is induced from that on $\tilde \Sigma$.  Here, the 
original spacetime $M$ is represented as the quotient $\tilde M/K$ 
of $\tilde M$ by a discrete subgroup $K$ of $G$. Hence $M$ is 
written as $\Sigma\times \RF$ with $\Sigma=\tilde\Sigma/K$. Since we 
are considering a space-orientable spacetime, $\Sigma$ is assumed to 
be an orientable 3-manifold, and $G$ is contained in the 
orientation-preserving diffeomorphism group of $\Sigma$, 
$\DiffG^+(\Sigma)$.

In terms of an appropriate set of variables on $\tilde\Sigma$, 
$X(t)$, constructed from the values of the metric, the matter 
variables and their time derivatives on $\Sigma(t)$, the Einstein 
equations are represented as a set of evolution equations, 
first-order in time, supplemented with the Hamiltonian and the momentum 
constraints. Since $X(t)$ is invariant under $G_s$, its expansion 
coefficients in terms of the invariant basis with respect to $G_s$ are 
functions only of $t$, and the Einstein equations reduce to a set of 
first-order ordinary differential equations for them. Hence, the whole 
solution on $\tilde M$ is uniquely determined by the 
initial data for $X$ at some time $t_0$. Therefore, when the pair 
$(X,K)$ of a $G$-invariant initial data set $X$ on $\tilde \Sigma$ 
satisfying the constraints and a discrete subgroup $K$ of $G$ such that 
$\tilde\Sigma/K\approx \Sigma$ is given, a locally $G$-homogeneous 
solution to the Einstein equations on $M$ is uniquely specified. This 
specification, however, is not complete because two solutions connected 
by a diffeomorphism should be identified physically in general 
relativity. 

As discussed in Paper I, when the topology of $\Sigma$ is given, a 
maximal geometry $(\tilde \Sigma,\Gmax)$ giving $\Sigma$ as a compact 
quotient is uniquely determined up to diffeomorphism, and when a 
standard realization of the maximal geometry is fixed, the invariance 
group $G$ can be always chosen as a subgroup of the fixed maximal 
symmetry group $\Gmax$ by choosing an appropriate projection from 
$\tilde \Sigma$ to $\Sigma$. Further, if $G'$ is a subgroup of $\Gmax$ 
which is conjugate to $G$ in $\DiffG^+(\tilde\Sigma)$, i.e., if there 
is a diffeomorphism $f\in\DiffG^+(\tilde\Sigma)$ such that 
$G'=fGf^{-1}$, a $G'$-invariant system and a $G$-invariant system are 
equivalent. Hence, we can fix $G$ to one representative group in its 
conjugate class. 

Under this setting, we only have to consider diffeomorphisms preserving 
the invariance group $G$ and the decomposition of the spacetime $\tilde 
M$ into space and time $\tilde\Sigma\times\RF$ in the identification of 
solutions. They are further classified into two types. First one is the 
one-parameter family of diffeomorphisms corresponding to time 
translations. Since the evolution equations for $X$ are first-order in 
time and autonomous, time translations are represented as a 
one-dimensional transformation group in the initial data space.  Second 
one is the diffeomorphisms which preserve each $G$-orbit, i.e., 
$\tilde\Sigma(t)$. Such diffeomorphisms are induced from $f$'s on 
$\tilde\Sigma$ such that $fGf^{-1}=G$. We call such a transformation 
{\it a homogeneity-group preserving diffeomorphism} or simply a HPD of 
$G$. All HPDs of $G$ form a group $\N(G)$ called {\it the normalizer of 
$G$}, whose subgroup consisting of orientation-preserving HPDs is 
denoted by $\N^+(G)$. In the initial data space, each HPD $f$ acts on 
$(X,K)$ as $(f_*X,fKf^{-1})$. We call the space obtained from the space 
of the initial data $(X,K)$ by the identification in terms of HPDs {\it 
the diffeomorphism-invariant phase space for the system $(\Sigma,G)$}, 
and denote it by $\Gamma_{\rm inv}(\Sigma,G)$. Here, we only impose the 
diffeomorphism constraints on $X$ to construct $\Gamma_{\rm 
inv}(\Sigma,G)$ as in Paper I, because the canonical structure always 
becomes degenerate in the subspace determined by the Hamiltonian 
constraint if the canonical description is possible as in the vacuum 
case.  This degeneracy is removed only by imposing a gauge-condition on 
the time coordinate. Hence, $\Gamma_{\rm inv}(\Sigma,G)$ is expressed 
as%
\Eq{
\Gamma_{\rm inv}(\Sigma,G)=\left(\Gamma_D(\tilde\Sigma,G)\times 
\M(\Sigma,G)\right)/\N^+(G),
\label{InvPhaseSpace:general}}
where $\Gamma_D(\tilde\Sigma,G)$ is the set of $G$-invariant 
covering data $X$ on $\tilde\Sigma$ satisfying the diffeomorphism 
constraints, and $\M(\Sigma,G)$ is the set of subgroups $K$ of $G$ such 
that $\tilde\Sigma/K\approx \Sigma$.%
\footnote{Some of these notations are different from those adopted in 
Paper I. There, $\Gamma_{\rm inv}(\Sigma,G), \Gamma_D(\tilde\Sigma,G)$ 
and $\N^+(G)$ were denoted by $\Gamma^+_{\rm LH,inv}(\Sigma,G), 
\Gamma^+_{\rm H,D}(\tilde \Sigma,G)$ and ${\rm 
HPDG}^+(\tilde\Sigma,G)$, respectively. Further, the symmetry group 
$G$ of a system was regarded as an abstract group, and $G$ and its 
realization $\tilde G$ as a transformation group on $\tilde\Sigma$ 
were distinguished in Paper I. We do not do such distinction in the 
present paper, and regard $\tilde G$ as the invariance group of the 
locally homogeneous system, denoting it simply as $G$.} 
In order to determine the dimension of the space of physically distinct 
solutions, $N_s$, we must subtract two from the dimension of 
$\Gamma(\tilde \Sigma, G)$, taking into account the identification by 
time translations and the Hamiltonian constraint:
\Eq{
N_s(\Sigma,G)=\dim\Gamma_{\rm inv}(\Sigma,G)-2.
}

As mentioned above, the homogeneous data set $X$ on $\tilde \Sigma$ is 
represented by a set of components in an invariant basis 
$\chi^I$($I=1,2,3$) with respect to a Bianchi group $G_s$. For fluid 
systems considered in the present paper, they are given by the matrix  
representing the space metric $Q=(Q_{IJ};1\le I,J\le3)$, its conjugate 
momentum $P=(P^{IJ};1\le I,J\le3)$, and the spatial velocity $u_I(1\le 
I \le 3)$ and the energy density $\rho$ of the fluid. For the invariant 
basis satisfying the Mauer-Cartan equation
\Eq{
d\chi^I=-\frac{1}{2}C^I{}_{JK}\chi^J\wedge\chi^K,
}
the momentum constraints are expressed in terms of these variables as
\Eq{
H_I\equiv 2C_JP^J_I +2C^K{}_{JI}P^J_K+c u_I=0,
\label{MomentumConstraints:general}}
where $C_I=C^J{}_{IJ}, P^I_J=P^{IK}Q_{KJ}$, and $c$ is expressed as 
$c_0 (\rho+p)u_0$ in terms of a constant $c_0$ and the total energy 
density $\rho$, the total pressure $p$ and the time component of the 
center of mass 4-velocity $u_0$ of the fluid.

When we construction the moduli space $\M(\Sigma, G)$ in the following 
sections, we start from the set of monomorphisms $\psi: 
\pi_1(\Sigma)\maps G$ whose image $K$ is a discrete subgroup consisting 
of transformations without a fixed point, $\Mono_*(\pi_1(\Sigma),G)$. 
Then, since $\psi_1,\psi_2\in \Mono_*(\pi_1(\Sigma),G)$ give the same 
$K$ if and only if they are related by an automorphism of $\pi_1(M)$, 
i.e., a modular transformation, $\M(\Sigma,G)$ is given by
\Eq{
\M(\Sigma,G)=\Mono_*(\pi_1(\Sigma),G)/\text{Modular\ transformations}.
}
Further, in order to parametrize the invariant phase space, we select a 
subspace of $\Mono_*(\pi_1(\Sigma),G)$, called {\it the reduced moduli 
space} or {\it the moduli sector of the invariant phase space}, which 
intersects with each $\N^+(G)$-orbit at a single point, and denote it 
by $\M_0(\Sigma,G)$. Let $\N_{\Sigma0}(G)$ be the istropy group of the 
action of $\N(G)$ on  $\M_0(\Sigma,G)$, which consists of 
transformations in $\N^+(G)$ induced from those in the maximal 
connected subgroup of $\DiffG^+(\Sigma)$. Then, the invariant phase 
space is expressed as
\Eq{
\Gamma_{\rm inv}(\Sigma,G)=\left(
\Gamma_{\rm dyn}(\Sigma,G)\times\M_0(\Sigma,G)
\right)/H_{\rm mod},
\label{InvPhaseSpace:reduced}}
where 
\Eq{
\Gamma_{\rm dyn}(\Sigma,G)=\Gamma_D(\tilde\Sigma,G)/\N_{\Sigma0}(G),}
and $H_{\rm mod}$ is the discrete group consisting of the 
combinations of HPDs and modular transformations which map 
$\M_0(\Sigma,G)$ onto itself. We call $\Gamma_{\rm dyn}(\Sigma,G)$ 
{\it the dynamical sector of the invariant phase space} because the 
variables describing the moduli sector $\M_0(\Sigma,G)$ become 
constants of motion (see Theorem 2.3 and the arugment following it 
in Paper I). In the present paper, when $(\Gamma_{\rm 
dyn}\times\M_0)/H_{\rm mod}$ is isomorphic to $\Gamma_{\rm 
dyn}\times(\M_0/H_{\rm mod})$, we write $\M_0(\Sigma,G)/H_{\rm mod}$ 
simply as 
$\M_0(\Sigma,G)$. As we will see in the following sections, 
$\N_{\Sigma0}(G)$ becomes trivial in many cases for an appropriate 
choice of $\M_0(\Sigma,G)$. 
  
To summarize, the invariant phase spaces for compact Bianchi models 
whose space $\Sigma$ is modelled on a maximal geometry $(\tilde 
\Sigma,\Gmax)$ is determined by the following steps:
\begin{itemize}
\item[1.] Determine $\Mono_*(\pi_1(\Sigma),\Gmax)$. 
\item[2.] Make the list of all conjugate classes of possible invariance 
groups  which contain a simply transitive subgroup and are contained in 
$\Gmax$, and select a representative element $G$ from each conjugate 
class.
\item[3.] Determine the normalizer $\N^+(G)$ of each representative 
invariance group $G$.
\item[4.] Pick up an appropriate reduced moduli space $\M_0(\Sigma,G)$ 
and determine the isotropy group $\N_{\Sigma0}(G)$ and the discrete 
transformation group $H_{\rm mod}$ consisting of modular$*$HPD 
transformations preserving $\M_0(\Sigma,G)$.
\item[5.] Construct the dynamical sector $\Gamma_{\rm dyn}(\Sigma,G)$ 
taking account of the momentum constraints.
\end{itemize}
Among these steps, the most cumbersome ones are the second and the 
third. In Appendix \ref{appendix:normalizer}, we have given some 
propositions which are useful in executing these tasks.

Finally we give some comments on notations. In all of the cases 
cosidered in the present paper, the simply connected space 
$\tilde\Sigma$ of each maximal geometry can be identified with some 
group $G_s$ which acts simply transitively on it. Under this 
identification, the group structure of $G_s$ naturally defines two 
types of transformations on $\tilde\Sigma$. First one is induced from 
the left multification of a fixed element $a$ in $G_s$:
\Eq{
L_a: G_s\ni g \mapsto ag \in G_s,
}
which we call {\it a left transformation}. Second one is {\it a right 
transformation} induced from the right multiplication
\Eq{
R_a: G_s\ni g \mapsto ga \in G_s.
}
$L_{a}$ and $R_{b}$ commute with each other for any 
$a,b\in G_s$. We denote the group of left transformations and that of 
right transformations by attaching the subscripts $L$ and $R$ to the 
group symbol as $G_L$ and $G_R$, respectively. These two transformation 
groups are isomorphic to the original group $G_s$. Throughout the 
paper, we regard $G_s$ itself as the left transformation group, and we 
omit the subscript $L$ when the distinction is not important.

\section{$E^3$}

\subsection{Maximal geometry}

The maximal geometry $E^3$ is given by the isometry group of the 
standard 3-dimensional Euclidean space
\Eq{
ds^2=dx^2+dy^2+dz^2,
}
that is $\IO(3)$ on $\RF^3$. Its orientation preserving part is 
given by $\ISO(3)$, and its Lie algebra is generated by
\Eq{
T_i=\partial_i,\ 
J_i=\epsilon_{ijk} x^j\partial_k
}
with the commutation relations
\Eqr{
&& [\bm{a}\cdot T,\bm{b}\cdot T]=0,\nonumber\\ 
&& [\bm{a}\cdot J,\bm{b}\cdot T]=-(\bm{a}\times\bm{b})\cdot T,
\nonumber\\ 
&& [\bm{a}\cdot J,\bm{b}\cdot J]=-(\bm{a}\times\bm{b})\cdot J.
}
%

\subsection{Invariance groups}

\begin{table}
\caption{\label{table:R3:normalizer} Normalizers for the invariance 
groups $G$ containing $\RF^3$. $Oct$ is the 
octahderal group whose element is given by the matrix $B_{ij}=\pm 
\delta_{j,\sigma(i)}$ for some permutation $\sigma$ of 
$(1,2,3)$.}\begin{tabular}{lcll}
\hline\hline
$G$ &o/u& $f\in \N(G)$ &$fgf^{-1}$ \\\hline
&&&\\
$\RF^3$       &o
& $(\bm{a},A)\in \IGL(3)$ 
& $fL_{\bm{c}}f^{-1}=L_{A\bm{c}}$ \\
$\RF^3\sdp\Set{1,-1}$ &u
& ibid  
& $f(-1)f^{-1}=(-1)L_{-2\bm{a}}$ \\
$\RF^3\sdp\Set{1,I_3}$ & o
& $A=D(\hat A,b)$ 
& $fI_3f^{-1}=I_3L_{-2\hat{\bm a}}$\\
$\RF^3\sdp\Set{\pm1,\pm I_3}$& u
& ibid   
& ibid \\
$\RF^3\sdp D_2$ & o 
& $A=D(p,q,r)B;$
& $fI_if^{-1}=I_{\sigma^{-1}(i)} L_{\bm{c}_i};$\\
&
&$B\in Oct,\ p,q,r>0 $  
& $\bm{c}_i=((I_{\sigma^{-1}(i)}-1)\bm{a}$\\ 
$\RF^3\sdp D_4$ & u 
& ibid 
& ibid\\
&&&\\
$\ISO(2)$       &o
& $A=\Set{1,I_1}R_3(\theta)$
& $f(\bm{c},R_3(\phi))f^{-1}=(\bm{c}',R_3(\pm\phi));$ \\
&
&$\quad \times D(k_1,k_1,k_2)$& 
$\bm{c}'=(1-R_3(\pm\phi))\bm{a}+A\bm{c}$\\
$\IO(2)$ &u
& ibid 
& $f(-1)f^{-1}=(2\bm{a},-1)$ \\
&&&\\
$\ISO(2)\sdp\Set{1,I_1}$ & o
& ibid 
& $fI_1f^{-1}=((1-B)\bm{a},B);$ \\
&&& $B=I_1R_3(\mp2\theta))$\\
$\IO(2)\sdp\Set{1,I_1}$& u
& ibid  
& ibid \\
&&&\\
$\ISO(3)$       &o
& $A=kO;$
& $f(\bm{c},R)f^{-1}=(kO\bm{c}+(1-B)\bm{a},B);$ 
\\
&
&$k>0,O\in \OG(3)$
& $B=ORO^{-1}$
\\
$\IO(3)$ &u
& ibid 
& ibid
\\
&&&\\
\hline\hline
\end{tabular}
\end{table}

As is shown in Table \ref{table:ThurstonBianchi}, $\IO(3)$ contains 
the two simply transitive groups of the Bianchi types $\BTI$ and 
$\BTVII_0$, up to conjugations with respect to $\N(\IO(3))$. 

The Bianchi type $\BTI$ subgroup $\RF^3$ is 
generated by $\xi_i=T_i$($i=1,2,3$) and the invariant basis is given 
by
\Eqr{
&& \chi^1=dx,\ \chi^2=dy,\ \chi^3=dz;
\label{InvariantBasis:R3}\\
&& d\chi^1=0,\ d\chi^2=0,\ d\chi^3=0.
}
For the translation $L_{\bm{a}}: \bm{x} \maps \bm{x}+\bm{a} 
(\bm{a}\in\RF^3)$, the condition $fL_{\bm{a}}f^{-1}\in \RF^3$ is 
equivalent to the condition $f(\bm{x}+\bm{a})=f(\bm{x})+\bm{b}$ for 
any $\bm{x}$ and $\bm{a}$, where $\bm{b}$ depends only on $\bm{a}$. 
From this it follows that $f\in \N(\RF^3)$ is given by the 
linear transformation 
$f(\bm{x})=A\bm{x}+\bm{a}\in \IGL(3)$, under which the invariant 
basis transforms as
\Eq{
f^*\chi^i=A^i{}_j\chi^j.
}

As was shown in Paper I, any invariance group $G$ whose maximal 
connected subgroup $G_0$ is given by $\RF^3$ is conjugate to a 
semi-direct product of $\RF^3$ and a subgroup of $D_4=\Set{\pm1,\pm 
I_1,\pm I_2,\pm I_3}$. Further, $I_1$ and $I_2$ are conjugate to 
$I_3$ by rotations. Hence, conjugate classes of $G$ other than 
$\RF^3$ are represented by the semi-direct product of $\RF^3$ with 
$\Set{1,-1}$, $\Set{1,I_3}$, $\Set{\pm1,\pm I_3}$, $D_2$ and $D_4$, 
where $D_2$ is the dihedral group 
consisting of the orientation preserving elements of $D_4$. The 
normalizers of these groups are listed in Table 
\ref{table:R3:normalizer}. In this table, o/u in the second column 
indicates whether $G$ consists of only orientation-preserving 
transformations (o) or contains orientation-reversing ones (u). Note 
that all invariance groups in this table are invariant under the space 
reflection, hence the conjugate class of each group in 
$\DiffG^+(\Sigma)$ and that in $\DiffG(\Sigma)$ conicide.

In contrast, subgroups of the type $\BTVII_0$ are not invariant under 
the space reflection, and if we restrict the arguments to orientation 
preserving transformations, the conjugate class of $\BTVII_0$ is 
divided into two subclasses $\BTVII_0^\epsilon$($\epsilon=\pm$), whose 
canonical generators are given by
\Eqr{
&& \xi_1=T_1,\ \xi_2=T_2,\ 
\xi_3=-\epsilon T_3-J_3,
\label{Generators:VII0}\\
&& [\xi_1,\xi_2]=0,\ 
   [\xi_3,\xi_1]=\xi_1,\ 
   [\xi_3,\xi_2]=-\xi_2.
}
A natural invariant basis is given by
\Eqr{
&& \begin{pmatrix}{c}\chi^1\\ \chi^2\end{pmatrix}=
R(-\epsilon z)\begin{pmatrix}{c}dx \\ dy \end{pmatrix},\ 
\chi^3=dz;\\
&& d\chi^1=-\chi^2\wedge \chi^3,\ d\chi^2=\chi^2\wedge\chi^3,\ 
d\chi^3=0.
\label{InvariantBasis:VII0}
}

The normalizer of $\BTVII_0^\epsilon$ does not depend on $\epsilon$ 
and consists only of orientation preserving transformations: 
$\N^+(\BTVII_0^\epsilon)=\N(\BTVII_0^\epsilon)=\N(\BTVII_0)$.  Its 
generic element is expressed as
\Eq{
f=\Set{1,I_1}f_{\bm{a},\hat{\bm{b}},k,\theta};\  
f_{\bm{a},\hat{\bm{b}},k,\theta}
:=L_{\bm{a}}R_{\hat{\bm{b}}}D(k,k,1)R_3(\theta),
}
where $L_{\bm{a}}$ and $R_{\bm{a}}$ are the transformations  
corresponding to the left and the right multiplications of the 
element $\bm{a}$ in $\BTVII_0$ respectively, 
\Eqr{
&& L_{\bm{a}}: \bm{x} \mapsto R_3(\epsilon a^3)\bm{x}+\bm{a},\\&& 
R_{\bm{a}}: \bm{x}\mapsto \bm{x}+R_3(\epsilon z)\bm{a},
}
$\hat{\bm{b}}=(\hat{\bm{b}},0)$ and $D(k,k,1)$($k>0$) is the linear 
transformation given by the same diagonal matrix.
Under this transformation, the invariant basis 
\eqref{InvariantBasis:VII0} transforms as
\Eq{
f^*\chi^i=F^i{}_j\chi^j: F=\Set{1,I_1}\times
\begin{pmatrix}{cc}
      kR(\theta) & \epsilon \hat{\bm{b}}^*\\
               0 & 1
\end{pmatrix},
\label{F:VII0}}
where $\hat{\bm{b}}^*=\Tp(-b^2,b^1)$. The conjugate transformation 
of $L_{\bm{c}}$ is given by
\Eqr{
&& fL_{\bm{c}}f^{-1}=L_{\bm{c}'};\nonumber\\
&& \bm{c}'=\Set{1,I_1}\left[
\begin{pmatrix}{cc} 
  kR(\theta+\epsilon a^3) & 0 \\ 
  0 & 1
\end{pmatrix}\bm{c}
+\left(1-R_3(\epsilon c^3)\right)\bm{a}\right].
\label{ConjTrf:VII0}}

An invariance group $G$ generated by $G_0=\BTVII_0^\epsilon$ and 
discrete transformations in $\IO(3)$ is conjugate to the 
semi-direct product of $\BTVII_0^\epsilon$ with one of 
$\Set{1,I_3}$, $\Set{1,I_1}$ and $D_2$ as discussed in Paper I. All of 
these groups are orientation preserving and their normalizers are the 
subgroups of 
$\N(\BTVII_0^\epsilon)$ listed in Table \ref{table:VII0:normalizer}.

\begin{table}[t]
\caption{\label{table:VII0:normalizer} Normalizers for the 
invariance groups with $G_0=VII^{\epsilon}(0)$.} 
\begin{tabular}{lcll}
\hline\hline
$G$ & o/u & $f\in\N(G)$ &$fgf^{-1}$ \\
\hline
&&&\\
$VII^\epsilon(0)$ & o 
& $\Set{1,I_1}f_{\bm{a},\hat{\bm{b}},k,\theta}$ 
& \eqref{ConjTrf:VII0}\\
$VII^\epsilon(0)\sdp \Set{1,I_3}$ & o 
& $\hat{\bm{b}}=0$ 
& $fI_3f^{-1}=I_3L_{{\bm{c}}}; {\bm{c}}=-2(a^1,\pm a^2,0)$\\
$VII^\epsilon(0)\sdp \Set{1,I_1}$ & o
& $R(\theta)=\pm 1, \hat{\bm{b}}=(b,0)$
& $fI_1f^{-1}=I_1L_{\bm{c}};\bm{c}=\left(I_1-R_3(-2\epsilon 
a^3)\right)\bm{a}$\\
$VII^\epsilon(0)\sdp D_2$ & o
& $\theta=\frac{n\pi}{2},\hat{\bm{b}}=0$ 
& $R_3(\pi/2)I_1R_3(\pi/2)=I_2$\\
&&&\\
\hline\hline
\end{tabular}
\end{table}

In addition to these subgroups, $\IO(3)$ has one conjugate class of 
4-dimensional connected subgroups, which is isomorphic to $\ISO(2)$ 
generated by $\RF^3$ and $R_3(\theta)$. It contains both $\RF^3$ and 
$\BTVII_0^\epsilon$ as subgroups. Since $\RF^3$ is the only 
3-dimensional Abelian subgroup of $\ISO(2)$, $\N(\ISO(2))$ preserves 
$\RF^3$. Hence it consists of $f\in\N(\RF^3)$ such that 
$fR_3(\theta)f^{-1}\in \ISO(2)$. The explicit form of $f$ is given 
in Table \ref{table:R3:normalizer}. An invariance group $G$ with 
$G_0=\ISO(2)$ is generated by $\ISO(2)$ and elements in 
$\N(\ISO(2))$, and is conjugate to one of the 4 groups listed in the 
same table. The normalizers of all these groups coincide with 
$\N(\ISO(2))$.

Finally, an invariance group $G$ with $G_0=\ISO(3)$ is given by 
$\ISO(3)$ or  $\IO(3)$. The normalizers of these groups are the same 
and are generated by $\IO(3)$ and dilation transformations.


\begin{table}[t]
\caption{\label{table:E3:Pi1}Fundamental groups and their 
representation in $\Isom(E^3)$ of compact, closed orientable 
3-manifolds of type $E^3$.}
\begin{tabular}{ll}
\hline
\hline
\bf Space & \bf Fundamental group and  representation \\
\hline
\\
$T^3$ 
& $ < \alpha,\beta,\gamma| [\alpha,\beta]=1,[\beta,\gamma]=1,
[\gamma,\alpha]=1>$ \\
\\
& $\alpha=(\bm{a},1),\beta=(\bm{b},1),\gamma=(\bm{c},1)$;\\
& $(\bm{a},\bm{b},\bm{c})\in\GL(3,\RF).$\\
\\
$T^3/\ZR_2$ 
& $<\alpha,\beta,\gamma| [\alpha,\beta]=1,
\gamma\alpha\gamma^{-1}\alpha=1,\gamma\beta\gamma^{-1}\beta=1>$\\
\\
& $\alpha=(\bm{a},1),\beta=(\bm{b},1),\gamma=(\bm{c},R_\gamma)$;
$R_\gamma=R_{a\times b}(\pi)$,\\
& $(\bm{a},\bm{b},\bm{c})\in\GL(3,\RF).$\\
\\
$T^3/\ZR_2\times\ZR_2$ 
& $<\alpha,\beta,\gamma|\alpha\beta\gamma=1,
\alpha\beta^2\alpha^{-1}\beta^2=1,\beta\alpha^2\beta^{-1}\alpha^2=1>$\\
\\
& $\alpha=(\bm{a},R_\alpha),\beta=(\bm{b},R_\beta),
\gamma=(\bm{c},R_\gamma)$;\\
&$R_\alpha,R_\beta$ and $R_\gamma$ are rotations of the angle $\pi$
about mutually \\
&orthogonal axes, such that
$R_\beta\bm{a}+R_\gamma\bm{b}+R_\alpha\bm{c}=0$, and\\
& $(\bm{a}+R_\alpha\bm{a},\bm{b}+R_\beta\bm{b},\bm{c}+R_\gamma\bm{c})
\in\GL(3,\RF)$\\
\\
$T^3/\ZR_3$ 
& $<\alpha,\beta,\gamma| [\alpha,\beta]=1,
\gamma\alpha\gamma^{-1}=\beta,
\gamma\beta\gamma^{-1}=\alpha^{-1}\beta^{-1}>$\\
\\
& $\alpha=(\bm{a},1),\beta=(\bm{b},1),\gamma=(\bm{c},R_\gamma)$;
$R_\gamma=R_{a\times b}({2\pi\over3})$,\\
& $ \bm{b}=R_\gamma\bm{a}$,
$(\bm{a},\bm{b},\bm{c})\in\GL(3,\RF).$\\
\\
$T^3/\ZR_4$ 
& $<\alpha,\beta,\gamma| [\alpha,\beta]=1,
\gamma\alpha\gamma^{-1}=\beta^{-1},
\gamma\beta\gamma^{-1}=\alpha>$\\
\\
& $\alpha=(\bm{a},1),\beta=(\bm{b},1),\gamma=(\bm{c},R_\gamma)$;
$R_\gamma=R_{a\times b}({\pi\over2})$, \\
& $\bm{b}=R_\gamma\bm{a}$,
$(\bm{a},\bm{b},\bm{c})\in\GL(3,\RF).$\\
\\
$T^3/\ZR_6$ 
& $<\alpha,\beta,\gamma| [\alpha,\beta]=1,
\gamma\alpha\gamma^{-1}=\beta,
\gamma\beta\gamma^{-1}=\alpha^{-1}\beta>$\\
\\
& $\alpha=(\bm{a},1),\beta=(\bm{b},1),\gamma=(\bm{c},R_\gamma)$;
$R_\gamma=R_{a\times b}({\pi\over3})$, \\
& $\bm{b}=R_\gamma\bm{a}$,
$(\bm{a},\bm{b},\bm{c})\in\GL(3,\RF).$\\
\\
\hline\hline
\end{tabular}
\end{table}

\subsection{Phase space}

There exist six diffeomorphism classes for the orientable 
compact closed 3-manifolds modeled on $E^3$ as listed in 
Table \ref{table:E3:Pi1}, where the fundamental group and 
its embedding into $\Isom^+(E^3)$ of each class are given. 
Although it is more natural from the geometrical point of 
view to discuss possible symmetries of the system and the 
structure of phase space for each space topology, we 
instead examine possible topologies and the structure of 
phase space for each invariance group as in Paper I, 
because it makes arguments simpler. In this section, we call the matrix 
$K=(\bm{a}\; \bm{b}\; \bm{c})$ consisting of the translational vectors 
associated with the three generators of the fundamental group in Table 
\ref{table:E3:Pi1} {\it the moduli matrix}, because the moduli sector 
of the invariant phase space is uniquely parametrized by that matrix 
after putting it into some canonical form by HPDs.

\subsubsection{$G\supset \RF^3$}

For an invariance group $G$ containing $\RF^3$, 
\eqref{InvariantBasis:R3} is the most natural invariant 
basis to represent homogeneous covering data on 
$\RF^3$. Since all the structure constants vanish for this 
basis, the momentum constraints for the single-component 
perfect fluid system are expressed as
\Eq{
H_I\equiv c u_I=0.
}
Hence, the fluid velocity is always orthogonal to the 
constant-time slices, and the phase space for the system 
is simply given by adding the energy density $\rho$ to 
that for the vacuum system discussed in Paper I. In 
particular, $G$ always contains $\RF^3\sdp D_2$ as a 
subgroup.

In contrast, for a multi-component system, the momentum 
constraints only restrict the center-of-mass velocity, and 
the 4-velocity of each component can be tilted. In this 
case, the system can have the lower symmetries, $\RF^3$ 
and $\RF^3\sdp\Set{1,I_3}$. However, the number of 
dynamical degrees of freedom in the gravitational sector 
for such cases is the same as that for $G=\RF^3\sdp D_2$. 

In fact, for $G=\RF^3$, only $T^3$ is allowed as the space 
topology. In this case, since $\N(\RF^3)=\IGL(3)$, the 
moduli matrix $K=(\bm{a}\;\bm{b}\;\bm{c})$ can be put to 
the unit matrix by a HPD and a modular transformation, and the isotropy 
group $\N_{\Sigma0}(\RF^3)$ of the action of $\N^+(\RF^3)$ at $K=1$ in 
the moduli space is trivial. Hence $N_Q=N_P=6$ and $N_m=0$, and the 
total parameter count $N_s$ is given by $10+3(n_f-1)+n_f=7+4n_f$, where 
$n_f$ is the number of fluid components. On the other hand, for 
$G=\RF^3\sdp\Set{1,I_3}$, $T^3$ and $T^3/\ZR_2$ are 
allowed. Now, from Table \ref{table:R3:normalizer}, $K$ 
for $T^3$ can be put to the form
\Eq{
K=\begin{pmatrix}{ccc}1 & 0 & c^1\\
                      0 & 1 & c^2\\
                      a^3&b^3&1
\end{pmatrix}
}
by a HPD and a modular transformation, for which $\N_{\Sigma0}(\RF^3)$ 
is trivial. On the other hand,  by the symmetry, 
$Q_{13}=Q_{23}=P^{13}=P^{23}=0$ and the second and the 
third components of all fluid velocities vanish. Hence $N_Q=N_P=N_m=4$ 
and $N_s=10+(n_f-1)+n_f=9+2n_f$. The argument for $T^3/\ZR_2$ is the 
same except that $K$ 
can be put to the unit matrix, hence $N_Q=N_P=4, N_m=0$ 
and $N_s=5+2n_f$.

Finally, note that for higher symmetries for which 
$G\supset \RF^3\sdp D_2$, the 4-velocity of every 
component must be orthogonal to the constant time slices 
and the dynamical degrees of freedom in the fluid sector 
are simply given by the energy density of each component. 
Hence $N_s$ is obtained by adding the number of components 
to the corresponding vacuum value. 
 

\begin{table}
\caption{\label{table:count:E3} The parameter counts for type $E^3$.
 }\begin{tabular}{llccccccc}
\hline
\hline
\bf Space &\bf Symmetry & $N_Q$ & $N_P$ & $N_m$ & 
$N_f$ & $N$ & $N_s$ & 
$N_s$(vacuum)\\
\hline\\
$\RF^3$
&$\RF^3\sdp D_2$& 0 & 3 & 0 & 1 & 4 & 2 & 1\\
&VII$_0$ 	& 2 & 3 & 0 & 4 & 9 & 7 & --\\
&VII$_0\sdp\ZR_2$	
		& 2 & 3 & 0 & 2 & 7 & 5 & --\\
&VII$_0\sdp D_2$& 2 & 3 & 0 & 1 & 6 & 4 &  3 \\
&$\ISO(2)\sdp\Set{1,I_1}$ 
		& 0 & 2 & 0 & 1 & 3 & 1 & 0 \\
&$\ISO(3)$	& 0 & 1 & 0 & 1 & 2 & 0 &  0 \\
&&&&&&&&\\
$T^3$
&$\RF^3\sdp D_2$& 3 & 3 & 6 & 1 & 13& 11& 10\\
&VII$_0$ 	& 3 & 3 & 4 & 4 & 14& 12& -- \\
&VII$_0\sdp\ZR_2$
		& 3 & 3 & 4 & 2 & 12& 10& -- \\
&VII$_0\sdp D_2$& 3 & 3 & 4 & 1 & 11& 9 &  8  \\
&$\ISO(2)\sdp\Set{1,I_1}$ 
		& 2 & 2 & 4 & 1 & 9 & 7 & 6 \\ 
&$\ISO(3)$	& 1 & 1 & 5 & 1 & 8 & 6 &  5  \\
&&&&&&&&\\
$T^3/\ZR_2$
&$\RF^3\sdp D_2$& 3 & 3 & 2 & 1 & 9 & 7 & 6 \\
&VII$_0$	& 3 & 3 & 2 & 4 & 12& 10& -- \\
&VII$_0\sdp\ZR_2$
		& 3 & 3 & 2 & 2 & 10& 8 & -- \\
&VII$_0\sdp D_2$& 3 & 3 & 2 & 1 & 9 & 7 &  6  \\
&$\ISO(2)\sdp\Set{1,I_1}$ 
		& 2 & 2 & 3 & 1 & 8 & 6 & 5 \\ 
&$\ISO(3)$	& 1 & 1 & 3 & 1 & 6 & 4 &  3  \\
&&&&&&&&\\
$T^3/\ZR_2\times\ZR_2$
&$\RF^3\sdp D_2$& 3 & 3 & 0 & 1 & 7 & 5 & 4 \\
&VII$_0\sdp\Set{1,I_1}$
		& 3 & 3 & 1 & 2 & 9 & 7 & --  \\
&VII$_0\sdp D_2$& 3 & 3 & 1 & 1 & 8 & 6 & 5 \\
&$\ISO(2)\sdp\Set{1,I_1}$
		& 2 & 2 & 1 & 1 & 6 & 4 & 3 \\ 
&$\ISO(3)$	& 1 & 1 & 2 & 1 & 5 & 3 & 2 \\
&&&&&&&&\\
$T^3/\ZR_k(k=3,4,6)$
&VII$_0$	& 3 & 3 & 0 & 4 & 10 & 8 & --  \\
&VII$_0\sdp\ZR_2$
		& 3 & 3 & 0 & 2 & 8 & 6 & --\\   
&VII$_0\sdp D_2$& 3 & 3 & 0 & 1 & 7 & 5 &  4  \\
&$\ISO(2)\sdp\Set{1,I_1}$
		& 2 & 2 & 0 & 1 & 5 & 3 & 2 \\ 
&$\ISO(3)$	& 1 & 1 & 1 & 1 & 4 & 2 &  1  \\
\\
\hline\hline
\end{tabular}
\end{table}

\subsubsection{$G_0=\BTVII^\epsilon_0$}

For $G_0=\BTVII^\epsilon_0$, the momentum constraint with 
respect to the invariant basis \eqref{InvariantBasis:VII0} 
is given by
\Eq{
H_1\equiv 2P^3_2+cu_1=0,\ 
H_2\equiv -2P^3_1+cu_2=0,\ 
H_3\equiv 2(P^2_1-P^1_2)+cu_3=0.
\label{MomentumConstraint:VII0}}
For the vacuum system, it follows from this constraint 
that the system always has the higher symmetry 
$\BTVII^\epsilon_0\sdp D_2$, but for the fluid system the 
lower symmetries listed in Table 
\ref{table:VII0:normalizer} are all allowed.

\subsubsection*{A. $G=\BTVII^\epsilon_0$}

\smallskip

Since the left transformation $L_{\bm{d}}$ is a linear 
transformation $(\bm{d},R_3(\epsilon d^3))\in \IGL(3)$, 
$\pi_1(\Sigma)$ can be embedded into $\BTVII^\epsilon_0$ 
only for $\Sigma=T^3$ or $T^3/\ZR_k$($k=2,3,4,6$) from 
Table \ref{table:E3:Pi1}. 

\paragraph{A-i) $T^3$:} $L_{\bm{d}}$ becomes a pure 
translation only when $d^3/2\pi$ is an integer. Hence the 
moduli matrix $K=(\bm{a}\;\bm{b}\;\bm{c})$ has the form
\Eq{
K=
\begin{pmatrix}{ccc}
  \hat{\bm{a}} & \hat{\bm{b}} & \hat{\bm{c}} \\
  2\pi l       & 2\pi m       & 2\pi n
\end{pmatrix}
\label{K:(T^3,VII0):generic}}
where $\hat{\bm{a}}, \hat{\bm{b}}$ and $\hat{\bm{c}}$ are 
two-component vectors, and $l,m$ and $n$ are integers. By 
a modular transformation $K \mapsto KZ$ with $Z\in 
\GL(3,\ZR)$, we can always put $l=m=0$ and $n>0$. Further, 
from \eqref{ConjTrf:VII0}, $f=f_{\bm{a},\hat{\bm{b}},k,\theta }\in 
\N(\BTVII_0)$ 
transforms $L_{\bm{d}}$ with 
$\bm{d}=(\hat{\bm{d}},2\pi p)$($p\in\ZR$) to $L_{\bm{d}'}$ with 
$\hat{\bm{d}}'=kR(\theta+\epsilon a^3)\hat{\bm{d}}$. Hence 
$K$ can be put by a HPD to
\Eq{
K=
\begin{pmatrix}{ccc}
  1 & X & V \\
  0 & Y & W \\
  0 & 0 & 2\pi n
\end{pmatrix},
\label{K:(T^3,VII0):canonical}}
where $Y>0$. Let us denote the set of $K$ with this form by $\M_1$. 
Then $\M_1$ intersects with each HPD orbit in the total space of the 
moduli matrix at a single point, and the isotropy group 
$\N_{T^30}(\BTVII_0^\epsilon)$ at $K$ consists of 
$f_{\bm{a},\hat{\bm{b}},k,\theta}$ with $k=1$ and $R(\theta+\epsilon 
a^3)=1$.

In the present case, a modular transformation represented by the matrix
\Eq{
Z=
\begin{pmatrix}{cc}
  \hat Z & \bm{p} \\
  0      &  1
\end{pmatrix}
}
maps $\M_1$ onto $\M_1$ when it is combined with an appropriate HPD, 
where $\hat Z\in \SL(2,\ZR)$ and $\bm{p}$ is a 2-vector 
with integer components. Hence $H_{\rm mod}$ in 
\eqref{InvPhaseSpace:reduced} is isomorphic to 
$\SL(2,\ZR)\sdp\ZR^2$. Although the HPDs associated with  each 
transformation in $H_{\rm mod}$ have rather complicated structures, 
their effective action on the invariant basis is simply given by the 
product of a $D(k,k,1)$ transformation with $k>0$ and a transformation 
induced from $f$ in $\N_{T^30}(\BTVII_0^\epsilon)$. Further, although 
the action of some transformations in $H_{\rm mod}$ on $\M_1$ have 
fixed points, the isotropy group at each fixed point induces 
transformations of the invariant basis with $k=1$. Hence, taking into 
account that a bundle with the structure group $\RF_+$ (the 
multiplicative group of positive numbers) is always trivial, the 
invariant phase space is simply given by the product $\Gamma_{\rm 
dyn}\times\M_0$ with $\M_0=\M_1/H_{\rm mod}$.

As mensioned above, the action of $H_{\rm mod}$ on $\M_1$ has 
non-trivial isotropy group $H_K$ at discrete points. Although there are 
infinite number of such points, they project to 6 points on $\M_0$ by 
the identification by the action of $H_{\rm mod}$:
\begin{itemize}
\item[i)] $(X,Y,V,W)=(0,1,1/2,0)$: $H_K=\ZR_2; \hat Z=-1, 
\bm{p}=(-1,0)$.
\item[ii)] $(X,Y,V,W)=(0,1,0,0)$: $H_K=\ZR_4; \hat Z=R(\pi/2), 
\bm{p}=(0,0)$.
\item[iii)] $(X,Y,V,W)=(0,1,1/2,1/2)$: $H_K=\ZR_4; \hat 
Z=R(\pi/2), \bm{p}=(0,1)$.
\item[iv)] $(X,Y,V,W)=(1/2,\sqrt{3}/2,1/2,0)$: $H_K=\ZR_2; \hat Z=-1, 
\bm{p}=(-1,0)$.
\item[v)] $(X,Y,V,W)=(1/2,\sqrt{3}/2,1/2,1/2\sqrt{3})$: $H_K=\ZR_3; 
\hat Z= \begin{pmatrix}{cc}-1&-1\\1&0\end{pmatrix}$,\\ 
$\bm{p}=(-1,0)$.
\item[vi)] $(X,Y,V,W)=(1/2,\sqrt{3}/2,0,0)$: $H_K=\ZR_6; \hat 
Z={\scriptstyle \begin{pmatrix}{cc}0&-1\\1&1\end{pmatrix}}, 
\bm{p}=(0,0)$.
\end{itemize}
Here $(\hat Z,\bm{p})$ represents the parameter of $Z$ generating 
$H_K$. The reduced moduli space has the orbifold singularities of the 
type specified by the isotropy groups at these fixed points. Except for 
these points, it is smooth and given by
\Eq{
\M_0(T^3,\BTVII^\epsilon_0)\approx \NN\times \M(T^2)\times T^2,
}
where $\NN$ is the set of positive integers, and $\M(T^2)$ is the 
standard moduli space of the flat 2-torus. 

As mentioned above, the moduli sector is invariant 
under the HPD $f=f_{\bm{a},\hat{\bm{b}},k,\theta }\in \N(\BTVII_0)$ 
with $k=1$ and $a^3=-\epsilon \theta$. From \eqref{F:VII0}, it is easy 
to see that in terms of this residual HPD we can 
transform the metric component matrix $Q$ into the 
diagonal form $Q=D(Q_1,Q_2,Q_3)$ with $Q_1\ge Q_2$. If 
$Q_1>Q_2$ after this diagonalization, the residual HPDs 
induce only the $I_3=R_3(\pi)$ transformation of the 
invariant basis, and the off-diagonal components of the 
$P$ matrix are determined by the fluid velocity $u_I$ 
through the momentum constraints 
\eqref{MomentumConstraint:VII0}. In particular, if two 
components of $u_I$ vanish, the system aquires has a 
higher symmetry. On the other hand, if $Q_1=Q_2$, $u_3$ 
must vanish and there remains the $R_3(\theta)$ HPD 
symmetry, in terms of which we can put $P^{12}=0$. If 
$u_1=0$ or $u_2=0$, $P^{23}=0$ or $P^{13}=0$ and the 
system has a higher symmetry again. Therefore, the 
dynamical sector of the invariant phase space is 
given by
\Eqr{
&\Gamma_{\rm 
dyn}(T^3,\BTVII^\epsilon_0)=
&\big\{(Q_1,Q_2,Q_3;P^{11},P^{22},P^{33};u_1,u_2,u_3,\rho)
\big| Q_1\ge Q_2>0,
\nonumber\\ 
&& Q_3>0,
u_1^2u_2^2+u_2^2u_3^2+u_1^2u_2^2\not=0\big\}/\Set{1,I_3},
}
from which we obtain the parameter count in Table \ref{table:count:E3}.

The extension of the argument to a multi-component fluid system is 
quite simple. The parameter count for $N_s$ is simply given by adding 
$4$ for each extra component to the value for the single component 
system.

\paragraph{A-ii) $T^3/\ZR_k$:} 
First we consider the case $k=2$. In this case two generators of the 
fundamental group, $\alpha$ and $\beta$, are represented by 
translations $\bm{a}$ and $\bm{b}$, and the third generator $\gamma$ is 
the combination of a translation $\bm{c}$ and the rotation of the angle 
$\pi$ around the axis $\bm{a}\times\bm{b}$. Since $\gamma$ belongs to 
$\BTVII_0^\epsilon$, this implies that $\bm{a}$ and $\bm{b}$ are 
orthogonal to the $z$-axis and $c^3/2\pi$ is an odd integer. Then, 
since $R_3(\epsilon c^3)=R_3(\pi)$, we can put 
$c^1=c^2=0$ by the conjugate transformation \eqref{ConjTrf:VII0}. 
 Hence, by the same argument as in the 
previous case, the moduli matrix $K$ can be put into the canonical form
\Eq{
K=
\begin{pmatrix}{ccc}
  1 & X & 0 \\
  0 & Y & 0 \\
  0 & 0 & (2n-1)\pi
\end{pmatrix},
}
where $n$ is a positive integer, and $Y>0$. The discrete group of 
transformations consisting of HPDs and modular transformations which 
preserve this form of $K$ is identical to the modular 
transformation group of the flat torus. Further, the HPD associated 
with a modular tranformation fixing $K$ is given by 
$f_{\bm{a},\hat{\bm{b}},k,\theta}$ with $k=1$. Hence, by the same 
argument as in the previous case, the invariant phase space is 
written as a product of the dynamical sector and the moduli sector, 
and the latter has the topological structure 
\Eq{
\M_0(T^3/\ZR_2,\BTVII^\epsilon_0)\approx\NN\times \M(T^2).
}

The argument for $k=3,4,6$ is the same except for two points. First, 
$c^3$ takes values of the form $2\pi(n+\epsilon/k)$ with integer 
$n$. Second, since $\bm{b}$ and $\bm{a}$ are related by 
$\bm{b}=R_3(2\pi/k)\bm{a}$, the value of $(X,Y)$ is fixed to 
$(\cos(2\pi/k),\sin(2\pi/k))$. Hence, the moduli sector becomes a 
set of discrete points and there exists no continuous moduli freedom:
\Eq{
\M_0(T^3/\ZR_k,\BTVII^\epsilon_0)=\ZR \quad (k=3,4,6).
}

Since the isotropy group of the action of $\N(\BTVII_0)$ in the moduli 
space does not depend on the space topology, the dynamical sector of 
the invariant phase space is the same as that for $\Sigma=T^3$. Hence 
the parameter count $N_s$ is simply given by replacing $N_m$ for $T^3$ 
by $N_m=2$ for $k=2$ and $N_m=0$ for $k=3,4,6$. The extension of the 
parameter count to multi-component systems is also the same.

\subsubsection*{B. $G=\BTVII_0^\epsilon\sdp\Set{1,I_3}$}

\smallskip

Since a transformation in $G$ is a translation or a linear 
transformation with a rotation around the $z$-axis, $\pi_1(\Sigma)$ can 
be embedded into $G$ only for $\Sigma=T^3$ or $T^3/\ZR_k$ as in the 
case $G=\BTVII_0^\epsilon$.

\paragraph{B-i) $T^3$:} In the present case, in addition to 
$I_{\bm{d}}$ with $d^3\equiv 0 (\mod 2\pi)$, $I_3L_{\bm{d}}$ also 
becomes a pure translation when $d^3/2\pi$ is an odd integer. Hence the 
generic moduli matrix is given by the expression 
\eqref{K:(T^3,VII0):generic} with $2l,2m$ and $2n$ replaced by 
integers $l,m$ and $n$, respectively. Apart from this difference, 
the same argument on the reduction of the moduli matrix $K$ into the 
canonical form as that for $(T^3,\BTVII_0^\epsilon)$ applies to the 
present case, because transformations in 
$\N(\BTVII_0^\epsilon\sdp\Set{1,I_3})$ are simple linear 
transformations, and their effective action on the moduli parameter 
$K$ is the same. Hence the canonical form for $K$ is given by 
replacing $2n$ by a positive interger $n$ in 
\eqref{K:(T^3,VII0):canonical} and the moduli sector of the 
invariant phase space has the same topological structure as that for 
$(T^3,\BTVII_0^\epsilon)$:
\Eq{
\M_0(T^3,\BTVII_0^\epsilon\sdp\Set{1,I_3})\approx 
\NN\times \M(T^2)\times T^2.
}

In contrast, the dynamical sector of the phase space has a different 
structure due to the higher symmetry. First, $I_3$ 
invariance requires that $Q_{13}=Q_{23}=P^{13}=P^{23}=0$ and 
$u_1=u_2=0$. The transformations of the invariant basis induced from 
the residual HPDs after fixing the moduli parameter are now given by 
rotations $R_3(\theta)$, by which we can diagonalize $Q$ to 
$D(Q_1,Q_2,Q_3)$ with $Q_1\ge Q_2$. After this diagonalization, the 
momentum constraints $H_1=0$ and $H_2=0$ become trivial. If 
$Q_1=Q_2$,  we can further diagonalize $P$. Since $u_3=0$ from the 
momentum constraint, this implies that the system has a higher 
symmetry in this case. Hence $Q_1>Q_2$ and $P^{12}$ is determined by 
$u_3$ via the momentum constraint $H_3=0$. Thus the dynamical sector 
of the invariant phase space is given by
\Eqr{
&\Gamma_{\rm 
dyn}(T^3,\BTVII_0^\epsilon\sdp\Set{1,I_3})=
&\big\{(Q_1,Q_2,Q_3;P^{11},P^{22},P^{33};u_3,\rho)
\big| 
\nonumber\\ 
&&\qquad Q_1> Q_2>0,Q_3>0,u_3\not=0\big\}.
}
%
 
\paragraph{B-ii) $T^3/\ZR_k$:} 
As in the previous case, the argument on the moduli sector is the same 
as that for $(T^3/\ZR_k,\BTVII_0^\epsilon)$ except for the range of the 
discrete parameter $c^3$: $c^3=n\pi$($n>0$) for $k=2$ and 
$c^3=\pi(n+2\epsilon/k)$ for $k=3,4,6$ with an interger $n$. 
Therefore the moduli sector of the invariant phase space is the same:
\Eq{
\M_0(T^3/\ZR_k,\BTVII_0^\epsilon\sdp\Set{1,I_3})
\cong \M_0(T^3/\ZR_k,\BTVII_0^\epsilon).
}
Further, since the residual HPDs after fixing the moduli parameter does 
not depend on the space topology in the present case, the dynamical 
sector of the phase space is the same as that for $T^3$.

Since the symmetry requires $u_1=u_2=0$ for any component, the 
parameter count for the multi-component system is simply obtained by 
adding $2$ for each extra component of the fluid, irrespective of the 
topology.

\subsubsection*{C. $G=\BTVII_0^\epsilon\sdp\Set{1,I_1}$}

\smallskip

For $G=\BTVII_0^\epsilon\sdp\Set{1,I_1}$, the fundamental group of any 
compact quotient of the Thurston type $E^3$ can be embedded in $G$.

\paragraph{C-i) $T^3$ and $T^3/\ZR_k$:} The image of $\pi_1(\Sigma)$ is 
contained in $\BTVII_0^\epsilon$ for these spaces except for 
$T^3/\ZR_2$. Further, the group of conjugate 
transformations of the moduli parameter induced from 
$\N(\BTVII_0^\epsilon\sdp\Set{1,I_1})$ is the same as that for 
$G=\BTVII_0^\epsilon$. Hence for $k\not=2$, the moduli sector of the 
invariant phase space is also the same as that for 
$G=\BTVII_0^\epsilon$:
\Eq{
\M_0(T^3/\ZR_k,\BTVII_0^\epsilon\sdp\Set{1,I_1})
=\M_0(T^3/\ZR_k,\BTVII_0^\epsilon)\quad (k=1,3,4,6).
}
On the other hand, for $\Sigma=T^3/\ZR_2$, embeddings for which the 
rotation axis of the generator $\gamma$ is orthogonal to the 
$z$-axis are allowed, because $R_1(\pi)R_3(\epsilon 
c^3)=R_3(-\epsilon c^3/2)R_1(\pi)R_3(\epsilon c^3/2)$. For such an 
embedding, the rotation axis of $\gamma$ can be rotated to any 
direction orthogonal to $z$-axis by the conjugate transformation 
with respect to 
$f=L_{\bm{d}}\in\N(\BTVII_0^\epsilon\sdp\Set{1,I_1})$, because from 
Table \ref{table:VII0:normalizer} the rotation matrix associated 
with $fI_1f^{-1}$ is given by $I_1R_3(-2d^3)$. Hence, by an 
appropriate HPD, $R_\gamma$ can be transformed to $R_1(\pi)$, for which 
$\gamma$ is represented by $I_1 L_{\bm{c}}$ and the moduli matrix $K$ 
takes the form
\Eq{
K=
\begin{pmatrix}{ccc}
  0    & 0    & c^1 \\
  a^2  & b^2  & c^2 \\
  2l\pi&2n\pi & 2m\pi
\end{pmatrix},
}
where $l,m$ and $n$ are intergers. By a modular transformation among 
the generators $\alpha$ and $\beta$, $l$ can be put to zero. Further, 
$m$ and $c^2$ can be put to zero by the HPD $f=L_{\bm{d}}$ with 
$d^3=m\pi$. Finally, by the HPD $f=D(\pm k,\pm k,1)$ and the modular 
transformations $\alpha\tend\alpha^{-1}, \beta\tend\beta^{-1}$ and 
$\gamma\tend\gamma^{-1}$, we can put $K$ to
\Eq{
K=
\begin{pmatrix}{ccc}
  0    & 0    & 1 \\
  X    & Y    & 0 \\
  0    &2n\pi & 0
\end{pmatrix},
}
where $n$ is a positive integer and $X>0$. The combinations of modular 
transformations and HPDs which preserve this form of $K$ are just the 
modular transformations $Y\tend Y+pX$ with $p\in\ZR$. Hence, the moduli 
sector of the invariant phase space is given by%
\Eq{
\M_0(T^3/\ZR_2,\BTVII_0^\epsilon\sdp\Set{1,I_1})
\approx\NN\times (\M(T^2)\cup \RF_+\times S^1).
}

The symmetry requires that $Q_{12}=Q_{13}=P^{12}=P^{13}=0$ and 
$u_2=u_3=0$. The transformations of the invariant basis induced by 
the residual HPDs are given by $F$ in \eqref{F:VII0} with 
$kR(\theta)=1$ and $\hat{\bm{b}}=(b,0)$. We can put $Q_{23}$ to zero 
and diagonalize $Q$ by these transformations. After this 
diagonalization, the momentum constraints $H_2=0$ and $H_3=0$ become 
trivial, and $P^{23}$ is determined by $u_1$ via $H_1=0$. Hence, the 
dynamical sector of the invariant phase space is given by
\Eqr{
&\Gamma_{\rm 
dyn}(\RF^3,\BTVII_0^\epsilon\sdp\Set{1,I_1})=
&\big\{(Q_1,Q_2,Q_3;P^{11},P^{22},P^{33};u_1,\rho)
\big|
\nonumber\\ 
&& Q_1,Q_2,Q_3>0,u_1\not=0\big\}.
\label{PhaseSpace:VII0x(1,I_1):dynamical}}
%
 
\paragraph{C-ii) $T^3/\ZR_2\times\ZR_2$:} In this case, the 
generators of the fundamental group are represented by three glind 
rotations, $\alpha=(\bm{a},R_\alpha), \beta=(\bm{b},R_\beta)$ and 
$\gamma=(\bm{c},R_\gamma)$, where $R_\alpha,R_\beta$ and $R_\gamma$ are 
rotations of angle $\pi$ around mutually orthogonal axes, one of which 
has to be the $z$-axis. As explained in the argument for $T^3/\ZR_2$, 
we can always set $R_\alpha=I_1, R_\beta=I_2$ and $R_\gamma=I_3$ by a 
HPD, for which $a^3=2l\pi, b^3=(2m-1)\pi,c^3=(2n-1)\pi$, and the 
relation $R_\beta\bm{a}+R_\gamma\bm{b}+R_\alpha\bm{c}=0$ gives the 
constraints%
\Eq{
a^1+b^2=c^1,\ 
b^2+c^2=a^1,\ 
l+n=m,
}
and $a^1b^2(2n-1)\not=0$. From the formula in Table 
\ref{table:VII0:normalizer}, the conjugate transformation by 
$L_{\bm{d}}\in \N(G)$ preserves the above form of the generators 
when $d^3=p \pi$ with an integer $p$. Under this condition, $\bm{a}$ 
and $\bm{c}$ transform as
\Eqr{
&& \bm{a}\mapsto \left((-1)^pa^1,(-1)^pa^2+2d^2,2(l+p)\pi\right),\\
&& \bm{c}\mapsto 
\left((-1)^pc^1+2d^1,(-1)^pc^2+2d^2,(2n-1)\pi\right).
}
Hence $a^2,l$ and $c^1$ can be put to zero. Further $a^1b^2$ can be 
made positive by the transformation $f=I_1$ if necessary, and then 
$b^1$ can be put to unity by $f=D(\pm k,\pm k,1)$. Hence, by taking 
account of the above constraints, the moduli matrix $K$ can be put into 
the canonical form
\Eq{
K=
\begin{pmatrix}{ccc}
  X  & -X  &  0 \\
  0  & 1   &  -1 \\
  0 &-(2n-1)\pi & (2n-1)\pi
\end{pmatrix},
}
where $X>0$. 
Since there remains no modular transformation freedom, the moduli 
sector of the invariant phase space is given by
\Eq{
\M_0(T^3/\ZR_2\times\ZR_2, \BTVII_0^\epsilon\sdp\Set{1,I_1})
\approx\ZR\times \RF_+.
}
Since the isotropy group of $\N(G)$ at $K$ is the same as that in the 
previous case, the dynamical sector of the invariant phase space is 
given by \eqref{PhaseSpace:VII0x(1,I_1):dynamical}.

As in the case of $G=\BTVII_0^\epsilon\sdp\Set{1,I_3}$, the parameter 
count for a multi-component system is given by adding two for each 
extra component to the value for the single component system, 
irrespective of the topology. 

\subsubsection*{D. $G=\BTVII_0^\epsilon\sdp D_2$}

\smallskip

For this invariance group, the symmetry requires that $u_I=0$. 
Hence, as in the case $G\supset\RF^3$, the moduli sector has the 
same structure as that for the vacuum system, and the invariant 
phase space is obtained simply by adding the fluid energy density to 
that for the vacuum system, both for the single and multiple 
component systems.

\section{$\Nil$}

\subsection{Maximal geometry and invariance groups}

$\Nil$ is the maximal geometry $(\RF^3,\Isom(\Nil))$ whose 
representative metric is given by
\Eq{
ds^2=dx^2+dy^2+Q_3\left[dz+\frac{1}{2}(ydx-xdy)\right]^2.
\label{MaximalGeometry:Nil}}
It contains the Bianchi type $\BTII$ group as the subgeometry. To see 
this, recall that the type $\BTII$ group is the Heisenberg group with 
the multiplication structure given by
\Eq{
(a,b,c)(x,y,z)=\left(a+x, b+y, c+z+{ay-bx\over2}\right).
}
The left transformation $L_{\bm{a}}$ and the right transformation 
$R_{\bm{a}}$ defined from this multiplication are represented by linear 
transformations in $\RF^3$:
\Eqr{
&& L_{\bm{a}}=(\bm{a}, \Tp N(\hat{\bm{a}}{}^*/2))\in \IGL(3),\\&& 
R_{\bm{a}}=(\bm{a}, \Tp N(-\hat{\bm{a}}{}^*/2))\in \IGL(3),
}
where $\hat{\bm{a}}{}^*=\Tp(-a^2,a^1)$, and $N(\bm{b})$ is the 
matrix 
\Eq{
N(\bm{b})=
\begin{pmatrix}{ccc}
	1 & 0 & b^1\\
	0 & 1 & b^2 \\
	0 & 0 & 1
\end{pmatrix},
}
which forms an Abelian group isomorphic to $\RF^2$ with respect to 
the muliplication. In particular, the infinitesimal left 
transformations are generated by
\Eq{\xi_1=\partial_x + {1\over2}y\partial_z,\quad
\xi_2=\partial_y - {1\over2}x\partial_z,\quad
\xi_3=\partial_z,
\label{Killing:II}}
with the commutation relations for the Bianchi type $\BTII$ group
\Eq{
[\xi_1,\xi_2]=-\xi_3,\ 
[\xi_1,\xi_3]=0, \ 
[\xi_2,\xi_3]=0.
\label{LieA:II}}
A natural invariant basis is given by
\Eqr{
&& \chi^1=dx, \quad \chi^2=dy,\quad \chi^3=dz+{1\over2}(ydx-xdy);
\label{InvariantBasis:II}\\
&& d\chi^1=0,\quad d\chi^2=0,\quad d\chi^3=-\chi^1\wedge\chi^2.
}

Since the metric \eqref{MaximalGeometry:Nil} is written in terms of 
this invariant basis as $ds^2=(\chi^1)^2+(\chi^2)^2+(\chi^3)^2$, it is 
clearly $\BTII_L$ invaraint. Further, it is easy to see that it is also 
invaraint under the rotation around the $z$-axis. Hence the metric has 
4-dimensional isometries generated by $L_{\bm{a}}$ and $R_3(\theta)$. 
Since the 3-space whose isometry group has a dimension greater than 4 
is always a constant curvature space, this 4-dimensional group gives 
$\Isom_0(\Nil)$:
\Eq{
\xi_4=-y\partial_x+x\partial_y,
}
\Eq{
[\xi_4,\xi_1]=-\xi_2,\quad [\xi_4,\xi_2]=\xi_1,\quad [\xi_4,\xi_3]=0.
\label{LieA:Isom(Nil)}
}
%


\begin{table}[t]
\caption{\label{table:Nil:normalizer} Normalizers for the invariance 
groups $G$ containing $\BTII_L$.} 
\begin{tabular}{lcll}
\hline\hline
$G$ & o/u & $f\in\N(G)$ &$fgf^{-1}$ \\
\hline\\
$\BTII_L$  & o
& $L_{\bm{a}}R_{\hat{\bm{b}}}D(\hat A,\Delta)$
& $fL_{\bm{c}}f^{-1}=L_{\bm{c}'}$
\\
&
& $\quad \Delta=\det\hat A$
& $\bm{c}'=(\hat A\hat{\bm{c}},\Delta c^3+\hat{\bm{a}}{}^*\cdot\hat 
A\hat{\bm{c}})$
\\
$\BTII_L\sdp \Set{1,I_3}$ &o
& $\hat{\bm{b}}=0$ 
& $fI_3f^{-1}=I_3L_{-2\hat{\bm{a}}}$ 
\\
$\BTII_L\sdp \Set{1,I_1}$ & o
& $\hat A=D(p,q)$,$\hat{\bm{b}}=(b,0)$
& $fI_1f^{-1}=I_1L_{\bm{c}}$
\\
&
& 
& $\bm{c}=(0,-2a^2,-2a^3-a^1a^2)$\\
$\BTII_L\sdp D_2$ & o
& $\Set{1,J}L_{\bm{a}}D(p,q,pq)$; 
& $L_{\bm{a}}I_iL_{\bm{a}}^{-1}=I_iL_{\bm{c}_i}$;
\\
&
& $J=R_3(\pi/2)I_1$
&$\quad \bm{c}_1=(0,-2a^2,-2a^3-a^1a^2)$,
\\
&
&&  $\quad \bm{c}_2=(-2a^1,0,-2a^3+a^1a^2)$,
\\
&
&&  $\quad \bm{c}_3=(-2a^1,-2a^2,0)$,
\\
&
&& $JI_1J=I_2,\ JI_3J=I_3$\\
$\Isom_0(\Nil)$ & o
& $\Set{1,I_1}L_{\bm{a}}R_3(\theta)D(k,k,k^2)$
& $I_1R_3(\phi)I_1=R_3(-\phi)$,
\\
&
&& 
$L_{\bm{a}}R_3(\phi)L_{\bm{a}}^{-1}=R_3(\phi)L_{(R(-\phi)-1)\hat{\bm{a}}}
$\\
$\Isom(\Nil)$ & o
& ibid
& ibid
\\
\\
\hline\hline
\end{tabular}
\end{table}

In order to determine extra discreate isometries, we need the 
information on the normalizer group $\N(\Isom_0(\Nil))$. Since 
$[\LieA(\Isom(\Nil)), \LieA(\Isom(\Nil))]=\LieA(\BTII_L)$, 
$\N(\Isom_0(\Nil))$ is a subgroup of $\N(\BTII_L)$ from 
Prop.\ref{prop:normalizer:invariantspace}. Hence, let us first 
determine $\N(\BTII_L)$, which is the set of transformations inducing 
linear transformations of the invariant basis. First, since $\xi_3$ is 
a generator of the center of $\BTII_L$, 
$f_*\xi_3=k\xi_3$, i.e., $\partial_z f^1=\partial_z f^2=0$ and 
$\partial_z f^3=k$ with a constant $k$, for 
$f=(f^1,f^2,f^3)\in\N(\BTII_L)$. Hence, $(f^1,f^2)$ are required to be 
a linear transformation $(\hat{\bm{b}}, \hat A)$ in the $(x,y)$ plane. 
Further, from the condition that $f^*\chi^3$ is a linear combination of 
$\chi^I$, it follows that $k=\det\hat A$. Since the linear 
transformation $D(\hat A,\det \hat A)$ belongs to 
$\N(\BTII_L)$, we can make $(f^1,f^2)$ a pure translation and 
put $\partial_z f^3$ to unity by combining $f$ with $D(\hat A,\det\hat 
A)^{-1}$. Then $f^3-z$ becomes linear in $x$ and $y$. This final form 
of $f$ is realized as an appropriate 
combination of the left and the right transformations. To summarize, a 
generic element $f$ of $\N(\BTII_L)$ is expressed as
\Eq{
f=L_{\bm{a}}R_{\hat{\bm{b}}}D(\hat A,\Delta); \Delta=\det\hat A.
}
The invariant basis \eqref{InvariantBasis:II} transforms by this HPD as
\Eq{
f^*\chi^i=F^i{}_j\chi^j:\ 
F=\Tp N(-\hat{\bm{b}}{}^*)D(\hat A,\Delta).
\label{F:II}}
It is directly checked that the conjugate transformation by $f$ maps a 
rotation around the $z$-axis into $\Isom_0(\Nil)$ if and only if 
$\hat{\bm{b}}=0$ and $\hat A$ is written as $\Set{1,I_1}kR_3(\theta)$. 
Hence $f\in \N(\Isom_0(\Nil))$ is written as 
$f=\Set{1,I_1}L_{\bm{a}}D(k,k,k^2)R_3(\theta)$. From this we find that
\Eq{
\Isom(\Nil)=\Isom_0(\Nil)\sdp \Set{1,I_1}
\cong \BTII_L\sdp O(2),
}
Further from the formulas on the conjugate transformations of $I_1$ 
given in Table \ref{table:Nil:normalizer} we can confirm that 
$\N(\Isom(\Nil))=\N(\Isom_0(\Nil))$. 

Next we determine the conjugate classes of invariance group $G$ such 
that $G_0=\BTII_L$ and their normalizers. Since $\Isom(\Nil)\subset 
\N(\BTII_L)$, possible discrete transformations to be added to 
$\BTII_L$ are $I_1$, $R_3(\theta)$ and their combinations. As in the 
type $E^3$ case, the invariance group contains $\Isom_0(\Nil)$ if it 
contains a transformation $R_3(\theta)$ with $\theta\not=0,\pi (\mod 
2\pi)$. Further $I_1R_3(\theta)$ is conjugate to $I_1$ for any 
$\theta$. Hence the conjugate classes of $G$ is given by $\BTII_L$ 
or the semi-direct product of $\BTII_L$ with one of the discrete 
groups $\Set{1,I_1}, \Set{1,I_3}$ and $D_2=\Set{1,I_1,I_2,I_3}$. By 
examaning the conjugate transformation of $I_i$ by 
$f\in\N(\BTII_L)$, one finds that the normalizer of these groups are 
given by those listed in Tabel \ref{table:Nil:normalizer}.

\begin{table}[t]
\caption{\label{table:Nil:Pi1}Fundamental groups and their
representation in $\Isom^+(\Nil)$ of compact closed orientable
3-manifolds of type $\Nil$.
In this table $\Delta(a,b)=a^1b^2-a^2b^1$.}
\bigskip
\noindent
\begin{tabular}{ll}
\hline
\hline
\bf Space & \bf Fundamental group and  representation \\
\hline
\\
$T^3(n)$ 
& $<\alpha,\beta,\gamma| [\alpha,\gamma]=1,[\beta,\gamma]=1,
[\alpha,\beta]=\gamma^n>  \quad (n\in\NN)$ \\
& $\alpha=L_{\bm{a}},\beta=L_{\bm{b}},
\gamma=L_{\bm{c}}$;\\
& $\bm{c}=(0,0,\Delta(a,b)/n)\not=0$\\
\\
$K^3(n)$ 
& $<\alpha,\beta,\gamma| [\alpha,\gamma]=1,
\beta\gamma\beta^{-1}\gamma=1,
\alpha\beta\alpha\beta^{-1}=\gamma^n>\quad(n\in\NN)$\\
& $\alpha=L_{\bm{a}},\beta=R_1(\pi)R_3(\theta)L_{\bm{b}},
\gamma=L_{\bm{c}};$;\\
& $R(\theta)\begin{pmatrix}{c}a^1\\ a^2\end{pmatrix}
=\begin{pmatrix}{c}-a^1\\a^2\end{pmatrix},
\bm{c}=(0,0,\Delta(a,b)/n)\not=0$\\
\\
$T^3(n)/\ZR_2$ 
& $<\alpha,\beta,\gamma| \gamma\alpha\gamma^{-1}\alpha=1,
\gamma\beta\gamma^{-1}\beta=1, [\alpha,\beta]=\gamma^{2n}>$\\
& $\alpha=L_{\bm{a}}, \beta=L_{\bm{b}}, \gamma=R_3(\pi)L_{\bm{c}};$\\ 
& $a^3=\Delta(a,c)/2, b^3=\Delta(b,c)/2),c^3=\Delta(a,b)/2n\not=0$\\ \\
$T^3(2n)/\ZR_2\times\ZR_2$ 
& $<\alpha,\beta,\gamma|
\alpha\gamma^2\alpha^{-1}\gamma^2=1,\gamma\alpha^2\gamma^{-1}\alpha^2=\gamma^{-2n}>
\quad (n\in \NN)$\\
& $\alpha=R_1(\pi)R_3(\theta)L_{\bm{a}},\gamma=R_3(\pi)L_{\bm{c}}$;\\
& $2n c^3=-(a^2,a^1)R(\theta)
\begin{pmatrix}{c}a^1-c^1\\ a^2-c^2\end{pmatrix}+\Delta(a,c)$\\
\\
$T^3(n)/\ZR_3$ 
& $<\alpha,\beta,\gamma| \gamma\alpha\gamma^{-1}=\beta,
\gamma\beta\gamma^{-1}=\alpha^{-1}\beta^{-1},[\alpha,\beta]=\gamma^{3n}> 
\quad(n\in\NN)$\\
& $\alpha=L_{\bm{a}},\beta=L_{\bm{b}},\gamma=R_3(\pm{2\pi\over3}) 
L_{\bm{c}}$;\\
&$\hat{\bm{b}}=R\hat{\bm{a}},
\hat{\bm{c}}
=\left({a^3-b^3\over\Delta(a,b)}R+{1\over2} 
-{a^3+2b^3\over\Delta(a,b)}\right)\hat{\bm{a}}$,\\
&$c^3={\Delta(a,b)\over 3n}+{1\over 6}\Tp\hat{\bm{c}}R
\hat{\bm{c}}{}^*$
with $R=R(\pm{2\pi\over3})$ \\
\\
$T^3(n)/\ZR_4$ 
& $<\alpha,\beta,\gamma| \gamma\alpha\gamma^{-1}=\beta^{-1},
\gamma\beta\gamma^{-1}=\alpha,[\alpha,\beta]=\gamma^{4n}> 
\quad(n\in\NN)$\\
& $\alpha=L_{\bm{a}},\beta=L_{\bm{b}},
\gamma=R_3(\pm{\pi\over2})L_{\bm{c}}$;\\
& $a^3=b^3=0, 
\hat{\bm{b}}=R(\pm{\pi\over2})\hat{\bm{a}},
\bm{c}=(0,0,\Delta(a,b)/4n)\not=0$\\
\\
$T^3(n)/\ZR_6$ 
& $<\alpha,\beta,\gamma| \gamma\alpha\gamma^{-1}=\beta,
\gamma\beta\gamma^{-1}=\alpha^{-1}\beta,[\alpha,\beta]=\gamma^{6n}> 
\quad(n\in\NN)$\\
& 
$\alpha=L_{\bm{a}},\beta=L_{\bm{b}},\gamma=R_3(\pm{\pi\over3})L_{\bm{c}}
$;\\
&$\hat{\bm{b}}=R\hat{\bm{a}},\hat{\bm{c}}=\left(-{a^3\over\Delta(a,b)}-{1
\over2} +{a^3-b^3\over\Delta(a,b)}R\right)\hat{\bm{a}}$, \\
&$c^3={\Delta(a,b)\over 6n}+{1\over 2}\Tp\hat{\bm{c}}R\hat{\bm{c}}{}^*$
with $R=R(\pm{\pi\over3})$\\
\\
\hline\hline
\end{tabular}
\end{table}

\subsection{Phase space}

Diffeomorphism classes of the orientable compact closed 3-manifold 
modeled on $\Nil$ are grouped into 7 subclasses, each of which is 
further classified by a positive integer, as listed in Table 
\ref{table:Nil:Pi1}. Although the representations of the fundamental 
groups have rather complicated structures for many of them, we have 
to treat only topologies with rather simple structures in the 
present paper because the phase space for the other complicated 
topologies is easily determined by that for the corresponding vacuum 
system discussed in Paper I.

To see this, first consider the case in which the invariance group 
$G$ contains $\BTII_L\sdp D_2$, i.e., $G=\BTII_L\sdp D_2$ or 
$G=\Isom(\Nil)$. In this case, all the spatial components of the 
fluid velocity must vanish. Hence, as in the corresponding case for 
the type $E^3$, the invariant phase space for the fluid system is 
simply given by adding the energy density of each component of the 
fluid to that for the vacuum system. Next, since the momentum 
constraint is expressed in terms of the invariant basis 
\eqref{InvariantBasis:II} as
\Eq{
H_1\equiv 2P^2_3+cu_1=0,\ 
H_2\equiv -2P^1_3+cu_2=0,\ 
H_3\equiv cu_3=0,
\label{MomentumConstraint:II}
}
$u_1=u_2=u_3=0$ if the system has $I_3$ symmetry. Hence, from the 
argument on the vacuum system in Papar I, the system is invariant 
under $\BTII_L\sdp D_2$ if the fluid has a single component. 

A similar situation arises for $G=\Isom_0(\Nil)$, which can be 
realized only for $\Sigma=T^3(n)/\ZR_k$($k=1,2,3,4,6$). Since 
$\N(\Isom_0(\Nil))=\N(\Isom(\Nil))$, the argument for the moduli 
sector for $G=\Isom_0(\Nil)$ is the same as that for 
$G=\Isom(\Nil)$, and the matrices $Q$ and $P$ become diagonal due to 
symmetry for both cases. Hence the difference in the structure of 
the  phase space comes from the fluid sector, for which $u_1=u_2=0$ 
by symmetry. In the single component case, $u_3=0$ is further 
required by $H_3=0$, hence the system is $\Isom(\Nil)$ invariant if 
it is invariant under 
$\Isom_0(\Nil)$. This does not hold for a multi-component system 
because $u_3$ of each fluid component may not be zero, but it does not 
affect the argument on the geometrical sector because HPDs fixing the 
moduli parameter does not affect $u_3$. Hence, the phase space for a 
multi-component system with $G=\Isom_0(\Nil)$ is obtained by adding the 
variables $(u_3,\rho)$ for each extra-component of fluid to that for 
the single component system with $G=\Isom(\Nil)$.  

Thus, we only have to consider in detail the cases $G=\BTII_L$ or 
$G=\BTII_L\sdp\Set{1,I_1}$ and a multi-component system with 
$G=\BTII_L\sdp\Set{1,I_3}$. From Table \ref{table:Nil:Pi1}, we 
immediately see that $\pi_1(\Sigma)$ can be embedded into $G$ only 
when $\Sigma=T^3(n)$ for $G=\BTII_L$, $\Sigma=T^3(n), K^3(n)$ for 
$G=\BTII_L\sdp\Set{1,I_1}$, and $\Sigma=T^3(n),T^3(n)/\ZR_2$ for 
$G=\BTII_L\sdp\Set{1,I_3}$. Since the moduli freedom is uniquely 
specified by the matrix $K=(\bm{a}\;\bm{b}\;\bm{c})$ in these cases, 
we call it the moduli matrix as before.

\subsubsection{$G=\BTII_L$}

In this case $\Sigma=T^3(n)$, and from the formula on 
$fL_{\bm{d}}f^{-1}$ in Table \ref{table:Nil:Pi1} it is easy to show 
that the moduli matrix $K$ can be put into the canonical form
\Eq{
K=\begin{pmatrix}{ccc}
  1 & 0 & 0 \\
  0 & 1 & 0 \\
  0 & 0 & 1/n
\end{pmatrix}.
\label{K:(T^3(n),II)}}
Hence the moduli sector is trivial. This matrix is invariant under 
$f\in\N(\BTII_L)$ only when $f=R_{\bm{b}}$. From \eqref{F:II}, we 
can put $Q_{13}=Q_{23}=0$ by this residual HPD. On the other hand, 
$u_3=0$, and $P^{13}$ and $P^{23}$ are determined by $u_1,u_2$ and 
$Q_{33}$. Since $Q$ can be always diagonalized by HPDs if we neglect 
the moduli sector, the system has a higher symmetry when $u_1$ or 
$u_2$ vanishes. Hence, the dynamical sector of the invariant phase 
space for the single component fluid system is given by
\Eq{
\Gamma_{\rm dyn}(T^3(n),\BTII_L)
=\left\{(\hat Q,Q_{33};\hat P,P^{33};u_1,u_2,\rho\mid
\det\hat Q>0,Q_{33}>0,u_1u_2\not=0 \right\}.
}
For a multi-component system, each extra component adds 4 to $N_s$.

\subsubsection{$G=\BTII_L\sdp\Set{1,I_1}$}

\paragraph{A. $T^3(n)$:} In this case, $\pi_1(\Sigma)$ is embedded 
in $\BTII_L$, and we can show that the moduli matrix $K$ can be 
always put by a HPD into the form 
\Eq{
K=\begin{pmatrix}{ccc}
   1 & X &  0 \\
   1 & Y &  0 \\
   0 & 0 & \frac{Y-X}{n}
\end{pmatrix},
}
where $X\not=Y$. The modular transformation group of $\pi_1(T^3(n))$ 
consists of the transformations $
(\alpha,\beta)\mapsto 
(\alpha^p\beta^q\gamma^{u},\alpha^r\beta^s\gamma^{v})$,
where $Z=\begin{pmatrix}{cc}p & r \\ q & s\end{pmatrix}$ is a matrix in 
$\GL(2,\ZR)$, and $u$ and $v$ are integers. By appropriate combinations 
with HPDs, they induce a discrete transformation group isomorphic to 
$\GL(2,\ZR)$ which preserves the above form of $K$. 

In the dynamical sector, the symmetry requires that 
$Q_{12}=Q_{13}=P^{12}=P^{13}=0$ and $u_2=u_3=0$. We can put $Q_{23}=0$ 
and diagonalize $Q$ with the help of the residual HPD 
$R_{\hat{\bm{b}}}$ with $\hat{\bm{b}}=(b,0)$. Then the momentum 
constraints $H_2=0$ and $H_3=0$ become trivial, and $H_1=0$ can be used 
to express $P^{23}$ by $u_1$. Hence, taking account of the residual 
discrete transformations, the invariant phase space is given by
\Eqr{
&&\Gamma_{\rm inv}(T^3(n),\BTII_L\sdp\Set{1,I_1})
=\big\{(Q_1,Q_2,Q_3;P^{11},P^{22},P^{33};u_1,\rho;X,Y)
\big|
\nonumber\\
&&\qquad\qquad Q_1,Q_2,Q_3>0,u_1\not=0,X\not=Y\big\}
/\GL(2,\ZR).
}
Here it can be checked that the action of $\GL(2,\ZR)$ is properly 
discontinuous and the invariant phase space has a smooth manifold 
structure. 

\paragraph{B. $K^3(n)$:} Since no transformation in $G$ contains the 
$R_3(\theta)$ factor, $R(\theta)$ in the representation of 
$\pi_1(K^3(n))$ in Table \ref{table:Nil:Pi1} must be unity, and hence 
$a^1=0$. Further, with the help of the formulas for the conjugate 
transformations in Table \ref{table:Nil:normalizer}, we can show that 
by a HPD $K$ can be put to the canonical form
\Eq{
K=
\begin{pmatrix}{ccc}
   0 & 1 & 0 \\
   1 & 0 & 0 \\
   0 & 0 & -1/n
\end{pmatrix}.
}
The argument on the dynamical sector is the same as that for $T^3(n)$. 
Hence the invariant phase space is given by
\Eqr{
&\Gamma_{\rm inv}(K^3(n),\BTII_L\sdp\Set{1,I_1})
=&\big\{(Q_1,Q_2,Q_3;P^{11},P^{22},P^{33};u_1,\rho)
\big|
\nonumber\\
&&\qquad Q_1,Q_2,Q_3>0,u_1\not=0\big\}.
}
%

\begin{table}[t]
\caption{\label{table:count:Nil} The parameter counts for type Nil.
 }\begin{tabular}{llccccccc}
\hline
\hline
\bf Space & \bf Symmetry & $N_Q$ & $N_P$ & $N_m$ & $N_f$ & $N$ & $N_s$ 
& $N_s$(vacuum)\\
\hline\\
$\RF^3$
&$\BTII_L$	& 1 & 3 & 0 & 3 & 7 & 5 & -- \\
&$\BTII_L\sdp\Set{1,I_1}$
		& 1 & 3 & 0 & 2 & 6 & 4 & -- \\
&$\BTII_L\sdp D_2$
		& 1 & 3 & 0 & 1 & 5 & 3 &   2 \\
&$\Isom(\Nil)$	& 1 & 2 & 0 & 1 & 4 & 2 &   1 \\
&&&&&&&&\\
$T^3(n)$
&$\BTII_L$  	& 4 & 4 & 0 & 3 & 11& 9 & -- \\
&$\BTII_L\sdp\Set{1,I_1}$
		& 3 & 3 & 2 & 2 & 10& 8 & -- \\
&$\BTII_L\sdp D_2$
		& 3 & 3 & 2 & 1 & 9 & 7 &  6 \\
&$\Isom(\Nil)$	& 2 & 2 & 2 & 1 & 7 & 5 &  4 \\
&&&&&&&&\\
$K^3(n)$ 
&$\BTII_L\sdp\Set{1,I_1}$
		& 3 & 3 & 0 & 2 & 8 & 6 & -- \\
&$\BTII_L\sdp D_2$
		& 3 & 3 & 0 & 1 & 7 & 5 &  4 \\
&$\Isom(\Nil)$ 	& 2 & 2 & 1 & 1 & 6 & 4 &  3 \\
&&&&&&&&\\
$T^3(n)/\ZR_2$
&$\BTII_L\sdp D_2$
		& 3 & 3 & 2 & 1 & 9 & 7 &  6 \\
&$\Isom(\Nil)$	& 2 & 2 & 2 & 1 & 7 & 5 &  4 \\
&&&&&&&&\\
$T^3(n)/\ZR_2\times\ZR_2$
&$\BTII_L\sdp D_2$
		& 3 & 3 & 0 & 1 & 7 & 5 &  4  \\
&$\Isom(\Nil)$	& 2 & 2 & 1 & 1 & 6 & 4 &  3  \\
&&&&&&&&\\
$T^3(n)/\ZR_k(k=3,4,6)$
&$\Isom(\Nil)$	& 2 & 2 & 0 & 1 & 5 & 3 &  2  \\
\hline\hline
\end{tabular}
\end{table}

\subsubsection{$G=\BTII_L\sdp\Set{1,I_3}$}

As discussed above, this case occurs only for a multi-component fluid 
system.

\paragraph{A. $T^3(n)$:} In this case $\pi_1(\Sigma)$ is contained 
in $\BTII_L$ and the effective action of 
$\N(\BTII_L\sdp\Set{1,I_3})$ is the same as that of $\N(\BTII_L)$. 
Hence the moduli sector is trivial. In the dynamical sector, the 
invariance and the momentum constraints require that 
$Q_{13}=Q_{23}=P^{13}=P^{23}=0$ and $u_I=0$. Therefore, 
$N_Q=4,N_P=4,N_m=0$ and $N_f=2(n_f-1)+1$, where $n_f$ is the number 
of the fluid components. Note that for $n_f=1$, the total degrees of 
freedom $N$ coincides with that for $(T^3(n),\BTII_L\sdp D_2)$ as 
expected, but the assignments of the degrees of freedom among 
$N_Q,N_P$ and $N_m$ are different. What is happening is that for 
$n_f=1$, the system really has the higher symmetry $\BTII_L\sdp D_2$ 
and two combinations of $Q$ and $P$ become constants of motion. By a 
HPD in $\N(\BTII_L\sdp\Set{1,I_3})$, these two constants of motion 
can be transfered to the moduli freedom.

\paragraph{B. $T^3(n)/\ZR_2$:} In this case the representation of 
the generator $\gamma$ contains the $I_3$ factor. From the formulas 
in Table \ref{table:Nil:normalizer}, we can show that $c^1$ and 
$c^2$ can be put to zero by $f=L_{\bm{d}}$. Then $a^3$ and $b^3$ 
also vanish. Further, we can put $(\hat{\bm{a}},\hat{\bm{b}})$ to 
the unit matrix by $f=D(\hat A,\Delta)$. Hence the moduli matrix $K$ 
can be transformed to the constant matrix \eqref{K:(T^3(n),II)}. The 
argument on the dynamical sector is the same as that for the 
previous case $T^3(n)$. Hence the degrees of freedom are again given 
by $N_Q=N_P=4,N_m=0$ and $N_f=2n_f-1$.

\section{$\Sol$}

\subsection{Maximal geometry and invariance groups}

$\Sol$ is the maximal geometry $(\RF^3,\Isom(\Sol))$ obtained by 
extension of the Bianchi type $\BTVI_0$ symmetry. In order to 
determine $\Isom(\Sol)$, let us first summarize basic properties of 
the Bianchi type $\BTVI_0$ group.

The $\BTVI_0$ group has the multiplication structure
\Eq{
(a,b,c)(x,y,z)=(a+e^{-c}x, b+e^c y, c+z).
}
The left and the right transformations on $\BTVI_0$ defined by this 
multiplication are expressed as
\Eqr{
&& L_{\bm{a}}=(\bm{a}, B(a^3)) \in \IGL(3),\\
&& R_{\bm{b}}(\bm{x})=\bm{x}+B(z)\bm{b},
}
where $B(c)=D(e^{-c},e^{c},1)$.
The Lie algebra of the left transformation group is generated by
\Eq{
\xi_1=\partial_x, \quad
\xi_2=\partial_y, \quad
\xi_3=\partial_z-x\partial_x+y\partial_y,
}
with the commutation relations
\Eq{
[\xi_1,\xi_2]=0,\quad
[\xi_3,\xi_1]=\xi_1,\quad
[\xi_3,\xi_2]=-\xi_2.
}
We adopt the following invariant basis in this paper:
\Eqr{
&& \chi^1=e^z dx+e^{-z}dy, \quad 
\chi^2=e^z dx - e^{-z}dy,\quad 
\chi^3=dz;
\label{InvariantBasis:VI0}\\
&& d\chi^1=\chi^3\wedge\chi^2,\quad
d\chi^2=\chi^3\wedge\chi^1,\quad
d\chi^3=0.
}

The normalizer of $(\BTVI_0)_L$ is determined in the following way. 
First, since $\xi_1$ and $\xi_2$ are the generators of linearly 
independent one-dimensional invariant subspaces of the adjoint 
representation of $\BTVI_0$, $f_*(\xi_1,\xi_2)$ is written as either 
$(k_1\xi_1, k_2\xi_2)$ or 
$(k_1\xi_2,k_2\xi_1)$ for $f\in\N((\BTVI_0)_L)$. The commutation 
relations require that $f_*\xi_3$ must be written as 
$\xi_3+a^1\xi_1+a^2\xi_2$ in the former case and as 
$-\xi_3+a^1\xi_2+a^2\xi_1$ in the latter case. Hence the 
automorphism induced by $f$ has the form
\Eq{
f_*(\xi_1,\xi_2,\xi_3)=(\xi_1,\xi_2,\xi_3)\Set{1,J}N(\hat{\bm{a}})
D(k_1,k_2,1),
}
where 
\Eq{
J=R_3(\pi/2)R_1(\pi)
=\begin{pmatrix}{ccc}
  0 & 1 & 0 \\
  1 & 0 & 0 \\
  0 & 0 & -1
\end{pmatrix}.
}
It is easiy to see that $f=L_{\hat{\bm{a}}}$ induces 
$N(\hat{\bm{a}})$, and the linear transformations represented by the 
matrices $J$ and $D(k_1,k_2,1)$ induce the automorphisms 
represented by the same matrices, respectively. Hence the generic 
element $f$ of  $\N((\BTVI_0)_L)$ is written as 
$f=\Set{1,J}L_{\hat{\bm{a}}}R_{\bm{b}}D(k_1,k_2,1)$, which can be 
rewritten as 
\Eq{
f=\Set{1,J}L_{\bm{a}}R_{\hat{\bm{b}}}D(k_1,k_2,1)
}
by appropriate redefinitions of the parameters. By this HPD the 
invariant basis \eqref{InvariantBasis:VI0} transforms as
\Eq{
f^*\chi^i=F^i{}_j\chi^j;\ 
F=\Set{1,I_1}
\begin{pmatrix}{ccc}
  k_+ & k_- & b_+ \\
  k_- & k_+ & b_- \\
  0   &  0  & 1
\end{pmatrix},
\label{F:VI0}}
where
\Eq{
k_\pm=\frac{1}{2}(k_1\pm k_2), \quad
b_\pm=-b^1\pm b^2.
}
The corresponding formulas for the conjugate transformaions are 
given in Appendix \ref{appendix:conj}.

\begin{table}[t]
\caption{\label{table:Sol:normalizer} Normalizers for the invariance 
groups with $G_0=\Sol$. The formulas for conjugate transformations 
are given in Appendix \ref{appendix:conj}} 
\begin{tabular}{lll}
\hline\hline
$G$ & u/o & $f\in\N(G)$ \\
\hline\\
$(\BTVI_0)_L$ & o 
& $\Set{1,J}R_{\hat{\bm{b}}}L_{\bm{a}}D(k_1,k_2,1)$ 
\\
$(\BTVI_0)_L\sdp \Set{1,I_3}$ & o
& $\hat{\bm{b}}=0$ 
\\
$(\BTVI_0)_L\sdp \Set{1,J}$  & o
& $b^1=b^2,\ k_1=k_2$
\\
$(\BTVI_0)_L\sdp \Set{1,I_3J}$ & o
& $b^1=-b^2,\ k_1=k_2$
\\
$\Isom^+(\Sol)$ & o 
& $\hat{\bm{b}}=0, k^1=k^2$
\\
$(\BTVI_0)_L\sdp\Set{1,I_3,-I_1,-I_2}$ & u
& $\hat{\bm{b}}=0$
\\
$(\BTVI_0)_L\sdp\Set{1,I_3,-I_1J,-I_2J}$ & u
& $\hat{\bm{b}}=0, k_1=k_2$
\\
$\Isom(\Sol)$  & u
& ibid
\\
$(\BTVI_0)_L\sdp\Set{1,-I_1}$ & u
& $R_{(0,b)}L_{\bm{a}}D(k_1,k_2,1)$ 
\\
\\
\hline\hline
\end{tabular}
\end{table}

We can show that there exists no higher dimensional isometry group 
containing $\BTVI_0$ by examining the structure of possible 
4-dimensional Lie algebra containing $\BTVI_0$. Hence $\Isom(\Sol)$ is 
the semi-direct product of $(\BTVI_0)_L$ with a discrete subgroup of 
$\N((\BTVI_0)_L)$. Such additional discrete symmetry is found in the 
following way. First, we transform the invariant metric 
$Q_{IJ}\chi^I\chi^J$ into some special form by HPDs in 
$\N((\BTVI_0)_L)$ because we are only interested in the conjugate class 
of the invariance group. From \eqref{F:VI0} we can make $Q_{13}$ and 
$Q_{23}$ vanish by $f=R_{\hat{\bm{b}}}$ to obtain $Q=D(\hat Q,Q_3)$. 
Then the tansformation matrix $F$ with $\hat{\bm{b}}=0$ acts on $\hat 
Q$ as the combination of a 2-dimensional Lorentz transformation and a 
dilation, in terms of which we can further put $Q=D(Q_1,1/Q_1,Q_3)$ 
(see the argument on the diagonalization of $Q$ by the Lorentz 
transformation in \S\ref{sec:SL2:PhaseSpace}). There remains no 
continuous HPD freedom preserving this metric form. Hence a canonical 
form of the $\Sol$ metric is given by
\Eq{
ds^2=Q_1(\chi^1)^2+Q_1^{-1}(\chi^2)^2+Q_3(\chi^3)^2.
\label{Q:Sol+}}
Clearly the transformation $F$ of the invariant basis preserving 
this form of the metric is either $I_1,I_3,-J$ or their 
multiplication and  is induced from the HPD $f=J, I_3,-I_1$ or their 
multiplication, respectively. These HPDs altogether form the 
discrete group
\Eq{
\bar D_4=\Set{1,I_3,J,I_3J,-I_1,-I_2,-I_1J,-I_2J}.
}
The above metric is invariant by this group, if and only if $Q_1=1$. 
Hence the maximal geometry including $\BTVI_0$ is given by
\Eq{
\Isom(\Sol)=(\BTVI_0)_L\sdp \bar D_4,
}
and the corresponding canonical metric is written as
\Eq{
ds^2=e^{2z}dx^2+e^{-2z}dy^2+Q_3dz^2.
}

Here note that $\Isom(\Sol)$ contains orientation reversing 
tranformations and that $Q_1=1$ is required by the invariance under 
such transformations. Hence, when we consider only the 
orientation-preserving transformations, the isometry group is 
reduced to%
\Eq{
\Isom^+(\Sol)=(\BTVI_0)_L\sdp\bar D_2,
}
where 
\Eq{
\bar D_2=\Set{1,I_3,J,I_3J},
}
and the most generic invariant metric is given by \eqref{Q:Sol+}.

Since the invariance group $G$ is always a semi-dicrect product of 
$\BTVI_0$ and a subgroup of $\bar D_4$, its conjugate class is easily 
classified. The result is given in Table \ref{table:Sol:normalizer} 
with the corresponding normalizer groups.

\begin{table}[t]
\caption{\label{table:Sol:Pi1}Fundamental groups and their
representation in $\Isom^+(\Sol)$ of compact, closed orientable
3-manifolds of type $\Sol$.}
\begin{tabular}{ll}
\hline
\hline
\bf Space & \bf Fundamental group and  representation \\
\hline
\\
$\Sol(n;\omega_1,\omega_2)$ 
& $[\alpha,\beta]=1,\gamma\alpha\gamma^{-1}=\alpha^p\beta^q,
\gamma\beta\gamma^{-1}=\alpha^r\beta^s;\quad
\begin{pmatrix}{cc}p&q\\r&s\end{pmatrix}\in\SL(2,\ZR)$\\
& \\
& $n=p+s, \omega_1={p-s+\sqrt{n^2-4}\over 2r},
\omega_2={p-s-\sqrt{n^2-4}\over 2r}$
\\
\\
$n>2:$ 
& $\alpha=L_{\bm{a}}; \bm{a}=(b^1\omega_1,b^2\omega_2,0)$\\
& $\beta=L_{\bm{b}};\bm{b}=(b^1,b^2,0)\quad(b^1b^2\not=0)$\\
& $\gamma=L_{\bm{c}}; \quad e^{c^3}={n+\sqrt{n^2-4}\over 2}$\\\\
$n<-2:$ 
& $\alpha=L_{\bm{a}}; \bm{a}=(b^1\omega_1,b^2\omega_2,0)$\\
& $\beta=L_{\bm{b}};\bm{b}=(b^1,b^2,0)\quad(b^1b^2\not=0)$\\
& $\gamma=R_3(\pi)L_{\bm{c}}; \quad e^{c^3}={|n|+\sqrt{n^2-4}\over 
2}$\\\\
\hline\hline
\end{tabular}
\end{table}

\subsection{Phase space}

A compact quotient of $\Sol$ is a torus bundle over a circle which 
is obtained by gluing the two boundaries of $T^2\times [0,1]$ by a 
hyperbolic large diffeomorphism of $T^2$\cite{Scott.P1983}. Hence 
its diffeomorphism class is classified by the $\SL(2,\ZR)$-conjugate 
class of the $\SL(2,\ZR)$ matrix 
$Z=\begin{pmatrix}{cc} p & q \\ r & s \end{pmatrix}$ with $|p+s|>2$ 
specifying the automorphism of $\pi_1(T^2)$ for the gluing. The 
fundamental groups with different values of $n=p+s$ are not 
isomorphic, and each class is uniquely specified by $n=p+s$ and the 
roots $\omega_1$ and $\omega_2$ of the characteristic equation 
$rx^2+(s-p)x-q=0$, but there exists finitely many different classes 
with the same $n$ in general (see the explanation in Paper I for 
further details).

As was discussed in the previous subsection, $Q$ can be diagonalized by 
a HPD in $\N(\BTVI_0)$. Further, with respect to the invariant basis 
\eqref{InvariantBasis:VI0}, the momentum constraints are written as
\Eq{
H_1\equiv -2P^3_2 +c u_1=0, \ 
H_2\equiv -2P^3_1 +c u_2=0, \ 
H_3\equiv 2(P^1_2+P^2_1)+cu_3=0.
\label{MomentumConstraint:VI0}}
Hence in the vacuum case for which $c=0$, $P$ also becomes diagonal, 
and the system always has the maximal symmetry $\Isom^+(\Sol)$.  
However, for the fluid system with $(u_I)\not=0$, this symmetry is 
broken, and the system can have lower symmetries listed in Table 
\ref{table:Sol:normalizer}. Conversely, if the system is invariant 
under the maximal group, all the velocities vanish and the structure 
of the phase space for a fluid system is easily determined by that 
for the vacuum system as in the other Thurston types discussed so 
far. 

\subsubsection{$G=(\BTVI_0)_L$}

From the structure of the fundamental group in Table 
\ref{table:Sol:Pi1}, only the space $\Sol(n)$ with $n>2$ can have this 
invariance group. From the formula for 
$L_{\bm{d}}L_{\bm{c}}L_{\bm{d}}^{-1}$ in the appendix, we see that 
$c^1$ and $c^2$ of $\gamma=L_{\bm{c}}$ can be put to zero by the 
transformation $f=L_{\bm{d}}$ because $c^3>0$. Further $b_1$ and $b_2$ 
can be put to $\pm1$ by the transformation $D(k_1,k_2,1)$($k_1k_2>0$) 
in $\N^+((\BTVI_0)_L)$. Hence the moduli matrix $K$ can be put to the 
canonical form
\Eq{
K=
\begin{pmatrix}{ccc}
  \omega_1  & 1   & 0 \\
  \pm\omega_2&\pm1& 0 \\
  0         & 6   &c^3
\end{pmatrix},
}
which implies that the moduli sector of the invariant phase space 
consists of two points representing two different orientations of the 
same space.

Because $L_{\bm{d}}$ with $\bm{d}=(0,0,d)$ induces the conjugate 
transformation $B(d)=D(e^{-d},e^d,1)$, the isotropy group  of the 
action of $\N((\BTVI_0)_L)$ at the above moduli matrix in the moduli 
space is generated by $R_{\bm{\hat a}}$ and $B(-d)$. In terms of the 
former transformation we can put $Q$ to the form $D(\hat Q,Q_3)$. 
Further, since the latter transformation induces a Lorentz boost, we 
can diagonalize $\hat Q$ by it. Then, by the momentun constraints, the 
non-diagonal entries of $P$ are expressed in terms of $u_I$. Therefore, 
taking into account that the system has a higher symmetry if two 
components of $(u_I)$ vanishes, we find that the invariant phase space 
is given by
\Eqr{
&&\Gamma_{\rm inv}(\Sol(n)(n>2),\BTVI_0)=
\big\{(Q_1,Q_2,Q_3;P^{11},P^{22},P^{33};u_1,u_2,u_3,\rho)
\big|
\nonumber\\ 
&& \qquad \qquad
Q_1,Q_2,Q_3>0,u_1^2u_2^2+u_2^2u_3^2+u_1^2u_2^2\not=0\big\}\times \ZR_2.
}

For a multi-component system, $N_f$ simply increases by 4 for each 
extra component.

\subsubsection{$G=(\BTVI_0)_L\sdp\Set{1,I_3}$}

First, let us consider the space $\Sol(n)$ with $n>2$. For this space, 
the fundamental group is still embedded into $(\BTVI_0)_L$, and the 
effective action of HPDs on the moduli is the same as for 
$G=(\BTVI_0)_L$. Hence the moduli sector is again given by two points. 
The arugment on the dynamical sector is also the same except that the 
symmetry requires that $Q=D(\hat Q,Q_3)$, $P=(\hat P,P^3)$ and 
$u_1=u_2=0$, for which the only non-trivial momentum constraint is 
$H_3=0$. 

Next for $\Sol(n)$ with $n<-2$, the generator $\gamma$ contains the 
$I_3$ transfomation. However, this does not introduce any problem and 
essensially the same argument as that for $n>2$ applies. Therefore, 
irrespective of the sign of $n$, the invariant phase space is given by
\Eqr{
&\Gamma_{\rm inv}(Sol(n),\BTVI_0\sdp\Set{1,I_3})
=&\big\{(Q_1,Q_2,Q_3,P^{11},P^{22},P^3:u_3,\rho)
\big|
\nonumber\\
&&\qquad Q_1,Q_2,Q_3>0,u_3\not=0 \big\}\times\ZR_2.
}
For a multi-component system, $N_f$ simply increases by 2 for each 
extra component.

\subsubsection{$G=(\BTVI_0)_L\sdp\Set{1,J}$}

This symmetry is realized only for $\Sol(n)$ with $n>2$. Now, 
$D(k_1,k_2,1)$ belongs to $\N(G)$ only when $k_1=k_2$. However, since 
$L_{(0,0,d)}$ induces the $D(e^{-d},e^d,1)$ transformation of the 
moduli vector, the moduli sector is reduced to two points as in the 
case $G=(\BTVI_0)_L$. 

In the dynamical sector, since $f=J$ induces the $I_1$ transformation 
of the invariant basis, the symmetry requires that $Q=(Q_1,\hat Q)$, 
$P=(P^1,\hat P)$ and $u_2=u_3=0$. $Q$ can be completely diagonalized in 
terms of the HPD $R_{(b,b)}$. Further, the non-trivial momentum 
constraint $H_1=0$ can be used to express $P^{23}$ by $u_1$. Hence the 
structure of the invariant phase space is the same as that for 
$G=(\BTVI_0)_L\sdp\Set{1,I_3}$ irrespective of the number of the fluid 
components:
\Eqr{
&\Gamma_{\rm inv}(Sol(n)(n>2),\BTVI_0\sdp\Set{1,J})
=&\big\{(Q_1,Q_2,Q_3,P^1,P^{22},P^{33}:u_1,\rho)
\big| \nonumber\\
&& \quad Q_1,Q_2,Q_3>0,u_1\not=0 \big\}\times\ZR_2.
}
%

\subsubsection{$G=(\BTVI_0)_L\sdp\Set{1,JI_3}$}

This symmetry is realized only for $\Sol(n)$ with $n>2$ again. The 
argument in the moduli sector is essentially the same as that for the 
previous case because the effective action of HPDs in the moduli sector 
is the same. Further, since $JI_3$ induces the $I_2$ transformation of 
the invariant basis and $R_{(b,-b)}$ shifts $Q_{13}$, the argument on 
the dynamical sector is just the repetition of the previous one with 
replacing the role of $\chi^1$ and $\chi^2$. Hence,
\Eqr{
&\Gamma_{\rm inv}(Sol(n)(n>2),\BTVI_0\sdp\Set{1,JI_3})
=&\big\{(Q_1,Q_2,Q_3,P^{11},P^2,P^{33}:u_2,\rho)
\big|\nonumber\\
&& \quad  Q_1,Q_2,Q_3>0,u_2\not=0\big\}\times\ZR_2.
}
%


\begin{table}[t]
\caption{\label{table:count:Sol} The parameter 
count for type Sol.}\begin{tabular}{llccccccc}
\hline
\hline
\bf Space & \bf Symmetry & $N_Q$ & $N_P$ & $N_m$ & $N_f$ & $N$ & $N_s$ & $N_s$(vacuum)\\
\hline\\
$\RF^3$ 
&$\BTVI_0$ 	& 2 & 3 & 0 & 4 & 9 & 7 & -- \\
&$\BTVI_0\sdp \ZR_2$
                & 2 & 3 & 0 & 2 & 7 & 5 & -- \\
&$\Isom^+(\Sol)$& 2 & 3 & 0 & 1 & 6 & 4 &   3 \\
&&&&&&&&\\
$\Sol(n)$ ($n>2$)
&$\BTVI_0$ 	& 3 & 3 & 0 & 4 & 10& 8 & -- \\
&$\BTVI_0\sdp \ZR_2$
                & 3 & 3 & 0 & 2 & 8 & 6 & -- \\
&$\Isom^+(\Sol)$& 3 & 3 & 0 & 1 & 7 & 5 &  4 \\
&&&&&&&&\\
$\Sol(n)$ ($n<-2$)
&$\BTVI_0\sdp \ZR_2$
                & 3 & 3 & 0 & 2 & 8 & 6 & -- \\
&$\Isom^+(\Sol)$& 3 & 3 & 0 & 1 & 7 & 5 &  4 \\
\\
\hline\hline
\end{tabular}
\end{table}

\section{$H^2\times\RF$}\label{sec:H^2xR}

\subsection{Invariance subgroups and normalizers}

\subsubsection{Bianchi type III group}\label{sec:BTIII}

Space $H^2\times\RF$ has the group structure of the Bianchi type III. 
To see this, let us embed the hyperbolic surface in the 
3-dimensional Minkowski space $E^{2,1}$,
\Eqr{
&& H^2: T=(1+X^2+Y^2)^{1/2},\\
&& ds^2=-dT^2+dX^2+dY^2,
}
into the space of symmetric $\SL_2\RF$ matrices with positive 
diagonal elements by
\Eq{
S=\begin{pmatrix}{cc}
  T+X & Y \\
  Y   & T-X
\label{embedding:H^2toSL2}
\end{pmatrix}.
}
This matrix can be always uniquely written  as
\Eq{
S=Z\Tp Z
}
in terms of a matrix $Z$ in $\HSL^+_2$, where
\Eq{
\HSL^+_2=\SetDef{
Z=\begin{pmatrix}{cc}
  p &  0 \\
  q &  1/p 
  \end{pmatrix}
}{p>0, q\in\RF}.
}
Conversely, for any $Z\in \HSL^+_2$, $S=Z\Tp Z$ is always written in 
the form \eqref{embedding:H^2toSL2}. Further, $\HSL_2^+$ forms a 
2-dimensional closed subgroup of $\SL_2\RF$. Hence the embedding gives 
a one-to-one correspondence between $H^2$ and the group 
$\HSL_2^+$ and defines a group structure in the hyperbolic space 
$H^2$.

In this and the next sections, we use the complex variable $\zeta$ 
defined by
\Eq{
\zeta =\frac{q}{p} + \frac{i}{p^2},
}
to parametrize the $\HSL_2^+$ matrix $Z$ as $Z(\zeta)$. This 
parametrization gives a one-to-one mapping from the hyperbolic 
surface to the upper half part of the complex plain 
$\SetDef{\zeta=x+iy}{x\in\RF,y>0}$. Through this mapping, we 
identify the hyperbolic surface $H^2$ with the upper half complex 
plain. Since the original Minkowski coordinates are expressed in 
terms of $\zeta$ as
\Eq{
T=\frac{1+|\zeta|^2}{2\Im \zeta},\ 
X=\frac{1-|\zeta|^2}{2\Im \zeta},\ 
Y=\frac{\Re \zeta}{2\Im \zeta},
}
the hyperbolic metric on $H^2$ is written in terms of $\zeta=x+iy$ as
\Eq{
ds^2=\frac{dx^2+dy^2}{4y^2}.
\label{metric:H^2}}

The group structure of $H^2$ defined in this way together with the 
natural Abelian group structure of $\RF$ makes $H^2\times \RF$ a 
3-dimensional group. In terms of the coordiantes $(x,y,z)\in 
H^2\times\RF$, the corresponding multiplication is written as 
\Eq{
(a,b,c)\cdot (x,y,z)=(a+bx,by,z+c).
\label{III:product}
}
Infinitesimal left transformations defined by this multiplication 
are generated by
\Eq{
\xi_1=\partial_x,\ 
\xi_2=x\partial_x + y\partial_y,\ 
\xi_3=\partial_z
\label{III:generators}
}
which form the Bianchi type III Lie algebra,
\Eq{
[\xi_1,\xi_2]=\xi_1,\ 
[\xi_3,\xi_1]=0,\ 
[\xi_3,\xi_2]=0.
}
We denote this left transformation group of $H^2\times \RF$ by 
$\BTIII_L$. The bases of $\BTIII_L$-invariant 1-forms are given by
\Eqr{
&& \chi^1=\frac{dx}{y},\ 
   \chi^2=\frac{dy}{y},\ 
   \chi^3=dz;
   \label{III:InvariantBasis}\\
&& d\chi^1=\chi^1\wedge\chi^2, \ 
   d\chi^2=0,\ 
   d\chi^3=0.
}

Let us detemine the normalizer of $\BTIII_L$ for later uses. Since 
the center of $\BTIII_L$ is generated by $\xi_3$, and 
$[\BTIII_L,\BTIII_L]$ is generated by $\xi_1$, from 
Prop.\ref{prop:normalizer:invariantspace}, 
$\phi\in\Aut(\LieA(\BTIII_L))$ has the form
\Eq{
\phi(\xi_1)=a\xi_1, \phi(\xi_2)=b^I\xi_I, \phi(\xi_3)=c\xi_3.
\label{III:AutLie}
}
From $[\phi(\xi_2),\phi(\xi_2)]=ab^2\xi_1=\phi(\xi_1)$, the 
condition for $\phi$ to be an automorphism is given  by $b^2=1$. HPDs 
generating these automorphisms are determined as follows. First, the 
discrete transformation $(-I_1): (x,y,z)\mapsto (-x,y,z)$ 
induces $(-I_1): (\xi_1,\xi_2,\xi_3) \mapsto (-\xi_1,\xi_2,\xi_3)$. 
Second, $Z_L(\alpha) \in H^2_L$ with $\alpha=\alpha_1+i\alpha_2$ as a 
transformation in $\BTIII_L$ induces the automorphism
\Eq{
(Z_L)_* (\xi_1,\xi_2)=(\xi_1,\xi_2)
\begin{pmatrix}{cc}
  1/\alpha_2 & \alpha_1/\alpha_2 \\
  0   & 1
\end{pmatrix}, (Z_L)_*(\xi_3)=\xi_3.
}
Third, the linear transformation
\Eq{
D(1,1,c): (x,y,z) \mapsto (x,y,cz)
}
gives the automorphism $(\xi_1,\xi_2,\xi_3)\mapsto 
(\xi_1,\xi_2,c\xi_3)$. Finally, the transformation
\Eq{
N'(b): (x,y,z) \mapsto (x,y, z + b \ln y)
}
induces the automorphism $(\xi_1,\xi_2,\xi_3)\mapsto 
(\xi_1,\xi_2+b\xi_3,\xi_3)$. Combining these transformations, we can 
generate all automorphisms of the form \eqref{III:AutLie}. Hence, from 
Prop.\ref{prop:normalizer:right}, $f\in \N(\BTIII_L)$ is written as
\Eq{
f=\Set{1,-I_1,I_2,-I_3}L_{\bm{a}}
N'(b)D(1,1,k)R_{\hat{\bm{b}}},
}
where $R_{\hat{\bm{b}}}=Z_R(b_1+ib_2)\in\BTIII_R$, and we have used the 
fact that the group consisting of translations in $\RF$ of 
$H^2\times\RF$ coincides with the center of $\BTIII$, $\BTIII_L\cap 
\BTIII_R$.

\subsubsection{Symmetry of $H^2$}\label{sec:H^2}

Since the hyperbolic surface is a maximally symmetric surface with 
the isometry group given by the Lorentz transformation group 
$\OG_+(2,1)$, the metric \eqref{metric:H^2} should have the same 
invariance. In order to express this isometry group 
$\Isom(H^2)$ in terms of the coordinates $\zeta=x+iy$, we utilize 
the standard representation of $\SL_2\RF$ onto $\SO_+(2,1)$ given by
\Eq{
\SL_2\ni V: S \mapsto VS\Tp V.
\label{rep:SLtoSO}}
For $S=Z(\zeta)\Tp Z(\zeta)$, after the decomposition
\Eq{
VZ(\zeta)=Z(\zeta')R;\ \zeta'\in H^2, R\in SO(2),
}
$VS\Tp V$ is expressed as $Z(\zeta')\Tp Z(\zeta')$. Hence each 
$\SO_+(2,1)$ transformation is represented by 
the $\PSL_2\RF$ transformaton $Z(\zeta)\maps Z(\zeta')$ determined 
by this decomposition. To be explicit, it is expressed as
\Eq{
\zeta'=V*\zeta := \frac{d \zeta + c}{b\zeta + a};\ 
V=\begin{pmatrix}{cc}
  a & b\\
  c & d
\end{pmatrix} \in \SL_2.
\label{transformation:PSL_2}}
Since this transformation is holomorphic in $\zeta$, it preserve the 
orientation of $H^2$.

By definition, this transformation coincides with a left 
transformation in $H^2$ for $V \in \HSL^+(2)$. Hence, 
$\Isom_0(H^2)\cong \PSL_2\RF$ is generated by this left 
transformation and the transformation $R_H(\theta)$ corresponding to 
$V=R(\theta) \in \SO(2)$: 
\Eqr{
& H^2_L: &\xi_1, \xi_2,\\
& \SO(2): & \xi_4=\frac{1}{2}(1+x^2-y^2)\partial_x + xy\partial_y.
}
The full isometry group $\Isom(H^2)$ is generated by $\Isom_0(H^2)$ 
and the space reflection $(-I_1)$.

Next we determine the normalizer of $\Isom(H^2)$ in $\DiffG(H^2)$. 
In terms of the basis 
$\eta_1=\xi_1-\xi_4,\eta_2=\xi_2,\eta_3=\xi_4$, the commutation 
relations of $\LieA(\Isom(H^2))\cong\LieA(\SO_+(2,1))$ are expressed 
in the standard form,
\Eq{
[\eta_1,\eta_2]=\eta_3,\ 
[\eta_2,\eta_3]=-\eta_1,\ 
[\eta_3,\eta_1]=-\eta_2.
}
Hence the adjoint representation of the algebra is given by
\Eq{
\Ad(X\cdot \eta)
=\begin{pmatrix}{ccc}
  0    &  X^3  & -X^2 \\
  -X^3 &   0   &  X^1 \\
  -X^2 &  X^1  & 0
\end{pmatrix}.
}
From this it follows that the Killing form is proportional to the 
metric of $E^{2,1}$,
\Eq{
\gamma(X,X)=2[(X^1)^2+(X^2)^2-(X^3)^2].
}
From Prop \ref{prop:normalizer:Cartan}, this implies that an 
automorphism $\phi$ of this algebra is represented by a matrix 
$\Lambda$ in $O(2,1)$ as $\phi(a\cdot\eta)=(\Lambda a)\cdot\eta$. 
This linear transformation belongs to $\LieA(\SO(2,1))$ when 
$\det\Lambda=1$ from
\Eq{
[\phi(a\cdot\eta), \phi(b\cdot \eta)]
=(\det\Lambda)\phi([(a\cdot\eta,b\cdot\eta]).
}
Hence
\Eq{
\Aut(\LieA(\Isom(H^2)))=\SO(2,1).
}

Clearly the maximal connected subgroup of this group, $\SO_+(2,1)$,  
is generated by the inner automorphisms, i.e., by the 
transformations in $\Isom_0(H^2)$. On the other hand, the space 
reflection $(-I_1): (x,y) \mapsto (-x,y)$ induces the automorphism 
$I_2: (\eta_1,\eta_2,\eta_3) \mapsto (-\eta_1,\eta_2,-\eta_3)$, which 
together with $\SO_+(2,1)$ generates $\SO(2,1)$. Therefore, from 
Prop.\ref{prop:normalizer:right}, $\N(\Isom_0(H^2))$ is 
contained in $ 
\Set{1,(-I_1)}\sdp(\Isom_0(H^2)\times H^2_R)$.  Here, from 
Eq.\eqref{III:product}, for $f=Z_R(\alpha)\in H^2_R$ and the isometry 
$R_H(\theta)\in \Isom_0(H^2)$ corresponding to $V=R(\theta)$, 
$fR_H(\theta)f^{-1}$ is given by the transformation
\Eq{
x+iy \mapsto a+bR(\theta)*(x+y \alpha'),
}
where $\alpha=a+ib$ and $\alpha'=(i-a)/b$. If this transformation 
belongs to $\Isom_0(H^2)$, it should be written as $\zeta=x+iy 
\mapsto V*\zeta$ with some $V\in\SL_2\RF$, which is holomorphic in 
$\zeta$. This implies that $\alpha'=i$, i.e., $f=Z_R(i)={\rm id}$. 
Hence
\Eq{
\N(\Isom_0(H^2))=\Isom(H^2)=\Set{1,(-I_1)}\sdp \PSL_2.
}
Since $\N(\Isom(H^2))\subset\N(\Isom_0(H^2))$, it follows from this 
that 
\Eq{
\N(\Isom(H^2))=\Isom(H^2).
}
%

\subsubsection{Maximal geometry}\label{sec:H^2xR:MaximalGeometry}

The hyperbolic metric \eqref{metric:H^2} on $H^2$ and the natural 
metric on $\RF$ defines the invariant metric on $H^2\times\RF$,
\Eq{
ds^2=Q(\chi_1^2+\chi_2^2)+\chi_3^2
=Q\frac{dx^2+dy^2}{y^2}+dz^2,
}
where $Q$ is an arbitrary positive constant. This is clearly invariant 
under the transformation group $\Isom(H^2)\times \IO(1)$, where 
$\IO(1)$ is the isometry group of the line $\RF$. Actually, it 
precisely gives the maximal symmetry of $H^2\times \RF$:
\Eq{
\Isom(H^2\times\RF)=\Isom(H^2)\times\Isom(\RF).
\label{Gmax:H^2xR}}
This follows from the following consideration.

First, note that $\Isom_0(H^2\times\RF)$ is given by
\Eq{
\Isom_0(H^2\times\RF)=\Isom_0(H^2)\times \RF,
}
because $H^2\times\RF$ is not a maximally symmetric space and 
therefore the dimension of its isometry group is not greater than $4$. 
Hence, $\Isom(H^2\times\RF)$ is generated by $\Isom_0(H^2\times\RF)$ 
and discrete transformations in $\N(\Isom_0(H^2\times\RF))$. 

The Lie algebra of $\Isom_0(H^2\times\RF)$ is generated by 
$\xi_1,\xi_2,\xi_3$ in Eq.\eqref{III:generators} and the infinitesimal 
rotation $\xi_4$ of $H^2$. Their commutators are given by
\Eq{
[\xi_4,\xi_1]=-\xi_2,\ 
[\xi_4,\xi_2]=\xi_1-\xi_4,\ 
[\xi_4,\xi_3]=0.
}
From these commutation relations, it follows that the center of this 
algebra is generated by $\xi_3$ and 
$[\LieA(\Isom(H^2\times\RF)),\LieA(\Isom(H^2\times\RF))]$ coincides 
with the subalgebra $\LieA(\PSL_2\RF)$ generated by 
$\xi_1,\xi_2$ and $\xi_4$. Hence, from 
Prop.\ref{prop:normalizer:invariantspace} and the argument on 
$\Aut(\LieA(\Isom(H^2)))$, the automorphism $\phi\in 
\Aut(\LieA(\Isom(H^2\times\RF)))$ has the structure
\Eqr{
&& \phi(\xi_3)=k\xi_3,\\
&& \phi(\xi_1-\xi_4,\xi_2,\xi_4)=(\xi_1-\xi_4,\xi_2,\xi_4)\Lambda;\ 
\Lambda\in \SO(2,1).
}
Since this automorphism is induced from a transformation in the 
group $\N(\Isom(H^2))$ $\times\N(\Isom(\RF))$, it follows from 
Prop.\ref{prop:normalizer:right} that for any 
$f\in\N(\Isom_0(H^2\times\RF))$, there is $f_0\in 
\N(\Isom(H^2))\times\N(\Isom(\RF))$ such that $g=f_0^{-1}f$ is some 
right transformation in $H^2\times\RF$. Here, since $\RF$ is an 
Abelian group, for which the right and the left transformations 
coincide, we can assume that $g$ belongs to $H^2_R$. Then the 
condition $g\in \N(\Isom_0(H^2\times\RF))$ is equivalent to the 
condition $g\in \N(\Isom_0(H^2))$. However, as shown before, 
$H^2_R\cap \N(\Isom_0(H^2))$ is trivial. This implies that $g$ is an 
identity transformation, and we find  
\Eq{
\N(\Isom(H^2\times\RF))=\N(\Isom_0(H^2\times\RF))
=\Isom(H^2)\times\N(\Isom(\RF)).
\label{Isom(H^2xR):normalizer}
}
It is clear that a transformation $\N(\Isom(\RF))$ induces an 
isometry of $H^2\times\RF$ only when it belongs to $\Isom(\RF)$. 
This proves the equality \eqref{Gmax:H^2xR}.

Finally we give one formula. A generic transformation in 
$\N(\Isom(H^2\times\RF))$ is written as
\Eq{
f=\Set{1,-I_1,I_2,-I_3}L_{(a,b,c)}R_H(\theta)D(1,1,k).
}
From \eqref{transformation:PSL_2}, we find that this induces the 
transformation of the invariant basis 
\eqref{III:InvariantBasis} given by
\Eq{
f^*\chi^I=F^I{}_J\chi^J:\ 
F=\Set{1,-I_1,I_2,-I_3}R_3(H(\theta,\zeta))D(1,1,k)
\label{F:Isom(H^2xR)}}
at the point $(x,y,z)$, where $\zeta=x+iy$ and $H(\theta,\zeta)$ is 
the angle defined mod. $2\pi$ by 
\Eq{
e^{-iH(\theta,\zeta)}=\frac{\cos\theta-\zeta\sin\theta}{|\cos\theta-\zeta\sin\theta|}.
\label{H-function:def1}
}
In the next section we will give a definition of $H$ without the 
ambiguity of mod. $2\pi$, but in the present section this ambiguity has 
no significance. 


\begin{table}[t]
\caption{\label{table:III:normalizer} Normalizers for the invariance 
groups with $G_0=\BTIII_L, \Isom_0(H^2\times\RF)$. The formulas for 
conjugate transformations are given in Appendix \ref{appendix:conj}} 
\begin{tabular}{lll}
\hline\hline\\
$G$ & o/u 
& $f\in\N(G)$ \\
\hline\\
$\BTIII_L$ & o  
& $\Set{1,-I_1,I_2,-I_3}L_{\bm{a}}N'(d)D(1,1,k)R_{\hat{\bm{b}}}(k>0)$
\\
$\BTIII_L\sdp \Set{1,-I_3}$ & u 
& $d=0$
\\
$\BTIII_L\sdp \Set{1,-I_1}$ & u 
& $\Set{1,-I_1,I_2,-I_3}L_{\bm{a}}N'(d)D(1,k_2,k_3) (k_2,k_3>0)$
\\
$\BTIII_L\sdp \Set{1,I_2}$ & o 
& $d=0$
\\
$\BTIII_L\sdp \Set{1,-I_1,I_2,-I_3}$ & u 
& ibid
\\
&&\\
$\Isom_0(H^2\times\RF)$ & o
& $\Set{1,-I_1,I_2,-I_3}L_{(a,b,c)}R_H(\theta)D(1,1,k)(k>0)$
\\
$\Isom_0(H^2\times\RF)\sdp\Set{1,I_2}$ & o
& ibid
\\
$\Isom_0(H^2\times\RF)\sdp\Set{1,-I_1}$ & u
& ibid
\\
$\Isom_0(H^2\times\RF)\sdp\Set{1,-I_3}$ & u
& ibid
\\
$\Isom(H^2\times\RF)$ & u
& ibid
\\
\\\hline\hline
\end{tabular}
\end{table}

\subsubsection{Invariance group}\label{sec:H^2xR:InvG}

We first determine the possible invariance groups $G$ with 
$G_0=\Isom_0(H^2\times \RF)$. Since the maximal isometry group of 
$H^2\times \RF$ is written as
\Eq{
\Isom(H^2\times\RF)=\Set{1,-I_1,I_2,-I_3}\sdp \Isom_0(H^2\times\RF),
}
where $-I_1, I_2, -I_3$ are the discrete linear transformations of 
$(x,y,z)$ represented by the matrices $-I_1,I_2,-I_3$, respectively, 
$G$ is given by $G_0$ or the semi-direct product of $G_0$ and one of 
the discrete groups $\Set{1,I_2}, \Set{1,-I_1}, \Set{1,-I_3}$, and 
$\Set{1,-I_1,I_2,-I_3}$. Among these, $\Isom_0(H^2\times\RF)$ and 
$\Set{1,I_2}\sdp\Isom_0(H^2\times\RF)$ are orientation preserving. 
It is easy to see that the normalizer of all of these groups 
coincide and is given by Eq.\eqref{Isom(H^2xR):normalizer}.

Next we consider the group $G$ with $G_0=\BTIII_L$. Since an 
invariance group with $G_0=\BTIII_L$ is a subgroup of 
$\N(\BTIII_L)\cap \Isom(H^2\times\RF)
=\Set{1,-I_1,I_2,-I_3}\BTIII_L$,
it is given by $G_0$ itself or the semi-direct product of $G_0$ with 
one of 
$\Set{1,-I_3}$, $\Set{1,I_2}$, $\Set{1,-I_1}$ and 
$\Set{1,-I_1,I_2,-I_3}$. Among these $\BTIII_L$ and 
$\BTIII_L\sdp\Set{1,I_2}$ are orientation preserving. 

Although we listed the normalizers of these invariance groups with 
$G_0=\BTIII_L$ in the table \ref{table:III:normalizer}, we do not need 
those because these groups do not appear as an invariance group of a 
compact Bianchi model for the following reason. The group $\BTIII$ is 
classified as class B and $c_I=c^J{}_{IJ}$ does not vanish. From this 
constant, we can construct an vector field $V=Q^{IJ}c_I X_J$, which is 
invariant under 
$\BTIII_L$. Since this vector field has a constant divergence $D_I 
V^I=Q^{IJ}c_Ic_J$, if the space $\Sigma$ is given by $\RF^3/K$ with a 
discrete group $K$ of isometries, there must be some transformation 
$f\in K$ such that $f_*V\not=V$, because otherwise $V$ is a pull back 
of some vector field $V'$ on $\Sigma$ with the same constant 
divergence, which leads to the contradiction
\Eq{
0=\int_\Sigma d^3v D_IV'{}^I=D_IV'{}^I \int_\Sigma d^3v.
}
In the present case, $(c_I)=(0,-1,0)$ and $V=-Q^{2I}X_I$. If we 
restrict to orientation preserving 
transformations, the transformation which may move this vector is 
only $I_2$. However, if the metric is $I_2$-invariant, $Q^{21}$ and 
$Q^{23}$ vanish, and $V$ is parallel to $X_2$, which is invariant 
under $I_2$. Therefore we cannot compactify $H^2\times\RF$ by a 
discrete group $K$ contained in a group $G$ with $G_0=\BTIII_L$. On 
the other hand, since $R_H(\theta)$ moves the vector $X_2$, we can 
compactify a system whose invariance group contains 
$\Isom_0(H^2\times\RF)$, as we see soon.

\subsection{Compact topologies}

\subsubsection{Fundamental group}\label{sec:FundamentalGroup}

As we saw in \S\ref{sec:H^2xR:MaximalGeometry}, any isometry of 
$H^2\times\RF$ is written as a product of isometries of $H^2$ and 
$\RF$. This implies that, if $H^2\times\RF$ is compactified to 
$\Sigma$ by some discrete group $K\subset \Isom(H^2\times\RF)$, each 
transformation in $K$ preserves the bundle structure $H^2\times\RF 
\maps H^2$. Hence $\Sigma$ has also a bundle structure $\pi: \Sigma 
\maps X=H^2/K'$ with fibre $S^1$, where $K'$ is a discrete 
transformation group of $H^2$ induced by $K$. To be precise, 
$\Sigma$ is not a bundle in the narrow sense in general, because a 
transformation in $K'$ may have a fixed point. If the fixed point is 
isolated, the isotropy group becomes an Abelian group of a finite 
order isomorphic to $\ZR_p$, and the base space $X$ has a conical 
singularity of the form $\RF^2/\ZR_p$ there. On the other hand, if 
the fixed point is not isolated, it belongs to a curve of fixed 
points. Such a situation arises when $K'$ contains a reflection with 
respect to a geodesic in $H^2$. In this case the reflection curve 
becomes a boundary of the base space. Thus the base space is 
represented by a surface with conical singilarities and reflection 
curves, which is called {\it an orbifold}. The integer $p$ 
characterizing the structure of a conical singularity is called {\it 
the orbifold index} of that point.

Let $D'$ be a small disk around a point $x$ in $X$ ( or a half disk 
in the case $x$ is on a reflector curve) and  $V$ be a 
connected component of the preimage in $H^2\times\RF$ of 
$U=\pi^{-1}(D')\subset \Sigma$. Then from the diagram,
\Eq{
\begin{array}{ccc}
  H^2\times\RF & \stackrel{/K}{\longrightarrow} & \Sigma \\
  \downarrow   &                                & \hbox to 
  0pt{$\downarrow\!\pi$} \\
  H^2          & \stackrel{/K'}{\longrightarrow}& X
\end{array}
}
$V$ is expressed as $V=D\times\RF$, where $D$ is a disk in $H^2$ 
projected onto $D'$ in $X$. Further, let $L$ be the kernel of the 
homomorphism $j: K \maps K'$, $K_y'$ be the isotropy group of $K'$ at 
the point $y$ in $D$ projected to $x$, and $K_y$ be the preimage of 
$K_y'$ in $K$ by $j$. Then, $L$ is isomorphic to $\ZR$ generated by a 
translation $\ell$ along $\RF$ in $H^2\times\RF$, and the following 
exact sequence holds:
\Eq{
0 \maps L \mapsnamed{i} K_y \mapsnamed{j} K'_y \maps 0\ 
\text{(exact)}.
}
Since $K_y$ coincides the set of transformations in $K$ which maps $V$ 
to $V$, $K_y$ is a subgroup of the isometry group of $V\cong 
D\times\RF$ which is generated by rotations of $D$, translations along 
$\RF$ and relections in $D$ and $\RF$, and $U$ is written as $V/K_y$. 
Since a combination of a translation and a reflection at a point in 
$\RF$ is always a reflection at some point, such a 
transformation always has fixed points in $V$ and cannot be 
contained in $K_y$. On the other hand, a combination of a 
translation in $\RF$ and reflection in $D$ does not have a fixed 
point. However, such an orientation-reversing transformation makes 
$\Sigma$ unorientable. Since we are only interested in orientable 
3-spaces, we will not consider such cases in the present paper. 
Under this restriction, the base orbifold has no reflector curve. 
Further, since a properly discontinuous  subgroup generated by 
translations and rotations in $D\times\RF$ is always an infinite 
cyclic group, $K_y$ is isomorphic to $\ZR$. This implies that $U$ is 
homeomorphic to a solid torus. 

Here note that the conjugate transformation $a\ell a^{-1}$ of $L$ by 
an element $a$ of $K_y$ depends only on $j(a)$ because $L$ is 
Abelian. Further, since the automorphism of a free cyclic group is 
given by either the identity transformation or the inversion 
transformation, $a\ell a^{-1}$ coincides with $\ell $ or $\ell 
^{-1}$. Since we are only considering the case $\Sigma$ is 
orientable, the former happens when $a$ projects to a orientation 
preserving transformation of $H^2$ and the latter when $a$ 
projects to a orientation reversing one.

If $x$ is a regular point, $K'_y$ is trivial and $K_y$ coincides 
with $L$. Hence, $U$ is written as $V/L\cong V/\ZR$ and has a 
regular bundle structure isomorphic to the solid torus $D'\times 
S^1$. On the other hand, if $x$ is a conical singularity with index 
$p$, $K'_y$ is isomorphic to $\ZR_p$ and $K_y$ does not coincide 
with $L$. Let $\gamma$ be a transformation in $K$ which projects to 
the generator $\bar \gamma$ of $K'_y$ corresponding to the 
rotation of angle $2\pi/p$ in $H^2$. Then, since 
$j(\gamma^p)=\bar\gamma^p=1$, there exists an integer $r$ such 
that $\gamma^p \ell ^r=1$ and $p$ and $r$ are coprime. We can 
make $r$ to be $0<r<p$ by replacing $\gamma$ by $\gamma \ell ^c$ with 
some integer $c$. Under this normalization, $\gamma$ 
represents the translation along $\RF$  corresponding to $-r/p$ times 
$\ell $ and rotation of angle $2\pi/p$ around the central line in 
$V=D\times \RF$. Here let $(q,s)$ be the pair of coprime integers 
uniquely determined by the condition that  $0<q<p$ and $ps+qr=1$. Then 
it is easy to show that for 
$\lambda=\gamma^{-q}\ell ^s$, $\ell =\lambda^p$ and 
$\gamma=\lambda^{-r}$. Hence $K_y$ is a free cyclic group 
generated by $\lambda$. This implies that $U$ is homeomorphic to the 
solid torus as before, but it is not a regular $S^1$ bundle over $D$ 
because $\lambda$ corresponds to a combination of the translation  
of $1/p$ times $\ell $ and the rotation $-2\pi q/p$ around the 
central line in $D\times\RF$. By this transformation $\lambda$, the 
cetral line is identified to $S^1$, but the other lines are 
identified to $S^1$ only by $\lambda^p$. The $S^1$ corresponding to 
this central line is called {\it a critical fibre}, and the others 
{\it regular fibres}. In the torus $U$, if one goes around the 
critial fiber once, regular fibres rotate by the angle $2\pi q/p$ 
aroung it. The latter close only by $p$ turns. 

Thus we have shown that a compact quotient of $H^2\times\RF$ is 
fibred by $S^1$, and a small neighborhood of each fibre is 
isomorphic to a solid torus with a natural fibring or a solid torus 
with a twisted fibring. The latter case occurs when the fibre is 
critical and projects to a conic singularity in the base orbifold. A 
manifold of this type of $S^1$-fibring structure is called {\it a 
Seyfert fibre space}. The pair of coprime integers 
$(p,q)$ or $(p,r)$ characterizing the twist around the critical 
fibre is called {\it the Seifert index} of that fibre. To be exact, 
a Seyfert fibre space is in general allowed to contains a critical 
fibre whose neighborhood is isomorphic to a solid Klein bottle. In 
such a case, the base orbifold contains a reflector curve and the 
total space becomes unorientable.

In general, a Seifert fibred space may allow a different fibring 
structure which is not isomorphic to the original one. 
Fortunately, however, it can be shown that a compact quotient of 
$H^2\times\RF$ or $\TSL$ admit a unique Seifert 
fibring\cite{Scott.P1983}. Hence we can determined the structure of 
the fundamental group $\pi_1(\Sigma)\cong K$ by the Seifert bundle 
structure of $\Sigma$ over the base orbifold $X$ described above in 
the following way.  

First, we consider the case in which the base orbifold is an 
orientable closed surface of genus $g$ with $k $ conical singular 
points $x_i$ with indices $(p_i,r_i)$. Let $D_i$ be a small disk 
around $x_i$ and $N$ be the closure of $X-\cup_i D_i$. Then, 
$\Sigma'=\pi^{-1}(N)$ contains only regular fibres and has a 
normal $S^1$-bundle structure over $N$.  Then the covering 
transformation in $H^2\times\RF$ corresponding to the $S^1$ fibre 
is given by the generator $\ell $ of $L$, and from the exact 
sequence for homotopies groups, we have
\Eq{
0 \maps L \mapsnamed{i} \pi_1(\Sigma') \mapsnamed{j} \pi_1(N) \maps 1.
}
Here, $\pi_1(N)$ is expressed in terms of  $g$-pairs of isometries 
$(\bar\alpha_a,\bar\beta_a)$ corresponding to generators of 
$\pi_1(X)$ and $\ell $ isometries $\bar\gamma_i$ corresponding to 
the boundary of $D_i$ as
\Eq{
\pi_1(N)=\langle 
\bar\alpha_1,\bar\beta_1,\cdots, \bar\alpha_g,\bar\beta_g,\bar\gamma_1,\cdots,\bar\gamma_k
\mid[\bar\alpha_1,\bar\beta_
1]\cdots[\bar\alpha_g,\bar\beta_g]\bar\gamma_1\cdots\bar\gamma_k=1\rangle.
}
Let $\alpha_a,\beta_a$ and $\gamma_i$ be generators of 
$\pi_1(\Sigma')$ which project to $\bar\alpha_a,\bar\beta_a$ and 
$\bar\gamma_i$, respectively. Then the relation among the generators 
in $\pi_1(N)$ is lifted to 
$[\alpha_1,\beta_1]\cdots[\alpha_g,\beta_g]
\gamma_1\cdots\gamma_k$ $ =\ell ^b$ 
with some integer $b$. We must add relations between $\ell $ and the 
other generators to specify $\pi_1(\Sigma')$. In the case $k>0$ it 
is simple. Since $N$ is homotopic to a join of circles in this case, 
and since an orientable $S^1$-bundle over each circle is always 
trivial, all elements in $\pi_1(\Sigma')$ commute with $\ell $. 
Therefore, 
$\pi_1(\Sigma')$ is given by 
\Eqr{
&\pi_1(\Sigma')= &\langle 
\alpha_1,\beta_1,\cdots,\alpha_g,\beta_g,\gamma_1,\cdots,\gamma_k, \ell  \mid 
[\alpha_a,\ell ]=[\beta_a,\ell ]=[\gamma_i,\ell ]=1, \nonumber\\
&& \qquad [\alpha_1,\beta_1]\cdots[\alpha_g,\beta_g]\gamma_1\cdots\gamma_k=\ell ^b\rangle.
\label{pi1:Sigma'}}

From the van Kampfen theorem, the fundamental group of the total space 
$\Sigma$ is obtained from this and the fundamental groups of solid tori 
around the critial fibres by adding relations of their generators in 
$\pi_1(\Sigma)$ to the free product of them. In the present case, since 
each critical fibre is homotopic to some product of $\ell$'s and 
$\gamma_i$'s as shown above, $\pi_1(\Sigma)$ is obtained from 
$\pi_1(\Sigma')$ by adding relations among their generators in 
$\pi_1(\Sigma)$. Since $\bar\alpha_a$ and 
$\bar\beta_a$ are independent in $\pi_1(X)$, such relations only 
contains $\ell$ and $\gamma_i$. From the explanation above on the 
homotopy relations among $\ell, \gamma$ and $\lambda$ in a twisted 
torus, we can adopt as these relations the normalization conditions 
$\gamma^{p_i}\ell ^{r_i}=1$ for the $\ell$-multiplication freedom in 
the choice of $\gamma_i$. Note that under these normalization 
conditions, $b$ becomes a topological invariant which is independent 
of rescaling of the other generators by $\ell $. Thus we obtain 
\Eqr{
& \pi_1(\Sigma)= &\langle 
\alpha_1,\beta_1,\cdots,\alpha_g,\beta_g,\gamma_1,\cdots,\gamma_k, \ell  \mid 
[\alpha_a,\ell ]=[\beta_a,\ell ]=[\gamma_i,\ell ]=1, \nonumber\\
&& \qquad \gamma_i^{p_i}\ell ^{r_i}=1,
[\alpha_1,\beta_1]\cdots[\alpha_g,\beta_g]\gamma_1\cdots\gamma_k=\ell ^b\rangle.
\label{Seifert:pi1:orientable}}

This expression also holds in the case $k=0$, i.e., when the 
orbifold $X$ is a regular surface and $\Sigma$ is a $S^1$ bundle 
over it. To see this, we take an arbitrary small disk in $X$ and 
define $N$ as the closure of $X-D$. Then $\pi_1(\Sigma')$ is given 
by the expression \eqref{pi1:Sigma'} with a single $\gamma$ 
generator. Now, the original bundle $\Sigma$ is obtained by gluing the 
solid torus $D\times S^1$, instead of a twisted torus, to $\Sigma'$ 
along their boundaries. After gluing, $\gamma$ becomes homotopic to 
$\ell^c$ with some integer $c$, because $\ell$ is the generator of the 
gluing torus in the present case. Hence, by eliminating $\gamma$ in 
terms of this relation and redefining $b$, we obtain the expression 
\eqref{Seifert:pi1:orientable} with no $\gamma$ generator. In this 
case, $-b$ coincides with the Euler number of the $S^1$ bundle.

Next we consider the case in which the base orbifold is 
unorientable. The derivation of the fundamental group in this case 
is similar to the above case except for the following two points. 
First, the fundamental group of an unorientable closed surface with 
genus $g$ is generated by $g$ loops $\bar\alpha_a$ in stead of $2g$ 
loops in the orientable case, and the relation for the generators of 
$\pi_1(N)$ are replaced by 
$\bar\alpha_1^2\cdots\bar\alpha_g^2\bar\gamma_1\cdots\bar\gamma_k=1$. 
Second, although the $S^1$ bundle over $N$ is still determined by 
its restriction on each loop $\bar\alpha_a$, this time we must 
choose the Klein bottle as the bundle over each $\bar\alpha_a$ to 
make the total space orientable, because a normal neighborhood of 
$\bar\alpha_a$ in $N$ is a M\"obius band. This implies that the 
relation between $\ell $ and $\alpha_a$ is given by $\alpha_a\ell 
\alpha_a^{-1}=\ell ^{-1}$. Taking these points, we find that the 
fundamental group of the Seifert bundle over an unorientable 
orbifold is given by%
\Eqr{
&\pi_1(\Sigma)=&\langle 
\alpha_1,\cdots,\alpha_g,\gamma_1,\cdots,\gamma_k, \ell  \mid 
\alpha_a\ell\alpha_a^{-1}=\ell^{-1}, [\gamma_i,\ell ]=1, 
\gamma_i^{p_i}\ell ^{r_i}=1,\nonumber\\
&& \qquad \alpha_1^2\cdots\alpha_g^2\gamma_1\cdots\gamma_k=\ell 
^b\rangle.
\label{Seifert:pi1:unorientable}}

The argument on the fundamental group so far is quite general and 
applies to any orientable Seifert bundle over an orbifold with no 
reflector curve. This implies that some additional constraints 
must be imposed on the indices appearing in the expressions for 
$\pi_1(\Sigma)$ above to obtain the fundamental groups for 
compact quotients of each Thurston type, because the fundamental 
groups for compact quotients of different Thurston types are not 
isomorphic due to Thurston's theorem\cite{Scott.P1983}. For the 
Thurston type $H^2\times\RF$, they are given by the conditions 
$\chi<0$ and $e=0$, where $\chi$ is the Euler characteristic of 
the orbifold defined in terms of the standard Euler 
characteristic $\chi_0(X)$ of the orbifold as a topological space  and 
the orbifold index $p_i$ by 
\Eq{
\chi(X)=  \chi_0(X)- \sum_i\left(1-\frac{1}{p_i}\right),
}
and $e$ is the Euler number of the Seifert bundle defined in 
terms of the Seifert index $(p_i,r_i)$ and $b$ as 
\Eq{
e=-b - \sum_i\frac{r_i}{p_i}.
}
From now on we generally denote the Seifert fibred spaces with the 
fundamental groups \eqref{Seifert:pi1:orientable} and 
\eqref{Seifert:pi1:unorientable} by 
$S^+(g,e;\{(p_1,r_1),\cdots,(p_k,r_k)\})$ and $S^-(g,e;\{(p_1,r_1)$, 
$\cdots, (p_k,r_k)\}$, respectively. They will be also written 
simply as $S^\pm(g,e;k)$, when we do not write the Seifert indices 
explicitly.

\subsubsection{Moduli freedom}

In order to determine the moduli freedom of each compact quotient of $H^2\times\RF$, we must classify the conjugate class of embeddings of $\pi_1(\Sigma)$ into each invariance group $G$ with respect to $\N^+(G)$. As explained in \S\ref{sec:H^2xR:InvG}, a possible invariance group $G$ is  $\Isom_0(H^2\times\RF)\cong \PSL_2\times \RF$ or $\Isom^+(H^2\times\RF)\cong \Isom_0(H^2\times\RF)\sdp\Set{1,I_2}$. 

For $\Sigma=S^+(g,0;k)$, in which the base orbifold is orientable, 
the image of $\pi_1(\Sigma)$ is always contained in 
$\Isom_0(H^2\times\RF)$. Further, $\N^+(G)$ is the same for both 
invariance groups. Hence we only have to consider the case 
$G=\Isom_0(H^2\times\RF)$. 

First, from the argument above, the 
generator $\ell$ must be represented by a translation $T_l: z 
\mapsto z+l$ along $\RF$. Then, from the relation 
$\gamma_i^{p_i}\ell^{r_i}=1$, $\gamma_i$ is uniquely determined if 
we specify the center position of the rotation in $H^2$ 
for $\bar\gamma_i$. Further, when $2g$ elements $\bar \alpha_a$ and  
$\bar\beta_a$ of $\Isom^+(H^2)$ satisfying the relation 
\Eq{
[\bar\alpha_1,\bar\beta_1]\cdots[\bar\alpha_g,\bar\beta_g]
\bar\gamma_1\cdots\bar\gamma_k=1
\label{rel:S+}
}
are given, $\alpha_a$ and $\beta_a$ are expressed as $T_{x_a 
l}\bar\alpha_a, T_{y_al}\bar\beta_a$, where $x_a$ and $y_a$ are 
arbitrary real numbers. Here, the number of degrees of freedom in the 
choice of the rotation center of each $\bar\gamma_i$ is two, that of 
$\bar\alpha_a$ or $\bar\beta_a$ as an element of $\PSL_2\RF$ is three, 
and the relation \eqref{rel:S+} gives three independent scalar 
equations. Hence, the number of the total degrees of freedom in the 
embedding $K$ is given by
\Eq{
1+2k+(3+1)\times2g-3=8g+2k-2.
}
On the other hand, $\N^+(\Isom(H^2\times\RF))$ is given by 
$\Set{1,I_2}\Isom^+(H^2)\times \Set{D(1,1,k)}\times\Set{T_c}$. Among 
the transformations in this group, $I_2$ changes $T_l$ to $T_{-l}$. 
Hence, this freedom is eliminated by requiring $l>0$. Further, 
$D(1,1,k)$($k>0$) transforms $T_l$ to $T_{kl}$, but does not 
affect the transformations in $\Isom(H^2)$. Finally, 
$\Isom^+(H^2)$ acts on the space of the set 
$(\bar\alpha_a,\bar\beta_b,\bar\gamma_i)$ freely at generic 
points, while $T_c$ does not affect the moduli. Hence, the dimension of 
each orbits of the action 
$\N^+(\Isom(H^2\times\RF))$ on the space of $K$ is $1+3=4$, and the 
number of the moduli degrees of freedom is given by
\Eqr{
&N_{m+}'
&=\dim\left( \Mono(\pi_1(S^+(g,0;k)),\Isom_0(H^2\times\RF))
/\ad(\N^+(\Isom_0(H^2\times\RF)))\right)
\nonumber\\
&&=8g+2k-6.
\label{H2xR+:ModuliFreedom}
}
The isotropy group $\N_{\Sigma0}(G)$ of the action of 
$\N^+(\Isom_0(H^2\times\RF))$ on the moduli space is generated by $T_c$ 
in $\BTIII_L$.

Here note that, since $\chi_0(X)$ for an orientable closed 
orbifold with genus $g$ is given by $2-2g$, the condition 
$\chi<0$ requires that for $g=0,1$ orbifolds always have conic 
singularities:  
\Eq{
k\ge\left\{\begin{array}{ll}3 &; g=0 \\ 1 &; g=1\end{array}\right..
\label{constr:k:+}}
Further, the condition $e=0$ yields the strong restriction on the 
possible values of the Seifert indices that $\sum_i r_i/p_i$ is an 
integer. It should be noted that this condition is essential to 
guarantee that $b$ becomes always an integer regardless of the 
choice of the parameter $l$.

Next let us consider the case $\Sigma=S^-(g,0;k)$ for which the 
base orbifold is unorientable. Since we need transformations 
which reverse the orientation of $H^2$ in this case, 
$\pi(\Sigma)$ can be embedded only in $\Isom^+(H^2\times\RF)$. 
$\ell$ is still represented as $T_l$, and $\gamma_i$ is uniquely 
specified by a rotation $\bar\gamma_i$ in $H^2$, but from the 
relation $\alpha_a\ell\alpha_a^{-1}=\ell^{-1}$, $\alpha_a$ is now 
represented as $I_2T_{x_a l}\bar\alpha_a$ with $x_a\in \RF$ and  
$\bar\alpha_a \in \Isom^+(H^2)$ satisfying the condition 
\Eq{
\bar\alpha_1^2\cdots\bar\alpha_g^2\bar\gamma_1\cdots\bar\gamma_k=1.
\label{rel:S-}
}
Hence, the number of the degrees of freedom in $K$ is now given by 
$4g+2k-2$. Further, since $T_c$ in $\N^+(\Isom^+(H^2\times\RF))$ 
transforms $I_2$ as $T_cI_2T_{-c}=I_2T_{-2c}$, the dimension of HPD 
orbits becomes 5. Hence the number of the moduli degrees of freedom for 
$S^-(g,0;k)$ is given by
\Eqr{
&N_{m-}'
&=\dim \left(\Mono(\pi_1(S^-(g,0;k)),\Isom^+(H^2\times\RF))
/\ad(\N^+(\Isom^+(H^2\times\RF)))\right)
\nonumber\\
&&=4g+2k-7,
\label{H2xR-:ModuliFreedom}}
and $\N_{\Sigma0}(G)$ becomes trivial.

We also obtain a constraint on the number of conic singularities 
like \eqref{constr:k:+}. Since $\chi_0(X)$ for the unorientable 
closed surface with genus $g$ is given by $2-g$($g\ge1$), the 
constraint is expressed as 
\Eq{
k\ge\left\{\begin{array}{ll}2 &; g=1 \\ 1 &; g=2\end{array}\right..
\label{constr:k:-}}
%

\subsection{Phase Space}


\begin{table}[t]
\caption{\label{table:count:H^2xR} The parameter count for the type 
$H^2\times\RF$.}
\begin{tabular}{llccccccc}
\hline
\hline
\bf Space & \bf Symmetry & $Q$ & $P$ & $N_m$ & $N_f$ & $N$ & $N_s$ & 
$N_s$(vacuum)\\\hline\\
$\RF^3$
&$\Isom(H^2\times\RF)$	& 1 & 2 & 0 & 1 & 4 & 2 &  1\\
&&&&&&&&\\
$M^\pm(g,k;0)$  
&$\Isom(H^2\times\RF)$	& 2 & 2 & $N'_{m\pm}$ & 1 & $5+N'_{m\pm}$ & 
$3+N'_{m\pm}$ & $2+N'_{m\pm}$\\
\\\hline\hline\\
\end{tabular}
\end{table}

As we have shown in the previous subsection, the isotropy group 
$\N_{\Sigma0}(G)$ of the action of $\N^+(G)$ on the moduli space is 
given by a 
subgroup of $\BTIII_L$. However, since the action of $\BTIII_L$ 
on the space of homogeneous data on the covering space 
$\tilde\Sigma=H^2\times\RF$ is trivial, the degrees of freedom 
for the homogeneous data is simply determined by the invariance 
of the data by $G$. 
  
First, note that if the system is invariant under 
$\Isom_0(H^2\times\RF)$, the components of any quantity with respect 
to the invariant basis are invariant under $R_3(\theta)$ for any 
$\theta$ from Eq.\eqref{F:Isom(H^2xR)}. Hence $Q_{IJ}$, $P^{IJ}$ 
must have the diagonal forms $Q=D(Q_1,Q_1,Q_3)$ and 
$P=(P^1,P^1,P^3)$, respectively, and $u_1=u_2=0$. Further from the 
diffeomorphism constraint 
\Eq{
H_3\equiv c u_3=0,
}
$u_3$ also vanishes. Hence, the system with a single-component fluid 
always has the maximal symmetry 
$G=\Isom^+(H^2\times\RF)=\Isom_0(H^2\times\RF)\sdp\Set{1,I_2}$, and the 
phase space for the homogeneous data is given by
\Eq{
\Gamma_{D}(H^2\times 
\RF,G)=\SetDef{(Q_1,Q_3;P^1,P^3;\rho)}{Q_1,Q_3>0}.
}
Thus, putting together this result with the argument on the 
moduli freedom in the previous subsection, we find that the 
number of the total dynamical degrees of freedom is given by 
\Eq{
N=5+N_{m\pm}'.
}

For a multi-component system, $\Isom_0(H^2\times\RF)$ is also 
allowed as the invariance group, for which $N$ is increased by 2 for 
each extra component of fluid. For $G=\Isom^+(H^2\times\RF)$, the 
contribution of each extra component to $N$ is just one.

\section{Type $\TSL$}

\subsection{Invariance subgroups and normalizers}

\subsubsection{Bianchi type $\BTVIII$ group}

Any matrix $V\in \SL_2\RF$ is uniquely decomposed into the product of a 
lower triangle matrix $Z(\zeta)\in \HSL_2^+$ and a $\SO(2)$ matrix 
$R(z)$ as
\Eq{
V=Z(\zeta)R(z).
}
From this Iwazawa decomposition, we see that $\SL_2\RF$ is 
homeomorphic to $H^2\times S^1$, and that an element of the universal 
covering group $\TSL$ of this group is parametrized by $\zeta\in H^2$ 
and $z\in \RF$ as $\tilde V(\zeta,z)$.  This 
element is mapped to $Z(\zeta)R(z)$ by the natural projection $p: 
\TSL\maps \SL_2\RF$. Let us denote $\tilde V(\zeta,0)$ by $\tilde 
V(\zeta)$ and $\tilde V(i,z)$ by $\tilde R(z)$. Then a 
general element of $\TSL$ is expressed as
\Eq{
\tilde V=\tilde Z(\zeta)\tilde R(z).
}

Here note that $\tilde Z(\zeta)$ and $Z(\zeta)$ are in one-to-one 
correspondence, while $\tilde R(z+2\pi)\not=\tilde R(z)$ and 
$\tilde R(z)$ is in one-to-one correspondence to $z\in\RF$. Thus 
$\tilde Z(\zeta)$'s form a group isomorphic to $H^2\cong \HSL_2^+$, and 
$\tilde R(z)$ form a group isomorphic to $\RF$. However, 
$\TSL$ is not isomorphic to $H^2\times\RF$ because $\tilde Z$ and 
$\tilde R$ do not commute. Using the fact that $\SL_2\RF$ and $\TSL$ 
have the same group structure near the unit element, we can show that 
the 
commutation relation is given by 
\Eqr{
&& \tilde R(z)\tilde Z(\zeta)=\tilde Z(\zeta')\tilde R(z');\\
&& \zeta'=R(z)*\zeta,\\
&& z'=H(z,\zeta),
}
where the function $H(z,\zeta)$ is defined by
\Eq{
H(z,\zeta)=\Im \zeta \int_0^z 
\frac{d\phi}{|\cos\phi-\zeta\sin\phi|^2}.
}
This function satisfies the relation \eqref{H-function:def1} and
\Eqr{
&& H(n\pi,\zeta)=n\pi,\quad \forall n\in\ZR, \forall\zeta\in H^2,\\&& 
H(z,i)=z,\ \forall z\in\RF.
}

From this it follows that the group multiplication law is given by
\Eq{
(a+ib,c)\cdot (x+iy,z)=\left(a+b R(c)*(x+iy), z+H(c,x+iy)\right).
}
In particular, the left transformation $L_{(a+ib,c)}$ 
corresponding to the left multiplication of $\tilde V(a+ib,c)$ is 
expressed as
\Eq{
L_{(a+ib,c)}: (\zeta,z) \mapsto \left(a+b R(c)*\zeta, 
z+H(c,\zeta)\right),
\label{VIII:LT}}
and the right transformation $R_{(a+ib,c)}$ as
\Eq{
R_{(\alpha,c)}: (x+iy,z) \mapsto \left(x+yR(z)*\alpha,
c+H(z,\alpha)\right).
\label{VIII:RT}}
From these expressions, we see that 
$L_{(\alpha,c)}=R_{(\alpha',c')}$ if and only if 
$\alpha=\alpha'=i$ and $c=c'=n\pi$ for some $n\in\ZR$. This 
implies that the center of $\TSL$ is given by
\Eq{
C(\TSL)\equiv \BTVIII_L\cap\BTVIII_R
=\SetDef{\tilde R(n\pi)}{n\in\ZR}.
}
Here note that $R_{(i,c)}$ gives just the translation of $z$,%
\Eq{
T_{c}=R_{(i,c)}: (x+iy,z) \mapsto (x+iy, z+c),
}
because $R(z)*i=i$. The center is contained in the intersection 
of this subgroup of right transformations with the corresponding 
subgroup of left transformations $\Set{L_{(i,c)}}$.

From \eqref{VIII:LT}, the infinitesimal left transformations of $\TSL$ 
are generated by
\Eqr{
&& \xi_1=\frac{1}{2}(1-x^2+y^2)\partial_x -xy\partial_y
 -\frac{1}{2}y\partial_z,\\
&& \xi_2=x \partial_x + y \partial_y,\\
&& \xi_3=\frac{1}{2}(1+x^2-y^2)\partial_x +xy\partial_y
 +\frac{1}{2}y\partial_z,
\label{VIII:generators}}
which form the Bianchi type VIII Lie algebra
\Eq{
[\xi_1,\xi_2]=\xi_3,\ 
[\xi_3,\xi_1]=-\xi_2,\ 
[\xi_3,\xi_2]=\xi_1.
}
From \eqref{VIII:RT}, the invariant vector fields, i.e., the 
infinitesimal right transformations of $\TSL$, are generated by
\Eqr{
&& X_1=y(\cos 2z\partial_x + \sin 2z \partial_y)
-\frac{1}{2}\cos2z\partial_z,
\\
&& X_2=y(-\sin 2z\partial_x + \cos 2z \partial_y)
+\frac{1}{2}\sin2z\partial_z,
\\
&& X_3=\frac{1}{2}\partial_z,
}
and their dual basis is given by
\Eqr{
&& \chi^1=\frac{1}{y}(\cos2z dx + \sin2z dy),\\ 
&& \chi^2=\frac{1}{y}(-\sin2z dx + \cos2z dy),\\ 
&& \chi^3=\frac{dx}{y} + 2dz;
\label{VIII:InvariantBasis}\\
&& d\chi^1=-\chi^2\wedge\chi^3, \ 
   d\chi^2=\chi^1\wedge\chi^3,\ 
   d\chi^3=\chi^1\wedge\chi^2.
}

Finally, we determine the normalizer of 
$\BTVIII_L$. Since $\LieA(\BTVIII_L)$ is isomorphic to 
$\LieA(\SL_2\RF)\cong\LieA(\Isom(H^2))$, its automorphism group is 
given by $\SO(2,1)$, whose connected component $\SO_+(2,1)$ is 
generated by inner-automorphisms of $\TSL$, as was shown in the 
previous section. The remaining 
discrete transformation is induced from $I_2: (x,y,z) \mapsto 
(-x,y,-z)$. Hence, from Prop.\ref{prop:normalizer:right}, the 
normalizer of $\BTVIII_L$ is given by 
\Eq{
\N(\BTVIII_L)=\BTVIII_L\cdot\BTVIII_R\sdp\Set{1,I_2}.
\label{III:normalizer}}
Here we have used the symbol $\cdot$ instead of $\times$ to 
emphasize that $\BTVIII_L$ and $\BTVIII_R$ share the non-trivial 
center. These HPDs induce the following transformations of the 
invariant basis. First, for $f=R_{(\alpha,c)}\in\BTVIII_R$, since the 
action of $f_*$ on $\xi_I$ and that of $f^*$ on $\chi^I$ is expressed 
by the same matrix, we obtain
\Eq{
R_{(\alpha,c)}^*\chi^I=\Lambda(\tilde V(\alpha,c))^I{}_J \chi^J,
\label{F:VIII:1}
}
where $\Lambda(\tilde V)$ is the representation of $\TSL$ to 
$\SO_+(2,1)$ induced from \eqref{rep:SLtoSO}. Second, $I_2$ 
transforms the basis as
\Eq{
(I_2)^*: (\chi^1,\chi^2,\chi^3)\mapsto (-\chi^1,\chi^2,-\chi^3).
\label{F:VIII:2}
}
%

\subsubsection{Maximal geometry}

The most symmetric metric with $\BTVIII_L$ invariance is given by
\Eq{
ds^2=Q_1\left[(\chi^1)^2+(\chi^2)^2\right]+Q_3 (\chi^3)^2
=Q_1\frac{dx^2+dy^2}{y^2}+Q_3\left(2dz+\frac{dx}{y}\right)^2.
\label{SL_2:metric}}
It is easy to see that this metric is also invariant under a 
translation in $z$, i.e., $T_c$. Since $\TSL$ is not a constant 
curvature space, this implies that
\Eq{
\Isom_0(\TSL)=\BTVIII_L\cdot \RF \ni L_{(\alpha,c)}T_{\theta},}%
which is generated by $\xi_1,\xi_2,\xi_3$ and 
\Eq{
\xi_4=\partial_z
}
with
\Eq{
[\xi_4,\xi_1]=0,\ 
[\xi_4,\xi_2]=0,\ 
[\xi_4,\xi_3]=0.
\label{Isom(SL_2):LieA}
}

Since $[\LieA(\Isom(\TSL)),\LieA(\Isom(\TSL))]=\LieA(\TSL)$, $f\in 
\N(\Isom_0(\TSL))$ preserves $\TSL$, i.e., 
$\N(\Isom_0(\TSL))\subset\N(\BTVIII_L)$. Among transformations in 
$\N(\BTVIII_L)$, $I_2$ transforms $T_\theta$ as
\Eq{
I_2 T_\theta I_2=T_{-\theta}.
}
Hence $I_2\in \N(\Isom_0(\TSL))$. On the other hand, if 
$R_{(\alpha,c)}T_\theta R_{(\alpha,c)}^{-1}=T_{\theta'}$ holds 
for any $\theta$, its projection to $\SL_2\RF$ gives
\Eq{
R(-c)Z(\alpha)^{-1}R(\theta)Z(\alpha)R(c)=R(\theta')
\equivalent
Z(\alpha)^{-1}R(\theta)Z(\alpha)=R(\theta').
}
This holds only when $Z(\alpha)\in\SO(2)$, i.e., $\alpha=i$. 
Therefore the normalizer of $\Isom_0(\TSL)$ is given by
\Eq{
\N(\Isom_0(\TSL))=\Set{1,I_2}\sdp \BTVIII_L\cdot \RF,
\label{Isom(SL_2):normalizer}}
where $\RF$ is the group $\SetDef{T_c}{c\in\RF}$.  Since the 
metric \eqref{SL_2:metric} is clearly invariant under this group, 
we obtain
\Eq{
\Isom(\TSL)=\N(\Isom(\TSL))=\N(\Isom_0(\TSL)).
}
%

\begin{table}[t]
\caption{\label{table:VIII:normalizer} Normalizers for the invariance 
groups with $G_0=\BTVIII_L, \Isom_0(\TSL)$.} 
\begin{tabular}{lll}
\hline\hline
$G$ & o/u & $f\in\N(G)$ \\
\hline\\
$\BTVIII_L$ & o  
& $\Set{1,I_2}L_{(\alpha,c)}R_{(\beta,d)}$
\\
$\BTVIII_L\sdp \Set{1,I_2}$ & o 
& $\Set{1,I_2}\Set{1,T_{\pi/2}}L_{(\alpha,c)}\tilde Z_R(ie^\gamma)$
\\
$\BTVIII_L\cdot \Set{1,T_{\pi/2}}$ & o 
& $\Set{1,I_2}L_{(\alpha,c)}T_\theta$
\\
$\BTVIII_L\cdot \tilde D_2$ & o
& $\Set{1,I_2}\Set{1,T_{\pi/2}}L_{(\alpha,c)}$
\\
$\Isom_0(\TSL)$ & o
& $\Set{1,I_2}L_{(\alpha,c)}T_\theta$
\\
$\Isom(\TSL)$ & o
& ibid
\\
\\\hline\hline
\end{tabular}
\end{table}

\subsubsection{Invariance groups}

From the argument in the previous subsection, invariance groups with 
$G_0=\Isom_0(\TSL)$ are given by $\Isom_0(\TSL)$, and $\Isom(\TSL)$.

The invariance subgroups $G$ with $G_0=\BTVIII_L$ are determined 
as follows. Since $\Isom(\TSL)\subset \N(\BTVIII_L)$, candidates 
of discrete isometries to be added are $I_2$ and $T_\theta$. 
Clearly $\BTVIII_L\sdp\Set{1,I_2}$ is a subgroup of 
$\Isom(\TSL)$. On the other hand, since $T_\theta$ transforms the 
invariant basis as
\Eq{
T_\theta^* \chi^I=R_3(2\theta)^I{}_J \chi^J,
}
any $\BTVIII_L$-invariant tensor quantity becomes 
$\Isom_0(\TSL)$ invariant if it is invariant under $T_\theta$ 
unless $2\theta=n\pi (n\in\ZR)$. Further, $T_{n\pi}=L_{(i, n\pi)}$. 
Hence, the only possible element of the form $T_\theta$ in $G$ which 
does not belong to $\BTVIII_L$ is $T_{\pi/2}$. It commutes with the 
left transformations and with $I_2$ mod. $\BTVIII_L$, 
$I_2T_{\pi/2}=T_{\pi/2}I_2T_{\pi}$. Hence the possible invariance 
groups with $G_0=\BTVIII_L$ are given by $\BTVIII_L, 
\BTVIII_L\sdp\Set{1,I_2}, \BTVIII_L\cdot\Set{1,T_{\pi/2}}$, and 
$\BTVIII_L\cdot \tilde D_2$, where
\Eq{
\tilde D_2=\Set{I,I_2,T_{\pi/2}, I_2T_{\pi/2}}.
}
The normalizers of these groups are given in Table 
\ref{table:VIII:normalizer}. 

Finally note that $\Isom_0(\TSL)$ also contains $\BTIII_L$ as a 
transitive subgroup, say 
$\{L_{\hat{\bm{a}}}T_c|\hat{\bm{a}}\in\RF^2,c\in\RF\}$. However, we do 
not consider this subgroup separately, because a system with the local 
$\BTIII_L$ invariance on a compact quotients always has an invariance 
group of higher dimension as shown in \S\ref{sec:H^2xR:InvG}. Thus, in 
the 
present case, it is contained in the cases with 
$G\supset\Isom_0(\TSL)$. 

\subsection{Compact topologies}

\subsubsection{Fundamental group}\label{sec:FundamentalGroup2}

Through the natural homomorphisms
\Eq{
\TSL \maps \SL_2\RF \maps \Isom_0(H^2)\cong \PSL_2\RF,
}
$\TSL$ acts transitively on $H^2$. Hence, $\TSL$ can be regarded 
as a fibre bundle over $H^2$ with a fibre $\RF$, which is 
isomorphic to the isotropy group $\SetDef{\tilde R(z)}{z\in\RF}$. 
In terms of the parametrization $\tilde V(\zeta,z)\in\TSL$, the 
projection of this bundle structure is expressed as
\Eq{
\tilde \pi: \tilde V(\zeta,z) \mapsto \zeta\in H^2.
}
Since the action of $L_{(\alpha,c)}T_\theta\in\Isom(\TSL)$ on 
$\TSL$ is written as
\Eq{
L_{(\alpha,c)}T_\theta: (\zeta,z)
\maps \left(a+bR(c)*\zeta,c+\theta+H(z,\alpha)\right),
}
$\Isom(\TSL)$ preserves this bundle structure. Hence, by the same 
arguments in \S\ref{sec:FundamentalGroup}, we can show that any 
compact quotient $\Sigma$ of $\TSL$ has a Seifert bundle structure 
over an orbifold $X$ covered by $H^2$. Further, it can be shown that 
$\Sigma$ has a unique Seifert bundle structure up to 
isomorphism\cite{Scott.P1983}. Therefore, the fundamental group 
of $\Sigma$ is given by Eq.\eqref{Seifert:pi1:orientable} or 
Eq.\eqref{Seifert:pi1:unorientable}, depending on whether the 
base orbifold is orientable or not. The only difference from the 
$H^2\times\RF$ type is that now the condition on $e$ is given by 
$e<0$. 

\subsubsection{Moduli freedom}

In the compactification of $H^2\times\RF$, discrete transformations in 
$\Isom(H^2)$ to compactify $H^2$ did not belong to the group 
$\HSL_2^+\cong H^2$ and were associated with rotations in $H^2$. This 
feature required the invariance group $G$ to include 
$\Isom_0(H^2\times\RF)$. In contrast, in the present case, 
$\Isom^+(H^2)$ is covered by the left translation group $\BTVIII_L$ 
itself and no rotation in $\TSL$ is required to compactify the base 
space $H^2$. Further, although the generator $\ell$ must be 
represented by a translation along the fibre, $T_l\in \BTVIII_R$, in 
the present case as well from the argument in 
\S\ref{sec:FundamentalGroup}, we can now represent $\ell$ as a 
transformation in $\BTVIII_L$, because $\BTVIII_L$ and $\BTVIII_R$ 
share the non-trivial center. Hence, in the present case, the 
fundamental group can be embedded even into the smallest group 
$\BTVIII_L$.%
\footnote{In the appendix of Ref. 
\citen{Koike.T&Tanimoto&Hosoya1994} it was claimed that $\TSL$ 
cannot be compactified by a discrete group $K$ contained in 
$\BTVIII_L$. But the proof there is not correct because they assumed 
that the transformation group of the base space $H^2$ induced from 
the action of $K$ through the bundle structure $\TSL\maps H^2$ was 
contained in $H^2_L$. In reality, $\BTVIII_L$ induces 
$\Isom_0(H^2)$, as shown above.}

First we consider the embedding of $\pi_1(S^+(g,e;k))$ in 
$G=\BTVIII_L$ or $G=\BTVIII_L\sdp\Set{1,I_2}$. The latter case is 
the same as for the former, because the transformation must 
preserve the orientation of the orbifold, hence the fundamental group 
is contained in $\BTVIII_L$. In the embedding of 
$\pi_1(\Sigma)$ given by Eq.\eqref{Seifert:pi1:orientable} into 
$\BTVIII_L$, the freedom in the choice of the generators 
$\bar\alpha_a,\bar\beta_a,\bar\gamma_i$ is the same as in the 
case of $H^2\times\RF$. However, the freedom in taking their 
lifts to $G$ and in the choice of $\ell$ is different. In fact, 
from the exact sequence
\Eq{
0 \maps \{T_{n\pi}\} \mapsnamed{i} \BTVIII_L
 \mapsnamed{j} \Isom_0(H^2) \maps 1,
}
$l$ for $\ell=T_l$ is restricted to $l=n\pi$ with non-zero integer $n$. 
Similarly,  the freedom in the lifts of the generators $\bar\alpha_a, 
\bar\beta_a\in 
\Isom_0(H^2)$ are represented by the multiplication of 
translations of the form $T_{m \pi}$ with $m\in\ZR$. This 
restriction to discrete values of the freedom in the lifts 
decreases the number of the moduli degrees of freedom by $1+2g$, 
compared with the 
$H^2\times\RF$ case. On the other hand, $\N(G)$ is now contained in  
$\Set{1,I_2}\BTVIII_L\cdot\BTVIII_R$, in which the action of the 
subgroup $\BTVIII_R$ on the moduli is trivial, while the action of 
$\BTVIII_L$ on the moduli has no continuous isotropy group. Hence the 
effective HPD freedom in the moduli is smaller by 1 than the 
$H^2\times\RF$ case. Thus, the number of the moduli freedom is 
given by
\Eq{
N_{m+}=N_{m+}'-2g =2k + 6g -6.
\label{VIII+:ModuliFreedom}}
Since $I_2$ reverses the direction of $\ell$ and $R_{\bm{a}}$ commutes 
with the action of $\BTVIII_L$, the isotropy group $\N_{\Sigma0}(G)$ of 
the action of $\N(G)$ in the moduli space is given by 
$\N(G)\cap\BTVIII_R$.

So far we have not touched upon the lift of $\bar\gamma_i$. 
Because the normalization condition $\gamma_i^{p_i}\ell^{r_i}=1$ always 
fixes the lift uniquely, it does not affect the 
argument on the degrees of freedom. However, it is not guaranteed 
that there exits a lift satisfying this condition in general. In 
fact, in the present case, this condition yields a restriction on 
topologies whose fundamental group can be embedded in 
$\BTVIII_L$. To see this, note that $\bar\gamma_i$ is represented 
by the left multiplication of an element in $\SL_2\RF$ of the 
form $Z(\zeta_i)R(\pi/p_i)Z(\zeta_i)^{-1}$ by the $I_2$ 
transformation if necessary, whose lift is given 
by $\tilde Z(\zeta_i)\tilde R(\pi/p_i+m_i\pi)\tilde 
Z(\zeta_i)^{-1}$ with some $m_i\in\ZR$. Thus the normalization condition is 
expressed as $1+m_i p_i=-nr_i$. This equation has an integer 
solution for $m_i$ only when the condition 
\Eq{
(1+n r_i)/p_i \in \ZR
\label{VIII:restriction}}
is satisfied for all $i=1,\cdots, k$. In terms of the pair of 
coprime integers $(q_i,s_i)$ such that $p_is_i+q_ir_i=1$, this 
condition is expressed as 
\Eq{
(n+q_i)/p_i \in \ZR.
}
It can be shown by induction that there exists an integer $n$ 
satisfying this condition if and only if the greatest common  
divisor of $p_i$ and $p_j$ divides $q_i-q_j$ for any pair 
$i\not=j$:
\Eq{
(p_i,p_j)|(q_i-q_j).
\label{VIII+:CondTop1}
}
Further if $n=n_0$ is the smallest positive integer satisfying the 
condition, $n=n_0+mL$ satisfies the condition for any integer $m$, 
where $L$ is the least common multiple of $p_i$. For example, $n=1$ 
satisfies the condition if and only if $r_i=p_i-1$ for all 
$i=1,\cdots,k$.

Further, the condition that $b$ is integer also restricts possible 
topologies.  From the structure of the invariance group, 
$[\alpha_1,\beta_1]\cdots[\alpha_g,\beta_g]\gamma_1\cdots\gamma_k$ 
is always written as $T_{h\pi}$ with some integer $h$.  Let 
$h_0$ be the value of $h$ for $\gamma_i=\tilde Z(\zeta_i)\tilde 
R(\pi/p_i)\tilde Z(\zeta_i)^{-1}$, and let $n=n_0+mL$ be integers 
satisfying the above condition. Then $h$ for $n=n_0+mL$ is 
written as $h=h_0-\sum_i(1+nr_i)/p_i$, and $b$ is given by 
$h=bn$. Hence, when an integer $b$ with $e<0$ is given, we can 
find an embedding of the corresponding fundamental group only 
when the following condition is satisfied:
\Eq{
\frac{h_0-\sum_i1/p_i+n_0e}{eL} \in \ZR,
\label{VIII+:CondTop2}
}
where $e=-(b+\sum_i r_i/p_i)$ is the Euler number. Since $n_0$ 
and $h_0$ are determined by the genus of the orbifold and the 
Seifert indices, the allowed values of $b$ ( or $e$) are also 
determined by them. It is very difficult to find a general 
expression for $h_0$. Hence, we cannot give an explicit criterion 
on the indices for the fundamental group to be embedded in 
$\BTVIII_L$, but it is certain that for any genus and any set of 
orbifold indices $p_i$ there exists at least one choice of $r_i$ 
such that the embedding is possible. It is because $n_0=1$ for 
$r_i=p_i-1$ and $b=h_0-k$ satisfies the above condition.

Next we consider the case $G=\BTVIII_L\cdot\Set{1,T_{\pi/2}}$ or 
$G=\BTVIII_L\cdot \tilde D_2$, the latter of which is reduced to the 
former case. The only difference in the case 
$G=\BTVIII_L\cdot\Set{I,T_{\pi/2}}$ from the previous case is that from 
the exact sequence
\Eq{
0 \maps \{T_{n\pi/2}\} \mapsnamed{i} G
 \mapsnamed{j} \Isom_0(H^2) \maps 1,
}
translations in the $z$ direction must be integer multiples of $\pi/2$ 
instead of $\pi$. So the generator $\ell$ is represented as $\ell=T_l$ 
with $l=\pi n/2$($n\in\ZR$), and the normalization 
condition for $\gamma_i$ gives the restriction
\Eq{
(2q_i+n)/p_i \in \ZR,
}
which has a solution for $n$ if and only if the greatest common 
divisor of $p_i$ and $p_j$ divides $2(q_i-q_j)$ for any pair 
$i\not=j$,
\Eq{
(p_i,p_j)|2(q_i-q_j),
\label{VIII-:CondTop1}
}
and $b$ becomes an integer if and only if
\Eq{
\frac{2h_0-2\sum_i1/p_i+n_0e}{eL} \in \ZR.
\label{VIII-:CondTop2}
}
Both of these conditions are weaker than the previous ones, hence a 
slightly wider class of topologies are allowed. Apart from this 
difference in the restriction on the topology, the count of the moduli 
degrees of freedom is the same as in the previous case and given by 
Eq.\eqref{VIII+:ModuliFreedom}. The isotropy group $\N_{\Sigma0}(G)$ is 
again given by $\N(G)\cap\BTVIII_R$.

The argument on the embedding of $\pi_1(S^-(g,e;k))$ into $G$ with 
$G_0=\BTVIII_L$ is quite 
similar. Now the invariance group $G$ should contain a 
transformation which reverses the orientation of $H^2$, hence 
$G=\BTVIII_L\sdp\Set{1,I_2}$ or $G=\BTVIII_L\cdot \tilde D_2$. In the 
former case, $\ell$ is represented as $T_{n\pi}$ with an 
non-vanishing integer $n$, and the lift of $\bar\alpha_a$ is 
represented by $I_2\tilde Z(\zeta_a)\tilde R(\theta_a)T_{x_a\pi}$ with 
some integer $x_a$. Now, there is no HPD which changes $x_a$ 
continuously unlike in the case of $H^2\times\RF$. Hence, the number of 
the moduli degrees of freedom is smaller by $g-1$ than that for the 
$S^-(g,0;k)$ case of $H^2\times\RF$:
\Eq{
N_{m-}=N_{m-}'-g+1=2k+3g-6.
\label{VIII-:ModuliFreedom}
}
The restriction on the indices characterizing the fundamental 
group is the same as that for the embedding of $S^+(g,e;k)$ in 
$G=\BTVIII_L$ and is given by \eqref{VIII+:CondTop1} and 
\eqref{VIII+:CondTop2}. For $G=\BTVIII_L\cdot \tilde D_2$, the number 
of the moduli degrees of freedom is given by the same expression 
\eqref{VIII-:ModuliFreedom}, but the possible topologies are 
restricted by \eqref{VIII-:CondTop1} and \eqref{VIII-:CondTop2}. Since 
$T_{\pi/2}$ in $\N(G)$ transforms $I_2$ non-trivially as 
$T_{\pi/2}I_2T_{-\pi/2}=I_2T_{-\pi}$, $\N_{\Sigma0}(G)$ is given by 
$\{\tilde Z_R(ie^\gamma)\}$ for $G=\BTVIII_L\sdp\Set{1,I_2}$, while it 
becomes trivial for $G=\BTVIII_L\cdot \tilde D_2$.

Finally we consider the case $G=\Isom_0(\TSL)$ or 
$G=\Isom(\TSL)$. First, the image of $\pi_1(S^+(g,e;k))$ is 
always contained in $\Isom_0(\TSL)$. Since $\N(G)$ is the same for 
$G=\Isom_0(\TSL)$ and $G=\Isom(\TSL)$, the argument for the latter is 
reduced to that for the former as for $G_0=\BTVIII_L$. Now $G$ contains 
translations in $z$-direction with distances represented by arbitrary 
real numbers. Hence, $l$ for $\ell=T_l$ can take arbitrary non-zero 
real number, the  degrees of freedom in the lifts of $\alpha_a$ and 
$\beta_a$ increase by $2g$ compared to the 
case $G=\BTVIII_L$, and no restriction like 
\eqref{VIII+:CondTop1} and \eqref{VIII+:CondTop2} arises. In fact, if 
we start from the choice $T_{x\pi}$ for $\ell$ and $(\gamma_i)_0=\tilde 
Z(\zeta_i)\tilde R(\pi/p_i)\tilde Z(\zeta_i)^{-1}$, 
$\gamma_i=(\gamma_i)_0T_{x_i\pi}$ with $x_i=-(1+xr_i)/p_i$ 
satisfies the normalization condition for $\gamma_i$. Further, after 
this rescaling, 
$[\alpha_1,\beta_1]\cdots[\alpha_g,\beta_g]\gamma_1\cdots\gamma_k$ 
is given by $T_{(h_0+\sum_i x_i)}$. Hence, for a given $b$, we 
obtain the equation $h_0+\sum_i x_i=xb$, or
\Eq{
xe = \sum_i\frac{1}{p_i}-h_0.
}
Since $e<0$, this equation always has a unique solution. Thus the 
fundamental group of any $S^+(g,e;k)$ can be embedded into $G$. Since 
$T_\theta$ commutes with the action of $\Isom_0(\TSL)$, the dimension 
of $\N(G)$-orbits in the moduli space is 3. Hence, the number of the 
moduli degrees of freedom is given by 
\Eq{
N''_{m+}=(2k+6g-3)+(2g+1)-3=2k+8g-5,
\label{TSL2+:ModuliFreedom}}
and the isotropy group $\N_{\Sigma0}(G)$ is again given by 
$\N(G)\cap\BTVIII_L$.

The argument for $\Sigma=S^-(g,e;k)$ is almost the same. Now the 
fundamental group can be embedded only into 
$G=\Isom(\TSL)$ and the $\alpha_a$ is represented by a 
transformation with $I_2$. Due to this, the action of $T_\theta$ on the 
moduli becomes effective, and the dimension of $\N(G)$-orbits in the 
moduli space is given by 4. Apart from this difference, the 
argument is just a combination of those given so far. Hence, the number 
of the moduli degrees of freedom is given by 
\Eq{
N''_{m-}=(2k+3g-3)+(g+1)-4=2k+4g-6,
\label{TSL2-:ModuliFreedom}}
and the isotropy group $\N_{\Sigma0}(G)$ becomes trivial. 
No restriction on topology exists apart from the conditions $\chi<0$ 
and $e<0$.

\subsection{Phase Space}\label{sec:SL2:PhaseSpace}

From the general argument in \S\ref{sec:InvPhaseSpace}, the 
invariant phase space is expressed as \eqref{InvPhaseSpace:reduced}. 
The dimension of the reduced moduli space $\M_0(\Sigma,G)$ was 
determined in the previous subsection, but its topological structure 
was not discussed there. It is because that problem is too difficult 
to address. For the same reason, we do not discuss the structure and 
the action of the possible residual discrete transformation group 
$H_{\rm mod}$ in the expression \eqref{InvPhaseSpace:reduced}, and 
only determine the structure of the dynamical sector $\Gamma_{\rm 
dyn}(\Sigma,G)$.

First we consider the case $G=\BTVIII_L$, which can be realized only 
for $\Sigma=S^+(g,e,;k)$. In this case 
$\N_{\Sigma0}(G)=\BTVIII_R\cap\N(G)$ coincides with  $\BTVIII_R$, which 
transforms the invariant basis for $\BTVIII_L$ as 
\eqref{F:VIII:1} and \eqref{F:VIII:2}. Let $\eta$ be the diagonal 
matrix $D(-1,1,1)$. Then, for the matrix $Q$ representing the 
components of the metric, $\eta Q$ always has an eigenvector $v_1$, 
$\eta Q v_1=\lambda_1 v$, in the complex 3-vector space. Since 
$v_1^\dagger Qv_1$ is positive, $\lambda_1 v_1^\dagger \eta 
v_1>0$, which implies that $\lambda_1$ is a non-vanishing real 
number. Hence we can take $v_1$ as a real 3-vector with unit 
length with respect to $\eta$. By repeating this argument in the 
subspace orthogonal to $v_1$ with respect to the metric $\eta$ in 
$\RF^3$, we finally obtain the set of three eigenvectors and 
three eigenvalues such that $\eta Q (v_1 v_2 v_3)=(v_1 v_2 
v_3)D(\lambda_1,\lambda_2,\lambda_3)$. From this we find that for 
$\Lambda=(v_1,v_2,v_3)\in\SO_+(2,1)$, $\Lambda Q 
\Tp\Lambda=D(Q_1,Q_2,Q_3)$, where $Q_i=\pm \lambda_i>0$. Hence, we can 
diagonalize $Q$ by $\BTVIII_R$. We can further put $Q_1\ge Q_2$ by 
$F=R_3(\pi/2)$ corresponding to the $T_{\pi/4}$ 
transformation if necessary. After this diagonalization, the 
momentum constraints are expressed as
\Eqr{
&& H_1=2P^{23}(Q_2+Q_3)+c u_1=0,\\
&& H_2=-2P^{13}(Q_1+Q_3)+c u_2=0,\\
&& H_3=2P^{12}(Q_1-Q_2)+c u_3=0.
\label{MomentumConstraint:VIII}}

For the vacuum system with $c=0$, we obtain the constraint 
$P^{13}=P^{23}=0$. This implies that the invariance group 
contains $T_{\pi/2}$ inducing $F=I_3$, hence is larger than 
$\BTVIII_L$. Further, when $Q_1>Q_2$, $P^{12}$ also vanishes and the 
invariance group $G$ of the system includes 
$\BTVIII_L\cdot \tilde D_2$. On the other hand, when $Q_1=Q_2$, $Q$ is 
invariant under $F=\Lambda=R_3(\theta)$, in terms of which $P$ can be 
diagonalized. Hence, $G\supset \BTVIII_L\cdot \tilde D_2$ again.

In contrast, for the fluid system, the constraints simply 
determine $P^{13}$ and $P^{23}$ by $u_1$ and $u_2$, and further 
$P^{12}$ by $u_3$ if $Q_1>Q_2$. In the special case $Q_1=Q_2$, for 
which we obtain the constraint $u_3=0$, we can put $P^{12}=0$ by the 
residual HPDs. Including this special case, $G=\BTVIII_L$ if and 
only if two of $u_i$ do not vanish. Hence, the dynamical sector of 
the invariance phase space for the single-component fluid system is 
given by
\Eqr{
&\Gamma_{\rm dyn}(S^+,\BTVIII)=
&\big\{(Q_1,Q_2,Q_3;P^{11},P^{22},P^{33};u_1,u_2,u_3,\rho) \big| 
\nonumber\\ 
&& Q_1\ge Q_2>0, Q_3>0,
u_1^2u_2^2+u_2^2u_3^2+u_1^2u_2^2\not=0\big\}/\ZR_2,}
where $\ZR_2$ represents the group $\Set{1,I_3}$ with $I_3$ 
induced from $T_{\pi/2}$. 

Next for $G=\BTVIII_L\sdp\Set{1,I_2}$, the symmetry requires that 
$Q_{12}=Q_{23}=P^{12}=P^{32}=u_1=u_3=0$, and a transformation in 
$\N_{\Sigma0}(G)$ is written as $\Set{1,T_{\pi/2}}\tilde 
Z_R(ie^\gamma)$ for $\Sigma=S^+(g,e;k)$ and by $\tilde 
Z_R(ie^\gamma)$ for $\Sigma=S^-(g,e;k)$. From the definition of the 
representation \eqref{rep:SLtoSO}, we find that $\Lambda(\tilde 
Z(ie^\gamma))$ is given by the Lorentz boost in the $T-X$ plain or 
the $\chi^1-\chi^3$ plain,
\Eq{
\begin{pmatrix}{c}
 \chi^1{}' \\ 
 \chi^3{}'
\end{pmatrix}
=\begin{pmatrix}{cc}
   \cosh\gamma & -\sinh\gamma\\
  -\sinh\gamma & \cosh\gamma
\end{pmatrix}
\begin{pmatrix}{c}
   \chi^1{} \\ 
   \chi^3{}
\end{pmatrix}\ .
}
We can diagonalize $Q$ by this transformation to 
$Q=D(Q_1,Q_2,Q_3)$. Then, the momentum constraint $H_2=0$ 
determines $P^{13}$ by $u_2$. If $u_2=0$, the system has a higher 
symmetry, so $u_2\not=0$. For $\Sigma=S^+(g,e;k)$, we can put 
$u_2>0$ by the $I_3$ transformation induced from $T_{\pi/2}$. Hence
\Eqr{
&\Gamma_{\rm dyn}(S^+,\BTVIII\sdp\Set{1,I_2})=
& \big\{(Q_1,Q_2,Q_3;P^{11},P^{22},P^{33};u_2,\rho) \big| \nonumber\\
&& \qquad Q_1, Q_2,Q_3>0,u_2>0 \big\}.
}
On the other hand, for $\Sigma=S^-(g,e;k)$, $T_{\pi/2}$ is not 
contained in $\N_{\Sigma0}(G)$. Hence,
\Eqr{
&\Gamma_{\rm dyn}(S^-,\BTVIII\sdp\Set{1,I_2})=
& \big\{(Q_1,Q_2,Q_3;P^{11},P^{22},P^{33};u_2,\rho) \big| \nonumber\\
&& \qquad Q_1, Q_2,Q_3>0,u_2\not=0 \big\}.
}
%


\begin{table}[t]
\caption{\label{table:count:SL2} The parameter count for type $\TSL$.}
\begin{tabular}{llccccccc}
\hline
\hline
\bf Space & \bf Symmetry & $Q$ & $P$ & $N_m$ & $N_f$ & $N$ & $N_s$ & $N_s$(vacuum)\\
\hline\\
$\RF^3$ 
&$\BTVIII$	& 3 & 3 & 0 & 4 & 10 & 8 &  --\\
&$\BTVIII\sdp\ZR_2$
                & 3 & 3 & 0 & 2 & 8 & 6 &  --\\
&$\BTVIII\cdot \tilde D_2$
                & 3 & 3 & 0 & 1 & 7 & 5 &  4\\
&$\Isom(\TSL)$	& 2 & 2 & 0 & 1 & 5 & 3 &  2\\
&&&&&&&&\\
$S^+(g,e;k)$
&$\BTVIII$	& 3 & 3 & $N_{m+}$ & 4 & $10+N_{m+}$ & 
$8+N_{m+}$ & --\\
&$\BTVIII\sdp\ZR_2$
                & 3 & 3 & $N_{m+}$ & 2 & $8+N_{m+}$ & $6+N_{m+}$ &--\\
&$\BTVIII\cdot \tilde D_2$
                & 3 & 3 & $N_{m+}$ & 1 & $7+N_{m+}$ & $5+N_{m+}$ & 
$4+N_{m+}$\\
&$\Isom(\TSL)$	& 2 & 2 & $N''_{m+}$ & 1 & $5+N''_{m+}$ & $3+N''_{m+}$ 
& $2+N''_{m+}$\\                
&&&&&&&&\\
$S^-(g,e;k)$
&$\BTVIII\sdp\ZR_2$
                & 3 & 3 & $N_{m-}$ & 2 & $8+N_{m-}$ & $6+N_{m-}$ & --\\
&$\BTVIII\cdot \tilde D_2$
                & 3 & 3 & $N_{m-}$ & 1 & $7+N_{m-}$ & $5+N_{m-}$ & 
$4+N_{m-}$\\
&$\Isom(\TSL)$	& 2 & 2 & $N''_{m-}$ & 1 & $5+N''_{m-}$ & $3+N''_{m-}$ 
& $2+N''_{m-}$\\                
\\
\hline\hline
\end{tabular}
\end{table}

The argument for $G=\BTVIII_L\cdot\Set{1,T_{\pi/2}}$ with 
$\Sigma=S^+(g,e;k)$ is quite similar to the previous case. Since 
$T_{\pi/2}$ induces the $I_3$ transformation of the invariant basis, 
the symmetry requires 
$Q_{13}=Q_{23}=P^{13}=P^{23}=u_1=u_2=0$. By HPD $T_\theta$ in 
$\N_{\Sigma0}(G)=\N(G)\cap\BTVIII_R$, we can diagonalize $Q$ as 
$Q=D(Q_1,Q_2,Q_3)$ with $Q_1\ge Q_2$. If $Q_1=Q_2$, we can further put 
$P^{12}=0$, which together with $H_3=0$ requires the fluid velocity to 
vanish. This implies that the system has an invariance group larger 
than $\BTVIII_L\cdot \tilde D_2$. Hence $Q_1>Q_2$, and $P^{12}$ is 
determined by $u_3$ through $H_3=0$. The dynamical sector of the 
invariant phase space is given by
\Eqr{
& \Gamma_{\rm dyn}(S^+,\BTVIII\cdot\Set{1,T_{\pi/2}})=
& \big\{(Q_1,Q_2,Q_3;P^{11},P^{22},P^{33};u_3,\rho) \big| 
\nonumber\\
&& \qquad Q_1>Q_2>0,Q_3>0, u_3\not=0\big\}.
}

The $G=\BTVIII_L\cdot \tilde D_2$ case is very simple. Since $\tilde 
D_2$ induces the transformation group $D_2=\Set{1,I_1,I_2,I_3}$ of the 
invariant basis, the invariance under $\tilde D_2$ requires that both 
$Q$ and $P$ are diagonal and the fluid velocity vanishes. Hence, the 
existence of fluid does not play an essential role in the analysis and 
simply adds the matter density freedom to the invariant phase space of 
the vacuum system. The system does not have a higher symmetry if and 
only if $Q_1\not=Q_2$ or $P^1\not=P^2$. Thus, for the single-component 
fluid system,
\Eqr{
& \Gamma_{\rm dyn}(S^\pm,\BTVIII\cdot\tilde D_2)=
& \big\{(Q_1,Q_2,Q_3;P^1,P^2,P^3;\rho) \big| 
\nonumber\\
&& \quad Q_1,Q_2,Q_3>0,Q_1\not=Q_2 \text{\ or\ } P^1\not=P^2\big\}.
}

Finally for $G\supset \Isom_0(\TSL)$, the argument is the same as 
that for the previous case apart from the symmetry conditions 
$Q_1=Q_2$ and $P^1=P^2$, and the system with a single-component 
fluid has always the maximal symmetry 
$\Isom(\TSL)$. Hence, the dynamical sector of its invariant phase 
space is given by
\Eq{
\Gamma_{\rm dyn}(S^\pm,\Isom(\TSL))=
\big\{(Q_1,Q_3;P^1,P^3;\rho) \big| Q_1,Q_3>0\big\}.
}

So far we have considered only the single-component fluid system, 
but the extension to a multi-component system is quite simple. 
First, for $G=\BTVIII_L$, each extra component of fluid adds the 
four variables $(u_I,\rho)$ to $\Gamma_{\rm dyn}$. Second, for 
$G=\BTVIII_L\sdp\ZR_2$, two variables $(u_2,\rho)$ or $(u_3,\rho)$ 
are added for each extra component. Third, for $G\supset 
\BTVIII_L\cdot\tilde D_2$, each extra component only add its energy 
density. Finally for $G=\Isom_0(\TSL)$, $\Gamma_{\rm dyn}(S^+,G)$ is 
obtained by adding $(u_3,\rho)$ for each extra component to 
$\Gamma_{\rm dyn}(S^+,\Isom(\TSL))$ of the single-component system.

The parameter counts for the vacuum and the single-component fluid 
system of the type $\TSL$ obtained in this section is summarized in 
Table \ref{table:count:SL2}.

\section{Summary and Discussion}

In this paper, we have systematically determined the structure of the 
diffeomorphism invariant phase space, i.e., the initial data space, of 
compact Bianchi models with fluid for all possible space topologies and 
local invariance groups except for those modeled on the Thurston types 
$S^3, H^3$ and $S^2\times\RF$ which have no moduli freedom. The main 
results are summarized in Tables 
\ref{table:count:E3}, \ref{table:count:Nil}, \ref{table:count:Sol}, 
\ref{table:count:H^2xR} and \ref{table:count:SL2}. Although these 
tables list only the parameter counts for the vacuum and the 
single-component fluid system, they can be easily extended to 
multi-component fluid systems as was described in detail in the 
arguments of individual systems. It is also easy to extend the 
analysis to systems with scalar fields. Since scalar fields have to 
be spatially constant if they are spatially homogeneous locally, 
their contribution to the phase space does not depend on the space 
topology or details of the invariance group and simply increases the 
parameter count by two times the number of fields.

We can read various things from these tables. First, although it is 
quite natural that the parameter count increases when the space is 
compactified due to the appearance of the moduli freedom, the degree 
of increase is quite sensitive to whether fluid exists or not as 
well as the local symmetry and the space topology of the system. In 
particular, although the parameter count $N_s$ for the locally 
$\RF^3$-symmetric system is larger than that for the locally 
$\BTVII_0$-symmetric system in the vacuum case, the latter becomes 
larger than the former for the system with fluid. Hence, the Bianchi 
type $\BTVII_0$ symmetry becomes more generic than the Bianchi type 
$\BTI$ symmetry for compact Bianchi models with fluid as for 
non-compact models. This inversion occurs because the locally 
$\RF^3$-symmetric system always has the higher local symmetry 
$\RF^3\sdp D_2$, while the minimum locally symmetry $\BTVII_0\sdp 
D_2$ of the compact vacuum Bianchi $\BTVII_0$ model is broken down 
to $\BTVII_0$ when fluid is introduced.

Second, the local spatial homogeneity often requires a local 
isotropy when the space has a complicated topology, even for the 
class A Bianchi models. For example, Bianchi-type $\BTI$ models with 
space $T^3/\ZR_k$($k=3,4,6$) and Bianchi-type $\BTII$ models with 
space $T^3(n)/\ZR_k$($k=3,4,6$) always have local rotational 
symmetry. There are also models for which additional discrete 
symmetries are required. They are the Bianchi-type $\BTVII_0$ model 
with fluid on $T^3/\ZR_2\times\ZR_2$, Bianchi-type $\BTII$ models on 
$K^3(n)$, $T^3(n)/\ZR_k$($k=2,3,4,6$) and $T^3(n)/\ZR_2\times\ZR_2$, 
Bianchi-type $\BTVI_0$ models on $\Sol(n)$ with $n<-2$ and 
Bianchi-type $\BTVIII$ models on $S^-(g,e;k)$. For these models, 
although additional discrete symmetries do not affect the 
geometrical degrees of freedom, they require the fluid velocity to 
align toward a special direction. 

This feature in general makes models with complicated topologies 
less probable if we assume that a model with a larger parameter 
count has a higher probability to be realized. However, there are 
very important exceptions to this general tendency. They are models 
with spaces covered by $H^2\times\RF$ or $\TSL$. For these models, 
the number of the moduli degrees of freedom increases without bound 
as the genus $g$ or the number $k$ of conical singularities in the 
base orbifold increases, as is clear from 
\eqref{H2xR+:ModuliFreedom}, \eqref{H2xR-:ModuliFreedom}, 
\eqref{VIII+:ModuliFreedom}, \eqref{VIII+:ModuliFreedom}, 
\eqref{TSL2+:ModuliFreedom} and \eqref{TSL2-:ModuliFreedom}. It 
should be noted here that although fully anisotropic Bianchi-type 
$\BTVIII$ models have larger  degrees of freedom in the fluid sector 
than locally rotationally symmetric models, the total parameter 
count $N_s$ for the latter becomes larger than that for the former 
when $g\ge2$ since $N_s(S^+,\TSL)-N_s(S^+,\BTVIII)=2g-2$ and 
$N_s(S^-,\TSL)-N_s(S^-,\BTVIII\sdp \ZR_2)=g-1$. Hence, the LRS 
Bianchi-type $\BTIII$ and $\BTVIII$ models are the most generic 
among all compact Bianchi models. 

This result has an important cosmological consequence. In most of 
the studies on cosmological models with compact space, in 
particular, in the investigations of the effect of space topology on 
the CMB anisotropy, space topologies of the Thurston types $E^3$ and 
$H^3$ have been considered. This is because many people believe that 
the Bianchi symmetries admitting the spatially homogeneous and 
isotropic models should be imposed to accord with the observed 
isotropy of the universe. From this view point, compact Bianchi 
models of the type $\BTIII$ or $\BTVIII$ should be rejected, because 
they only admit LRS models at best. However, if a sufficient 
inflation occurs in the early stage of the universe, a universe with 
any spatial symmetry can be made sufficiently flat and isotropic on 
the horizon scale at present. Hence, there is no reason to consider 
only models with spatial isotropy. From this view point, the above 
result rather suggests that the space topology of our universe may 
be of the Thurston type $H^2\times\RF$ or $\TSL$ if our universe has 
compact space.  

Of course, it is dangerous to say something definite about the real 
universe with infinite degrees of freedom simply based on the 
analysis of locally homogeneous systems with finite degrees of 
freedom. However, it is also hard to think that the strong influence 
of the space topology on the isotropy and the dynamical degrees of 
freedom found in the present paper is just an accidental feature of 
the locally homogeneous systems. We hope that this point will be 
clarified in future studies.


\section*{Acknowledgments}
The author would like to thank John D. Barrow for valuable and 
stimulating discussions. This work was supported by the JSPS grant No. 
05640340.

\appendix

\section{Propositions on the normalizer}\label{appendix:normalizer}

In this appendix we give some useful general propositions, which are 
used to determine the normalizer groups in the present paper.

In general a transformation group $G$ is not connected. The normalizer 
of such a group can be easily determined if one knows the normalizer of 
the maximal connected subgroup $G_0$, from the following proposition.

\smallskip

\begin{proposition}\label{prop:normalizer:discrete}
The normalizer of $G$ is contained in that of $G_0$. Hence, if $G$ is 
generated by $G_0$ and a set of discrete transformations $S$, $\N(G)$ 
consists of $f\in\N(G_0)$ such that $fgf^{-1}\in G$ for any $g\in S$.
\end{proposition}

\begin{proof}
For any $f\in\N(G)$, $fG_0f^{-1}$ is a connected subgroup of $G$ and 
$fG_0f^{-1}\cap G_0\not=\emptyset$. Hence, $fG_0f^{-1}\subset G_0$. 
This implies that $f\in\N(G_0)$.
\end{proof}

\smallskip

Each HPD $f \in \N(G)$ gives an isomorphism of $G$, which induces an 
isomorphism $f_*$ of $\LieA(G)$. That is, $f_*$ is contained in  
$\Aut(\LieA(G))$ which is the set of linear transformations $\phi$ 
satisfying the condition
\Eq{
[\phi(X),\phi(Y)]=\phi([X,Y])\quad \forall X,Y\in\LieA(G).
}
In general, it is easier to determine $\Aut(\LieA(G))$ than $\N(G)$, 
because the former is an algebraic problem. In particular, if the 
Killing form is non-degenerate, it can be used to restrict possible 
candidates for the linear transformations in $\Aut(\LieA(G))$. Let 
$\gamma(X,Y)$($X,Y\in\LieA(G)$) be the Killing form of $\LieA(G)$ 
defined by
\Eq{
\gamma(X,Y)=\Tr(\Ad(X) \Ad(Y)),
}
where $\Ad(X)$ is the linear transformation defined by 
$\Ad(X)Y=[X,Y]$. Then, since $\phi\in\Aut(\LieA(G))$ preserves the 
matrix representation of $\Ad(X)$, the following proposition holds.

\smallskip

\begin{proposition}\label{prop:normalizer:Cartan}
Each linear transformation of $\Aut(\LieA(G))$ preserves the Killing 
form. That is, $\Aut(\LieA(G))$ is a subgroup of the orthogonal 
transformation group with respect to the Killing form.
\end{proposition}

\smallskip

If there is a special subspace of $\LieA(G)$, it can be also used to 
restrict the form of $\phi$. For example, the center of $\LieA(G)$, 
$\C(\LieA(G))$, consisting of elements which commute with all elements 
of $\LieA(G)$, is preserved  by $\phi$ because $\C(\LieA(G))$ is the 
unique subspace with that property. Similarly, because 
$\phi(\LieA(G))=\LieA(G)$, 
$\phi([\LieA(G),\LieA(G)])=[\phi(\LieA(G)),\phi(\LieA(G))] 
=[\LieA(G),\LieA(G)]$. Thus

\smallskip

\begin{proposition}\label{prop:normalizer:invariantspace}
An automorphism of $\LieA(G)$ preserves its center and 
$[\LieA(G),\LieA(G)]$.
\end{proposition}

\smallskip

Let $\xi_I=\xi^i_I(x)\partial_i$ be a basis of $\LieA(G)$. Then the 
condition that a transformation $x'{}^i=f^i(x)$ induces the 
automorphism
\Eq{
\phi \xi_I=\xi_J\phi^J{}_I
}
is expressed as
\Eq{
\xi^j_I(x)\partial_j f^i(x)=\xi^i_J(f(x))\phi^J{}_I.
}
In general, it is not a easy task to solve this set of differential 
equations directly. However, special solutions are often found easily 
by guess or other methods. In such cases, the following proposition is 
very useful in determining the general solution to this equation.

\smallskip

\begin{proposition}\label{prop:normalizer:right} Let $G$ be a 
transformation group containing a simply transitive group $G_s$ on a 
connected manifold $M$.  If $f\in \N(G)$ induces the identity 
transformation of $\LieA(G_s)$, $f$ coincides with some right 
transformation $R_b$ with respect to $G_s$.
\end{proposition}

\begin{proof}
Let $g_t$ be a one-parameter subgroup of $G_s$. Then the infinitesimal 
transformation for $fg_tf^{-1}$ coincides with that for $g_t$ from the 
assumption. Hence $fg_tf^{-1}=g_t$. Since any element of the connected 
group $G_s$ is contained in such a subgroup, this implies that $f$ 
commute with any left transformation $L_a$. Let $o$ be a fixed point in 
$M$, and identify $G_s$ with $M$ by the diffeomorphism $F_o: G_s\maps 
M$ defined by $F_o(a)=L_a o$. Then, for any point $x$ in $M$, there 
exists a unique element $g_x$ in $G_s$ such that $F_o(g_x)=L_{g_x}o=x$. 
Now, let $b$ be an element in $G_s$ defined by $b=F_o^{-1}(f(o))$. 
Then, since the right transformation is defined by $R_bF_o(a)=F_o(ab)$, 
 
$R_b(o)=F_o(b)=f(o)$. Hence, $f(x)=fL_{g_x}( o)=L_{g_x} 
f(o)=L_{g_x}R_b(o)=R_bL_{g_x}(o)=R_b(x)$.
\end{proof}

\section{Conjugate transformations}\label{appendix:conj}

\subsection{$\Sol$}

\Eqr{
&& fL_{\bm{c}}f^{-1}=L_{f(\bm{c})} \text{\ for\ } 
f=\Set{1,J}D(k_1,k_2,1),\\
&& L_{\bm{a}}L_{\bm{c}}L^{-1}_{\bm{a}}=L_{\bm{c}'};
\bm{c}'=B(a^3)\bm{c}+\bm{a}-B(c^3)\bm{a},\\
&& J(-I_1)J=(-I_2),\ JI_3=I_3J, \\
&& L_{\bm{a}}(-I_1)L^{-1}_{\bm{a}}=(-I_1)L_{\bm{c}};\ 
\bm{c}=(-2a^1,0,0),\\
&& L_{\bm{a}}(-I_2)L^{-1}_{\bm{a}}=(-I_2)L_{\bm{c}};\ 
\bm{c}=(0,-2a^2,0),\\
&& L_{\bm{a}}I_3L^{-1}_{\bm{a}}=I_3 L_{-2\hat{\bm{a}}},\\
&& L_{\bm{a}}JL^{-1}_{\bm{a}}=JL_{\bm{c}};\ 
\bm{c}=(J-B(-2a^3))\bm{a},\\
&& R_{\bm{b}}(-I_1)R^{-1}_{\bm{b}}=(-I_1)R_{\bm{b}'};\ 
\bm{b}'= (-2b^1e^{b^3},0,0),\\
&& R_{\bm{b}}(-I_2)R^{-1}_{\bm{b}}=(-I_2)R_{\bm{b}'};\ 
\bm{b}'= (0,-2b^2e^{-b^3},0),\\
&& R_{\bm{b}}I_3R^{-1}_{\bm{b}}=I_3R_{\bm{b}'};\ 
\bm{b}'= (-2b^1e^{b^3},-2b^2e^{-b^3},0),\\
&& R_{\bm{b}}JR^{-1}_{\bm{b}}=JR_{\bm{b}'};\ 
\bm{b}'= (-(b^1-b^2)e^{b^3},(b^1-b^2)e^{-b^3},-2b^3).
}
%

\subsection{$H^2\times\RF$}

\paragraph{$\N(\BTIII_L)$}:
\Eqr{
&& L_{(a,b,c)}L_{(x,y,z)}L_{(a,b,c)}^{-1}= L_{(a+bx-ay,y,z)},\\&& 
(-I_1)L_{(x,y,z)}(-I_1)=L_{(-x,y,z)},\\
&& I_2L_{(x,y,z)}I_2=L_{(-x,y,-z)},\\
&& (-I_3)L_{(x,y,z)}(-I_3)=L_{(x,y,-z)},\\
&& N'(d)L_{(x,y,z)}N'(d)^{-1}=L_{(x,y,z+d\ln y)},\\
&& D(1,1,k)L_{(x,y,z)}D(1,1,k)^{-1}=L_{(x,y,kz)},\\
&& D(1,1,k)L_{\bm{a}}=L_{\bm{a}}D(1,1,k),\\ 
&& L_{(a,b,c)}(-I_1)L_{(a,b,c)}^{-1}=(-I_1)L_{(-2a,0,0)},\\
&& L_{(a,b,c)}I_2L_{(a,b,c)}^{-1}=I_2L_{(-2a,0,-2c)},\\
&& L_{(a,b,c)}(-I_3)L_{(a,b,c)}^{-1}=(-I_3)L_{(0,0,-2c)},\\
&& (-I_1)N'(d)=N'(d)(-I_1).
}

\paragraph{$\N(\Isom_0(H^2\times\RF))$}:
\Eqr{
&& 
L_{(a,b,c)}R_H(\theta)L_{(a,b,c)}^{-1}
=Z_L\left(R(\theta)*\gamma \right)
R_H\left(H(\theta,-\bar\gamma)\right);\nonumber\\
&&\qquad \gamma=\frac{-a+i}{b},\\ 
&& R_H(\theta)L_{(a,b,c)}R_H(\theta)^{-1}= 
Z_L(R(\theta)*(a+ib))R_H(H(\theta,\alpha)-\theta),\\
&& R_H(\theta)(-I_1)R_H(\theta)^{-1}=(-I_1)R_H(-2\theta),\\
&& R_H(\theta)I_2R_H(\theta)^{-1}=I_2R_H(-2\theta),\\
&& R_H(\theta)(-I_3)=(-I_3)R_H(\theta),\\
&& R_H(\theta)D(1,1,k)=D(1,1,k)R_H(\theta).
}
%

\subsection{$\TSL$}

\paragraph{$\N(\BTVIII_L)$}:
\Eqr{
&& L_{(\alpha,c)}L_{(\zeta,z)}L_{(\alpha,c)}^{-1}= L_{(\zeta',z')};\\
&& \quad 
\zeta'=Z(\alpha)*Z(R(c)*\zeta)*R(H(c,\zeta)+z-c)*\frac{i-a}{b},
\nonumber\\  
&& \quad z'=H\left(H(c,\zeta)+z-c, \frac{i-a}{b}\right),
\nonumber\\
&& I_2L_{(\zeta,z)}I_2=L_{(-\bar\zeta,-z)},\\
&& I_2R_{(\zeta,z)}I_2=R_{(-\bar\zeta,-z)},\\
&& L_{(\alpha,c)}I_2L_{(\alpha,c)}^{-1}=I_2L_{(\alpha',c')},\\
&& R_{(\alpha,c)}I_2R_{(\alpha,c)}^{-1}=I_2R_{(\alpha',c')};\\
&& \quad \alpha'=-a+bR(-2c)*\frac{-a+i}{b},\nonumber\\
&& \quad c'=H(-c,\gamma)-H(c,\gamma),\nonumber\\
&& \quad \gamma=R(-c)*\frac{-a+i}{b}.\nonumber
}

\paragraph{$\N(\Isom_0(\TSL))$}:
\Eq{
R_{(i,\theta)}I_2R_{(i,\theta)}^{-1}=I_2R_{(i,-2\theta)}.
}
%



\end{document}